\documentclass[a4paper,12pt]{article}

\pdfoutput=1

\usepackage{jcappub}
\usepackage[english]{babel}
\usepackage{graphicx,subfig}
\usepackage{a4wide}
\usepackage{amsfonts,amsmath,amssymb,euscript,bm}
\usepackage{color,comment}

\newcommand{\eq}{ \; = \; }
\newcommand{\eeq}{\; \equiv \;}
\newcommand{\defi}{ \; \equiv \; }
\newcommand{\cO}[1]{\mathcal{O}(#1)}

\newcommand{\dirac}{ \delta_{D}^{(3)} }
\newcommand{\hubble}{ \mathcal{H} }
\newcommand{\cH}{ \mathcal{H} }
\newcommand{\kNL}{ k_\mathrm{NL} }
\newcommand{\EV}[2]{ {\big\langle} #1 \, #2 {\big\rangle} }
\newcommand{\EVb}[3]{ {\big\langle} #1 \, #2 \, #3 {\big\rangle} }

\definecolor{orange}{rgb}{1,0.4,0}

\newcommand{\be}{\begin{eqnarray} }
\newcommand{\ee}{\end{eqnarray} }
\newcommand{\bs}{\begin{split} }
\newcommand{\es}{\end{split} }
\newcommand{\non}{\nonumber \\}
\newcommand{\dn}[1]{\delta_{(#1)} }
\newcommand{\tn}[1]{\theta_{(#1)} }
\newcommand{\kv}{{\bm k} }
\newcommand{\km}{k_\textit{max}}
\newcommand{\qv}{{\bm q} }
\newcommand{\vv}{{\bm v} }
\newcommand{\xv}{{\bm x} }
\newcommand{\yv}{{\bm y} }
\newcommand{\piv}{{\bm \pi} }
\newcommand{\intq}{\int_\qv }
\newcommand{\intqn}[1]{\int_{\qv_{#1} } }

\newcommand{\smooth}[1]{\left[ #1 \right]_\Lambda }
\newcommand{\hMpc}{h^{-1}\, \text{Mpc}}
\newcommand{\hMpcsq}{h^{-2}\, \text{Mpc}^2}
\newcommand{\ihMpc}{h\, \text{Mpc}^{-1}}
\newcommand{\Pl}{P_\textit{lin} }
\newcommand{\n}{ {\bm n} }

\def\d{\partial}
\def\lap{\bigtriangleup}
\def\O{\mathcal{O}}

\renewcommand{\comment}[1]{}
\newcommand{\expect}[1]{\left\langle #1 \right\rangle}

\allowdisplaybreaks[1]

\title{\boldmath The Bispectrum in the Effective Field Theory of Large Scale Structure}

\author[a]{Tobias Baldauf,}
\author[b]{Lorenzo Mercolli,}
\author[a]{Mehrdad Mirbabayi,}
\author[c]{and Enrico Pajer}

\affiliation[a]{Institute for Advanced Study, Einstein Drive, Princeton, NJ 08540, USA}
\affiliation[b]{Department of Astrophysical Sciences, Princeton University, \\ Peyton Hall, Princeton, NJ 08544, USA}
\affiliation[c]{Department of Physics, Princeton University, \\ Jadwin Hall, Princeton, NJ 08544, USA}

\abstract{We study the bispectrum in the Effective Field Theory of Large Scale Structure, consistently accounting for the effects of short-scale dynamics. We begin by proving that, as long as the theory is perturbative, it can be formulated to arbitrary order using only operators that are local in time. We then derive all the new operators required to cancel the UV-divergences and obtain a physically meaningful prediction for the one-loop bispectrum. In addition to new, subleading stochastic noises and the viscosity term needed for the one-loop power spectrum, we find three new effective operators. The three new parameters can be constrained by comparing with $N$-body simulations. The best fit is precisely what is suggested by the structure of UV-divergences, hence justifying a formula for the EFTofLSS bispectrum whose only fitting parameter is already fixed by the power spectrum. This result predicts the bispectrum of $N$-body simulations up to $k_{max}\approx0.22\, h\, \text{Mpc}^{-1}$ at $z=0$, an improvement by nearly a factor of two as compared to one-loop standard perturbation theory.}

\begin{document}

\maketitle

\newpage
\begin{table}[h!]
\resizebox{\textwidth}{!}{

\begin{tabular}{|c|l|l|c|}  \hline
Symbol    & Relation &	Meaning 	&  Equation\\ \hline\hline
$a$	  & 	& scale factor 	&		\\ \hline
$\tau$ & $a \, d\tau=dt$ & conformal time & \\ \hline
$\Omega_m^0$ &  & matter density in units of the critical density at $z=0$ & \\ \hline
$\hubble$ &$\equiv d \ln (a)/ d\tau$ & conformal Hubble parameter & \eqref{e:HubbleLCDM} \\ \hline
$ f $ & & phase space distribution & \eqref{e:Boltzmann}  \\ \hline
$ \dn{n} \;,  \tn{n} $ & & density contrast and velocity divergence in SPT at order $ n $ & \eqref{e:dn} \\ \hline
$ \phi $ & & Newtonian potential & \eqref{e:EOMrealspace} \\ \hline
$ \Phi $ & $\Delta \Phi=\delta  $  & rescaled Newtonian potential & \eqref{Phi} \\ \hline
$ u $ &$\partial^{i} u= v^{i} \;, \; \partial_i v^i = \theta  $ & velocity potential & \eqref{Phi} \\ \hline
$ s_{ij} $ & $\equiv \partial_i \partial_j \Phi - \frac{1}{3} \delta_{ij} \, \bigtriangleup \Phi   $& tidal tensor & \eqref{defs} \\ \hline
$ \tau_{\theta} $ & & EFT source in the Euler equation & \eqref{tautheta} \\ \hline
$ \mathcal{S}_{\alpha,\beta} $ & & SPT quadratic source terms & \eqref{e:Salphabeta} \\ \hline
$ D_1(a) $ & & linear growth factor & \eqref{e:linsol} \\ \hline
$\delta_{(1)}^{c}\;,\;  \delta_{(2)}^{c} $ & & leading- and next-to-leading order viscosity correction & \eqref{e:deltac2epsilongamma} \\ \hline
$ d^2 $ &  $\equiv c_2^2 + c_v^2  $  & leading order viscosity parameter  & \eqref{e:defd}, \eqref{e:tauvisLO} \\ \hline
$ e_{1,2,3} $ &   & next-to-leading order viscosity parameters  & \eqref{e:tauvisNLO} \\ \hline
$ \bar{d}^2 $ &  $\equiv d^2 /(D_1^m(a) \hubble_0^2 \Omega_m^0)$  & constant leading order viscosity parameter  &  \eqref{e:constpar} \\ \hline
$ \bar{e}_{1,2,3} $ &  $\equiv e_{1,2,3} /(D_1^m(a) \hubble_0^2 \Omega_m^0)$  & constant next-to-leading order viscosity parameters  & \eqref{e:constpar} \\ \hline
$ \gamma $ &  $=- g_{1}(a,m) \bar d^{2}  $  & leading order parameter  & \eqref{dtogamma}, \eqref{g1} \\ \hline
$ \epsilon_{1,2,3} $ & $ =-g_{2}^{e}(a,m) \bar e_{1,2,3}  $  & next-to-leading order parameters  & \eqref{e2e}, \eqref{g2e} \\ \hline
$ F_{2}^{c} $ & & total quadratic EFT kernel & \eqref{e:Fc2} \\ \hline
$ E_{1,2,3} $ & & new quadratic EFT kernels & \eqref{e:F2tau} \\ \hline
$ F_{2}^{\alpha\beta}, \bar F_{2}^{\alpha\beta} $ & & EFT kernel for $ d^{2} $ at second order & \eqref{e:Falphabeta} \\ \hline
$ F_{2}^{\delta}, \bar F_{2}^{\delta} $ & & EFT kernel for $ d^{2} $ at linear order & \eqref{e:Fdelta} \\ \hline
$ B_{c11,c21} $ &  & EFT contributions to the bispectrum & \eqref{e:bc11},\eqref{e:bc21} \\ \hline
\end{tabular}
}
\end{table}

\section{Introduction}

The Large Scale Structure of the universe (LSS), i.e.~the late time distribution of matter and galaxies on cosmological scales, has the potential to widen our knowledge about the origin and composition of the universe.
While the analysis of the temperature fluctuations in the Cosmic Microwave Background has provided us with a lot of invaluable information, the power of this linear observable of cosmic density fluctuations is limited due to its fixed emission time and its two dimensional nature.

The three dimensional distribution of matter at late times encodes information about all elements of our cosmological model. In particular it will be probed to unprecedented precision in the near future, potentially providing us the key to understand the nature of Dark Energy and Dark Matter and elucidate the mechanism that generate primordial initial conditions. Extracting information from the late time observables is hampered by their non-linear nature. 
While numerical simulations have been very successful in predicting the outcome of the non-linear gravitational clustering process, we would like to push analytic descriptions as far as possible. Analytical treatments both facilitate numerical inference algorithms and help us understand the clustering process and avoid possible confusions between mundane non-linearities and the necessity for extensions of the cosmological standard model.

Perturbative treatments have provided valuable insight into the gravitational clustering process on cosmological scales. Two notable implementations are the Eulerian Standard Perturbation Theory (SPT) and Lagrangian Perturbation Theory (LPT). For a thorough review on these two approaches see Ref.~\cite{Bernardeau2002}. However, both of these theories are incomplete since (i) they do not make explicit the expansion parameter, (ii) they typically are limited to the perfect, pressureless fluid approximation, and (iii) they are plagued by UV-divergent integrals (or equivalently they depend on a high scale cut-off). These issues result in fundamental limitations to reliably predict the dynamics of mildly non-linear scales, which encode lots of information about cosmology.

The powerful tool of Effective Field Theories (EFTs), widely used in high-energy physics, has been adapted to the situation of LSS in order to overcome the problems of the standard approaches. The strength of EFT is that it fully exploits the symmetries of a system in order to account for possible small scale effects. In Refs.~\cite{Baumann2012,Carrasco2012} the Effective Field Theory of Large Scale Structure (EFTofLSS) has been formulated (see Ref.~\cite{Pietroni2012} for a similar approach), allowing for a well-defined formulation of perturbation theory for LSS. The general idea is to restrict the validity of the perturbation theory to scales larger than the non-linear scale where the dynamics becomes fully non-linear and an analytical approach is bound to fail. In general, however, the small scale dynamics couples also to the long wavelength modes and we need to consistently integrate out the small scale modes. This generates an effective stress tensor which parametrizes the response of the long modes to short modes in a set of free (time-dependent) coefficients that are not determined by the theory itself and which encode our ignorance towards the non-linear physics. They need to be fixed by numerical simulations or observations.

After the original formulation, Refs.~\cite{Hertzberg2012,Pajer2013a,Carrasco2013,Mercolli2013,Carrasco2013a,Carroll2013,Porto2013,Assassi2014,Senatore2014} have been exploring various aspects of the EFTofLSS (see also Refs.~\cite{Uhlemann2014,Rigopoulos2014}). While previous work focusses on the two-point correlators, our aim is to extend the literature on the EFTofLSS by investigating the bispectrum. 
In this paper we focus on the bispectrum but we also deepen the general understanding of the EFTofLSS in various ways. The bispectrum contains a large amount of information thanks to its many independent configurations. Also, it is the main probe of primordial non-Gaussianity, which is one of the most promising ways of discriminating among different models of the very early universe. In this work we limit ourselves to Gaussian initial conditions, and differ the inclusion of primordial non-Gaussianity to future work. Our main results and findings are summarized in the following.

\paragraph{Locality in time:} It was noticed in Refs.~\cite{Carrasco2013a,Carroll2013} that since before virialization the short scales evolve on a time-scale comparable with that of large scales, namely the Hubble time, they have a long memory and their back-reaction on the large-scale dynamics leads to an EFT that is non-local in time. We show that, as long as the perturbation theory is valid, the theory can be reformulated in such a way that only local-in-time operators appear at each order. We also notice that the effect of the long memory seems to introduce spatially non-local terms. However, they appear only starting at cubic order, which is higher than what we need for the one-loop bispectrum.

\paragraph{EFTofLSS at the next-to-leading order:} The discussion of the density bispectrum in the EFTofLSS requires going beyond the leading order (LO) approximation of the effective stress tensor. Based on the above considerations we can restrict ourselves to a local expansion of the effective stress tensor. We derive the viscosity and noise terms at next-to-leading order (NLO) in the expansion of the effective stress tensor. At the order we are working and for $ \Lambda $CDM, three new operators become relevant, we choose them to be $ \bigtriangleup \delta^{2}$,  $\bigtriangleup s^{2} \equiv \bigtriangleup (s_{ij}s^{ij} )$ and $ \partial_{i} (s^{ij}\partial_{j}\delta) $, with $ s_{ij} $ the tidal field defined in \eqref{defs}.

\paragraph{The one-loop bispectrum:} We compute the one-loop as well as the full NLO EFT contributions to the bispectrum and show that all UV-divergences are cancelled through a renormalization of the free parameters in the effective stress tensor. Due to the shape dependence of the UV-divergences this is a non-trivial check for the NLO expression of the stress tensor. We show that the LO counterterm is exactly the same as predicted by the renormalization of the one-loop power spectrum. Furthermore, we discuss the interplay of LO and NLO counterterms and discuss how the renormalization of the EFTofLSS at higher orders can be conveniently organized.

\paragraph{Numerical simulations:} We numerically evaluate the one-loop power- and bispectrum for a $\Lambda$CDM cosmology and compare them with a suite of $N$-body simulations. The theory has four fitting parameters, one of which can be fixed by the power spectrum (as we also re-do using our simulations). Theoretical considerations suggest to relate the three new NLO parameters to the LO parameter using the structure of the UV-divergences. This leads to a formula for the EFTofLSS bispectrum (see Eqs.~\eqref{e:e123values}, \eqref{BEFT} and \eqref{magic}) with a single parameter $\gamma $ (corresponding to the LO speed of sound or viscosity) that can be determined through the power spectrum, i.e. without invoking the bispectrum data. This ``zero-parameter'' formula agrees with simulations up to  $\km \approx 0.22 \,\ihMpc$, a significant improvement with respect to $\km \approx 0.13\, \ihMpc$ in the case of SPT. If we let any subset of the four bispectrum fitting parameters float, we find that the fit does not improve compared to the zero-parameter prediction. This suggest that to push the fit to higher wavenumbers one needs to go to higher order in both SPT and the EFT corrections. \\[2ex]

This paper is structured as follows. In Sec.~\ref{s:oneloopbi}, we introduce our notation by recalling the equations of motion of the EFTofLSS. Although we consider the UV-limit of the one-loop bispectrum in some detail, the reader familiar with the topic might want to skip this section at first. Next, in Sec.~\ref{s:tau} we show how the theory can be recast in a local-in-time form and explicitly derive the second order effective operators.
Sec.~\ref{s:eftbispectrum} deals with the EFT contribution to the bispectrum and the renormalization of the UV-divergences. This is a somewhat technical section where we consider the subtleties involved in computing the EFT solution at second order. Finally, in Sec.~\ref{s:simulations} we present our results for the comparison of the EFTofLSS with numerical simulations. We conclude in Sec.~\ref{concl}.

\section{The one-loop bispectrum} \label{s:oneloopbi}

In this section, we review the equations of motion of the EFTofLSS and consider the bispectrum in SPT. We investigate the UV-limits of the one-loop integrals and derive the general form of the bispectrum in a scale free universe. This shows how the various contributions scale with the wavenumber $k$.

\subsection{Equations of motion and SPT}\label{s:SPT}

We will consider the bispectrum both in $\Lambda$CDM and Einstein-de Sitter (EdS) cosmologies. The conformal Hubble parameter $\cH$ is given through the Friedmann equation 

\begin{equation}\label{e:HubbleLCDM}
\cH^2 \eq  a^2 \cH^2_0 \, \left\{  \, \frac{\Omega_m^0}{a^3} + \Omega_\Lambda^0 \right\} \;,
\end{equation} 
for a flat Friedmann-Lema\^{\i}tre-Robertson-Walker universe with scale factor $a$ (normalized to $a_0=1$ at the present epoch) and filled with non-relativistic matter $\Omega_m^0$ and a cosmological constant $\Omega_\Lambda^0$. The index ``$0$" denotes the present epoch. 
Whenever we give quantitative results for the $\Lambda$CDM cosmology, we will assume the following set of cosmological parameters: $\Omega_m^0=0.272$, $\Omega_\Lambda^0=0.728$, $h=0.704$, 
$\sigma_8=0.81$ and $n_s=0.967$.
We shall often consider the limit of a matter dominated EdS universe with $\Omega_m=1$. In this case, $\cH$ reduces to the simple form 

\begin{equation}\label{e:hubbleEdS}
\cH \eq \frac{\cH_0}{\sqrt{a}} \;, 
\end{equation}
Note that when comparing EdS to $\Lambda$CDM one has to be careful with the definition of $\cH_0$. Whenever generalizing an expression from EdS to $\Lambda$CDM, $\cH_0$ has to be rescaled by a factor $\sqrt{\Omega_m^0}$ since we want $\cH$ to be the same at the present epoch. We shall use $\tau$ as the conformal time which is defined as $dt = a \,d\tau$ with respect to the physical time $t$.

Let us start by briefly reviewing the derivation of the equations of motion for the EFTofLSS and its relation to SPT (see Refs.~\cite{Baumann2012,Carrasco2012,Hertzberg2012} for more details). We assume that Dark Matter can be described as an ensemble of $N$ non-relativistic and collisionless point particles with equal mass $m$. We work in Newtonian cosmology, i.e.\ no relativistic effects are considered. The temporal evolution of this ensemble of point particles is described by the Boltzmann equation for the total phase space density $f(\xv, \kv)=\sum_{i=1}^N f_i(\xv, \kv)$

\begin{equation}\label{e:Boltzmann}
\partial_{\tau} f(\xv, \kv) + \frac{1}{a \, m} \, \kv \cdot \nabla_x \, f(\xv, \kv) - a\, m  \sum_{i, j ;i \neq j}^N  \nabla_{k} \, f_i(\xv, \kv) \cdot \nabla_x \, \phi_{j}(\xv)  \eq 0 \;,
\end{equation}
where $\xv$ are comoving coordinates. Taking the first two moments of $f(\xv, \kv)$, allows us to define the matter and momentum density $\tilde{\rho}$ and $\tilde{\piv}$ of the system $N$-particle system

\begin{equation}\label{e:unsmoothed}
\tilde{\rho} (\xv,\tau)\equiv \int d^3 q \; f(\xv,\qv) \;, \qquad \quad \tilde{\piv} (\xv,\tau) \equiv  \int d^3 q  \; \qv \, f(\xv, \qv)\;. 
\end{equation}
It is then straight forward to derive the equations of motion for $\tilde{\rho}$ and $\tilde{\piv}$ and it turns out that they are simply the continuity and the Euler equation of an imperfect fluid.

The ultimate goal is to find an analytic description of the dynamics at large scales, i.e.~in the mildly non-linear regime where linear dynamics are dominant. The non-linear scale $\kNL$ (in momentum space) acts as a scale that separates the fully non-linear physics at small scales from the large scale dynamics. The situation where one is interested only in a limited range of scales is encountered in many areas of physics and is most conveniently tackled using EFT techniques. EFT allows us to consistently compute non-linear corrections to the linear solution on large scales but without ignoring possible effects due to the short scale dynamics. The problem is that whenever we consider non-linear solutions of the equation of motion, all scales couple together. In particular, modes that are smaller than $\kNL$ can couple to modes that are larger than $\kNL$. The EFT framework helps us to account for these effects systematically without forcing us to consider the full non-linear solution of the equations of moiton.

The first step towards the equations of motion in the EFTofLSS is to introduce a smoothing procedure. The smoothing over some cut-off scale $\Lambda^{-1}$ can be done using a window function $W_\Lambda$ which is e.g.~a Gaussian or top hat function. This is commonly referred to as ``regularizing" the theory and it means that all loop integrals are made manifestly convergent. The cut-off scale $\Lambda$ is artificial and does not have any physical meaning. Upon renormalization, the dependence of all physical quantities on $\Lambda$ is absorbed in the free ``bare'' parameters of the theory (for details see Sec.~\ref{s:eftbispectrum} and Refs.~\cite{Pajer2013a,Baumann2012}). In the end, all physical predictions do not depend on $\Lambda$ nor on the specific form of $W_\Lambda$. After fixing the renormalized value of effective parameters, a new physical scale emerges: the non-linear scale $\kNL$. This scale indicates when the dynamics becomes fully non-linear and perturbation theory breaks down. The non-linear scale $ \kNL $ will be determined in Sec.~\ref{s:simulations} when we compare the theoretical predictions to numerical simulations. Applying the smoothing to the quantities in Eq.~\eqref{e:unsmoothed}

\begin{equation}
\begin{split}
& \rho(\xv,\tau) \eeq \smooth{\tilde{\rho}} \eq \int d^3y W_\Lambda ( \xv - \yv) \tilde{\rho}(\yv,\tau) \;,  \\[1.5ex]
&  \piv(\xv,\tau) \eeq \smooth{\tilde{\piv}} \eq \int d^3y W_\Lambda ( \xv - \yv) \tilde{\piv}(\yv,\tau) \;,
\end{split}
\end{equation}
we ensure that $\rho$ and $\piv$ now only depend on wave vectors that are smaller than $\Lambda$ (we use the same notation as Ref.~\cite{Baumann2012} for the smoothing of a quantity). Applying the smoothing to the Boltzmann equation, we can derive the equations of motion for the smoothed quantities

\begin{equation}\label{e:EOMrealspace}
\begin{split}
& \partial_{\tau}\delta + \partial_i  \left[ (1+\delta)  v^i \right] \eq 0 \;, \\[1.5ex]
& \partial_{\tau}v^i + \hubble \, v_{l}^i + \partial^i \phi + v_{l}^j \, \partial_j v^i  \eq  - \frac{1}{a \, \rho }\, \partial_j  \tau^{ij} \;, \\[1.5ex]
& \bigtriangleup \phi  \eq \frac{3}{2} \hubble^2 \Omega_m \, \delta \;.
\end{split}
\end{equation}
$\phi$ is the Newtonian potential which is generated by the (smoothed) density contrast $\delta \equiv \rho/\bar{\rho} - 1$, where $\bar{\rho}$ is the time dependent background density, and the velocity field $\vv$ is defined as $\vv \equiv \piv/\rho + \textit{counterterms}$  \footnote{As discussed in Ref.\ \cite{Mercolli2013}, the definition of the velocity involves counterterms which are needed to renormalize the velocity correlators. The finite part of these counterterms, however, can be set to zero. We can therefore safely ignore these counterterms for the rest of this paper (see also Ref.\ \cite{Carrasco2013a}). 
}. Let us stress again that all quantities in Eq.~\eqref{e:EOMrealspace} contain only long wavelength modes. The stress tensor $\tau^{ij}$ that enters in the Euler equation is a complicated function of the Newtonian potential (see e.g.\ Eqs.\ $(34) - (36)$ of Ref.\ \cite{Hertzberg2012}) and, as we shall see, plays a crucial role in the renormalization of loop integrals. 
Throughout this work we will neglect the vorticity since at the linear order it decays as $\sim 1/a$ (see Refs.~\cite{Mercolli2013,Carrasco2013a} for a discussion of the vorticity in the EFTofLSS). It is therefore useful to rewrite the Euler equation for the velocity divergence $\theta \equiv \nabla \cdot \vv$ 

\begin{equation}\label{e:EOMtheta}
\partial_\tau \theta + \cH \, \theta+ v^j \partial_j \theta + \partial_i v^j \partial_j v^i + \bigtriangleup \phi \eq \tau_\theta \;, 
\end{equation} 
where we defined 

\begin{equation}\label{tautheta}
\tau_\theta \eeq  - \partial_i \left[ \frac{1}{a \, \rho} \, \partial_j \tau^{ij}  \right] 
\end{equation}
 and $\bigtriangleup \equiv \nabla \cdot \nabla = \partial_i \partial^i $.

The quest of finding a perturbative solution for the equations of motion in Eq.~\eqref{e:EOMrealspace} is not new, in particular in the approximation where the effective stress tensor is neglected. This is what is usually called SPT. For a thorough review on the subject see Ref.~\cite{Bernardeau2002}. Here, we shall merely recall well known results and introduce our notation. It is most convenient to rewrite Eqs.~\eqref{e:EOMrealspace} and \eqref{e:EOMtheta} in Fourier space 

\begin{equation}\label{e:EOM}
\begin{split}
& \partial_\tau \delta(\kv, \tau) + \theta(\kv, \tau) \eq \mathcal{S}_\alpha(\kv, \tau)  \;, \\ 
& \partial_\tau  \theta(\kv, \tau) + \cH \,  \theta(\kv, \tau) + \frac{3}{2} \Omega_m \cH^2  \delta (\kv, \tau) \eq \mathcal{S}_\beta (\kv , \tau) \;, 
\end{split}
\end{equation}
where, in a slight abuse of notation, we used the same notation for the fields in Fourier space as in real space. We will mostly work in Fourier space and we shall often drop the arguments of the fields for streamlining the notation. The two source terms $\mathcal{S}_\alpha$ and $\mathcal{S}_\beta$ contain the non-linear terms of the equations of motion as well as the effective stress tensor 

\begin{equation}\label{e:Salphabeta}
\begin{split}
\mathcal{S}_\alpha (\kv, \tau) & \eeq - \int_\qv \, \alpha(\qv , \kv - \qv) \, \theta(\qv, \tau) \delta(\kv -\qv, \tau) \;, \\[1.5ex]
\mathcal{S}_\beta (\kv, \tau) & \eeq - \int_\qv \, \beta(\qv , \kv - \qv) \, \theta(\qv, \tau) \theta(\kv -\qv, \tau) + \tau_\theta (\kv, \tau) \;. 
\end{split}
\end{equation}  
We use the abbreviation $\int_\qv \equiv \int d^3 q/ (2\pi)^3$ and denote the absolute value of a vector as $k = |\kv|$. The kernel functions $\alpha$ and $\beta$ are

\begin{equation}
\alpha(\qv_1, \qv_2) \eeq \frac{\qv_1 \cdot (\qv_1 + \qv_2) }{q_1^2} \;, \qquad \qquad \beta (\qv_1, \qv_2) \eeq \frac{1}{2} \, (\qv_1 + \qv_2)^2 \, \frac{\qv_1 \cdot \qv_2}{q_1^2 \, q_2^2}  \;.
\end{equation}
Replacing the time derivative by derivatives with respect to the scale factor $a$ ($\partial_\tau = \cH a \partial_a $), we can rewrite the differential equations for $\delta$ and $\theta$ for a $\Lambda$CDM universe as

\begin{equation}\label{e:EOMdiag}
\begin{split}
& \cH^2 \left\{ - a^2  \partial_a^2  + \frac{3}{2} \left( \Omega_m - 2\right) a \partial_a  + \frac{3}{2} \Omega_m \right\} \delta  \eq \mathcal{S}_\beta - \cH\,  \partial_a \, \big( a\, \mathcal{S}_\alpha  \big) \;, \\[1.5ex]
& \cH \left\{ a^2  \partial_a^2 +\left( 4-\frac{3}{2}\Omega_m \right) a\partial_a + \left(2-3\Omega_m \right) \right\} \theta \eq \partial_a \, \big( a\, \mathcal{S}_\beta  \big) - \frac{3}{2} \Omega_m \cH \mathcal{S}_\alpha \;.
\end{split}
\end{equation}
The equations of motion for the EdS case are easily recovered by setting $\Omega_m=1$. The linear solution of these equations is simply the linear combination of a growing and a decaying mode. Clearly, we are interested only in the growing mode

\begin{equation}
\dn{1} (\kv, a)  \eq D_1 (a)\, \delta_1 (\kv) \;, 
\end{equation}
where $\delta_1$ is a Gaussian random field that describes the initial conditions of the density field and $D_{1}$ is the usual growth factor 
\begin{equation}
D_1 (a) \eq  \frac{5}{2} \Omega_m^0 \cH^2_0  \, \frac{\cH}{a} \int_{0}^a da' \; \frac{1}{\cH^3} \;,
\label{e:linsol}
\end{equation} 
which reduces to $D_1 = a$ in EdS. It is useful to consider the Green's function of Eq.~\eqref{e:EOMdiag}, which is obtained by replacing the right hand side of Eq.~\eqref{e:EOMdiag} with a Dirac distribution $\delta_D(a-a')$. The Green's function for $\delta$ in $\Lambda$CDM is given by

\begin{equation}\label{e:greens}
G_{\delta} (a, a')  \eq \Theta (a - a') \, \frac{2}{5} \frac{1 }{ \cH_0^2 \Omega_{m}^0 } \frac{ D_1(a')}{a'}  \, \left\{  \frac{D_{-}(a)}{D_{-}(a')}- \frac{D_1(a)}{D_1(a')}\right\} \;, 
\end{equation}
The Green's function for $\theta$ in EdS is simply $G_\theta = - \hubble G_\delta $. The exact solution for the decaying mode $D_{-} = \cH/(a \cH_0)$ can in principle be approximated as $D_{-}  \approx D_1^{-3/2}$. The function $f(a)$ which will be used occasionally, is the logarithmic derivative of the growth factor $f(a) \equiv d \ln D_1 / d  \ln a$ and is $f(a) =1$ in EdS.

\begin{figure}
\centering
\includegraphics[width=5cm]{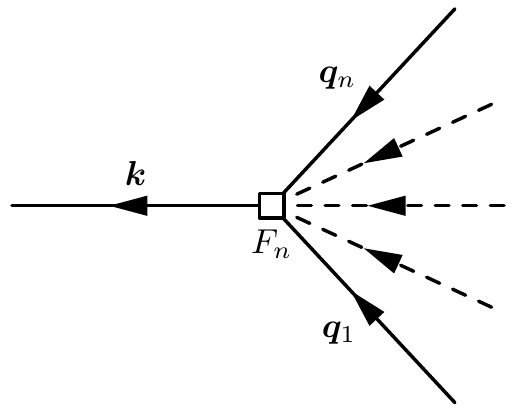}
\caption{SPT vertex.}
\label{fig:sptvertex}
\end{figure}

On large scales, it is safe to assume that the linear solution is dominant and that the density contrast is small, i.e.~$\delta_{(1)}<1$. In particular, this is the case for the smoothed fields. This allows us to solve the equations of motion perturbatively. In the absence of an effective stress tensor, we recover the results form SPT where the solution of the equations of motion is written as a series in powers of $\delta_1$

\begin{equation}\label{e:SPTsol}
\delta(\kv,a) \eq  \sum_{i=1}^\infty \delta_{(i)} (\kv,a) \;, \quad \qquad  \theta(\kv,a) \eq  - \cH \, f(a) \sum_{i=1}^\infty \theta_{(i)} (\kv,a) \;.
\end{equation}
In $\Lambda$CDM, the growth factor of the $n$-th order solution has to be computed at every given order using Eqs.~\eqref{e:linsol} and \eqref{e:greens}. However, it is possible to write the $n$-th order growth factor as the $n$-th power of $D_1$

\begin{equation}\label{e:dn}
\dn{n} (\kv,a) \eq D_1^n(a) \, \delta_{n}(\kv) \;, \qquad \quad \tn{n} (\kv,a) \eq D_1^n(a) \,\theta_{n}(\kv) \;.
\end{equation} 
In the limit of $\Omega_m =1$, the above solution is exact, i.e.~the $n$-th order solution scales exactly as $a^n$. The approximation in \eqref{e:dn} is valid at the $ 1\% $ level of accuracy up to third order as pointed out in Ref.~\cite{Takahashi:2008yk}.\footnote{We checked that with the Green's function of Eq.~\eqref{e:greens} the difference between the (exact) second order growth factor and $D_1(a)^2$ is at the $\sim 0.1\%$ level. However, replacing $D_{-}  \approx D_1^{-3/2}$ inside the Green's function increases this difference to $\sim 4\%$ at late times.} The momentum dependence is given in terms of a convolution of powers of $\delta_{1}$

\begin{equation}\label{e:fngn}
\begin{split}
& \delta_{n} (\kv)\eq \intqn{1} ... \intqn{n} (2\pi)^3 \dirac(\kv - \qv_1  ... - \qv_n ) \; F_n(\qv_1, ..., \qv_n) \, \delta_{1} (\qv_1) ... \delta_{1} (\qv_n) \;,  \\[1.5ex]
& \theta_{n} (\kv)\eq \intqn{1} ... \intqn{n} (2\pi)^3 \dirac(\kv - \qv_1  ... - \qv_n ) \; G_n(\qv_1, ..., \qv_n) \, \delta_{1} (\qv_1) ... \delta_1 (\qv_n) \;, 
\end{split}
\end{equation}
where the symmetric kernel functions $F_n$ and $G_n$ are known and given e.g.~in Ref.~\cite{Bernardeau2002}. Note that $F_n$ and $G_n$ only depend on ratios of the momenta. A diagrammatic representation of SPT has been discussed in the literature. One usually represents the kernels $F_n$ and $G_n$ as a vertex to which one can attach $n$ external legs as is shown in Fig.~\ref{fig:sptvertex}. Note that as opposed to the diagrammatic language of renormalized perturbation theory (RPT, see Refs.~\cite{Crocce2006,Crocce2006a,Crocce2008,Matarrese2007}), we use already the time integrated kernels as vertices.\footnote{Note that in renormalized perturbation theory ``renormalization'' does not refer to the cancellation of UV-divergences as in EFTofLSS but to a procedure to include higher order contributions in SPT.}

For the following discussion, it is important to know how the kernels scale if one of the momenta becomes very large. It was noted in Ref.~\cite{Goroff1986} (see also Ref.~\cite{Bernardeau2002}) that the kernels obey scaling laws such as

\begin{equation}\label{e:Fscaling}
\lim_{q\to \infty}F_n(\kv_1,\ldots,\kv_{n-2},\qv,-\qv) \; \propto \; \frac{k^2}{q^2} \;,
\end{equation}
where $\kv=\kv_1+\ldots \kv_{n-2}$. For us it will turn out to be important that also for $F_2$ and $F_3$ a similar scaling holds when the sum of the arguments remains finite while one of the momenta goes to infinity, i.e.

\begin{equation}
\begin{split}
& \lim_{q \to \infty} F_2(-\qv, \qv + \kv) \; \propto \; \lim_{q \to \infty} F_2(-\qv + \kv_1, \qv + \kv_2) \; \propto \; \frac{k^2}{q^2}  \;, \\[1.5ex]
& \lim_{q \to \infty} F_3(-\qv , \qv+ \kv_1, \kv_2 ) \; \propto \; \frac{k^2}{q^2} \;, 
\end{split}
\end{equation}
where we assumed that the momenta $k_1 \sim k_2 \sim k$ are of the same order.

\subsection{The bispectrum in SPT}\label{s:sptbispectrum}

\begin{figure}
\centering
\includegraphics[width=16cm]{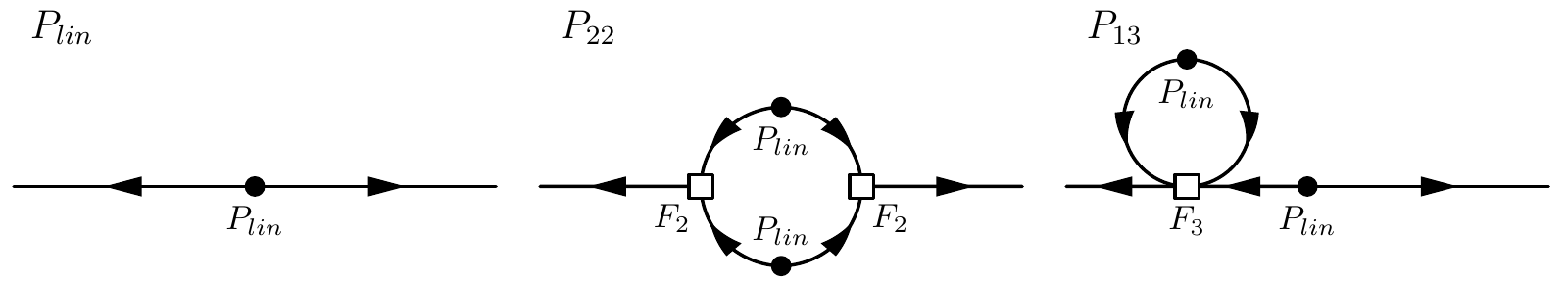}
\caption{Tree-level and one-loop power spectrum.} \label{f:ps}
\end{figure}

\begin{figure}
\centering
\includegraphics[width=16cm]{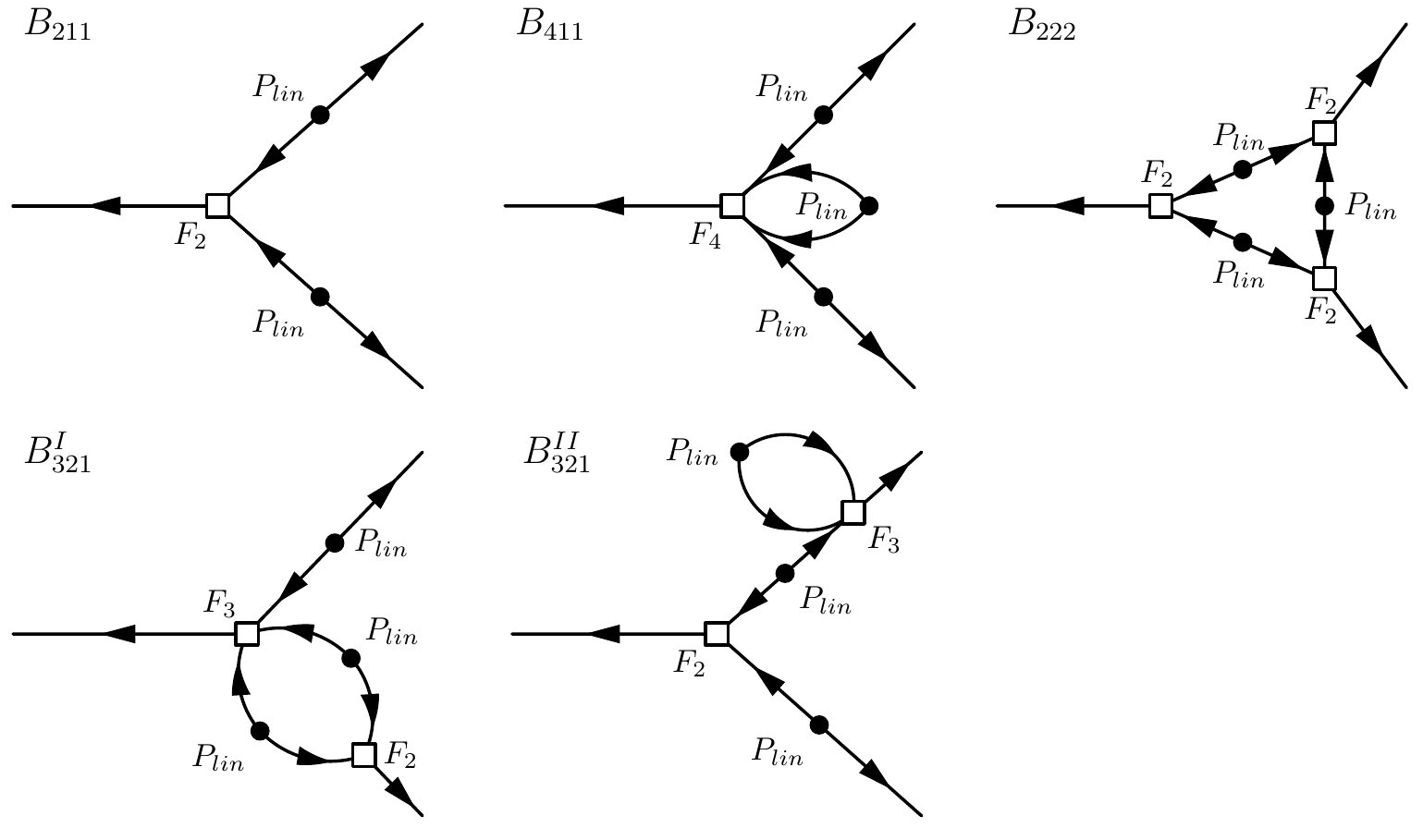}
\caption{Tree-level and one loop-bispectra.}\label{f:bs}
\label{fig:biloopdiag}
\end{figure}

Let us for the moment focus only on the SPT part of the equations of motion and postpone a detailed discussion of the effective stress tensor to Secs. \ref{s:tau} and \ref{s:eftbispectrum}. The two- and three-point connected correlators of the stochastic field $\delta$ are the quantities that we will consider in this paper. In Fourier space, the power- and bispectrum are defined as

\begin{equation}\label{e:p}
\langle \delta(\kv_1,a) \delta(\kv_2,a) \rangle \equiv (2\pi)^3 \dirac(\kv_1 + \kv_2) \, P(k_1,a) \;.
\end{equation}
and

\begin{equation}\label{e:defB}
\EVb{\delta(\kv_1,a)}{ \delta (\kv_2,a)}{ \delta (\kv_3,a) } \defi (2\pi)^3 \dirac (\kv_1 + \kv_2 + \kv_3 ) \, B(\kv_1,\kv_2,\kv_3, a)
\end{equation}
Because of the $\delta_D$-function, the bispectrum is not a function of three independent vectors. We will usually drop the time argument of $B$ and $P$ and write $B$ as a function of the three moduli of the momenta $B(k_1, k_2,k_3)$. The linear power spectrum $\Pl$ is then nothing but the two-point correlator of two $\dn{1}$ and it can be represented diagrammatically by a simple dot with two external lines as shown on the left in Fig.~\ref{f:ps}. The arrows show the direction of the momenta. Since we are considering only the case of Gaussian initial conditions, the correlator of three $\dn{1}$ is zero. The first non-trivial contribution stems from the first non-linear contribution to $\dn{1}$, i.e. $\dn{2}$, which gives us the tree-level bispectrum
\begin{equation}\label{e:blin}
B_{112}(k_1, k_2,k_3,a) \eq 2 F_2(\kv_1, \kv_2) \Pl(k_1,a) \Pl(k_2,a) + \mbox{2 cycl. perm.} 
\end{equation}
From a diagrammatic point of view, we can easily convince ourselves that there is no possibility to connect the three external points without invoking the three-point vertex of $F_2$. On the top left of Fig.~\ref{f:bs} the tree-level bispectrum is shown.

One can then start computing higher-order corrections to the power- and bispectrum. As shown in Fig.~\ref{f:ps} there are two possible one-loop corrections to the power spectrum and they take the rather simple form 

\begin{equation}\label{e:ploop}
\begin{split}
& P_{22} (k) \eq 2 \intq \Pl(q) \Pl(|\kv -\qv|) \, F_2^2(\qv , \kv - \qv) \;, \\[1.5ex]
& P_{13} (k) \eq 6 \Pl(k)  \intq \Pl(q) \, F_3(\kv, \qv, -\qv) \;,
\end{split}
\end{equation}
giving the SPT power spectrum

\begin{equation}
P_\textit{SPT}(k) \eq \Pl(k) + P_{22}(k) + P_{13}(k) + \; \mbox{higher order loops} \;.
\end{equation}
These integrals can be divergent when the loop momentum $\qv$ becomes large and the renormalization of these divergences has been discussed in the Ref.~\cite{Pajer2013a}. It is in fact one of the main shortcomings of SPT that depending on the initial conditions, i.e.~the form of the linear power spectrum, the perturbative expansion leads to divergent, non-physical results. 

At the one-loop level, the bispectrum receives contributions from correlating either three $\dn{2}$, one $\dn{3}$ with one $\dn{2}$ and one $\dn{1}$ or one $\dn{4}$ with two $\dn{1}$ (see Refs.~\cite{Scoccimarro1997,Scoccimarro1998,Bernardeau2002} for discussions of the one-loop bispectrum in SPT as well as Ref.~\cite{Bernardeau2008}). This is what is shown in Fig.~\ref{f:bs}. Translating the graphs of Fig.~\ref{f:bs} into mathematical expressions, the four one-loop contributions are

\begin{eqnarray}
 B_{222}  & = & 8 \intq F_2(-\qv,\qv+\kv_1) F_2(\qv+\kv_1, -\qv+\kv_2)F_2(\kv_2-\qv, \qv) \non
 & & \Pl(q) \Pl(|\qv +\kv_1|) \Pl(|\qv -\kv_2|)  \;, \label{e:b222}\\[1.5ex]
B_{321}^I & =& 6 \Pl(k_3) \intq F_3(-\qv, \qv - \kv_2, -\kv_3 ) F_2(\qv, \kv_2 - \qv) \, \Pl(q) \Pl(|\qv -\kv_2|) \non
&  & + \, \mbox{$5$ perm. } \;, \label{e:b321i} \\[1.5ex]
B_{321}^{II} & =& 6 F_2(\kv_2,\kv_3) \, \Pl(k_2) \Pl(k_3) \intq F_3(\kv_3,\qv, -\qv )\, \Pl(q) + \, \mbox{$5$ perm. }  \;,  \non
& = &  F_2(\kv_2,\kv_3) \, \Pl(k_2) P_{13}(k_3)+ \, \mbox{$5$ perm. }  \;, \label{e:b321ii} \\[1.5ex]
B_{411} & = & 12 \Pl(k_2) \Pl(k_3) \intq F_4(\qv,-\qv, -\kv_2, -\kv_3)\, \Pl(q) + \, \mbox{$2$ cyc. perm.} \; \label{e:b411}
\end{eqnarray}
Note that $B_{321}^{II}$ reduces to the one-loop contribution to the power spectrum stemming from the correlator $\langle \dn{3} \dn{1} \rangle$, i.e.~$P_{13}$. Again, these integrals can be divergent just as in the case of the one-loop power spectrum. An important part of this paper is dedicated to prove that these divergences can be cancelled. In sum, the SPT bispectrum at the one-loop level reads

\begin{equation}
B_\textit{SPT}(k_1,k_2,k_3) \eq B_{112} + B_{222} + B_{321}^I + B_{321}^{II} + B_{411}\;. 
\end{equation}
If properly regularized, the integrals in Eqs.~\eqref{e:b222}, \eqref{e:b321i}, \eqref{e:b321ii} and \eqref{e:b411} can be evaluated analytically for a power-law linear power spectrum $\Pl(k) \propto k^n$ in EdS as done in Ref.~\cite{Scoccimarro1997}. For a more realistic $\Lambda$CDM universe, these integrals have to be evaluated numerically since we do not have an analytic form of the linear power spectrum at the present epoch. Also, in this case we do not encounter formally divergent integrals since modes entering the horizon during radiation domination are suppressed. 

\begin{figure}

\subfloat{
\includegraphics[width=0.47\textwidth]{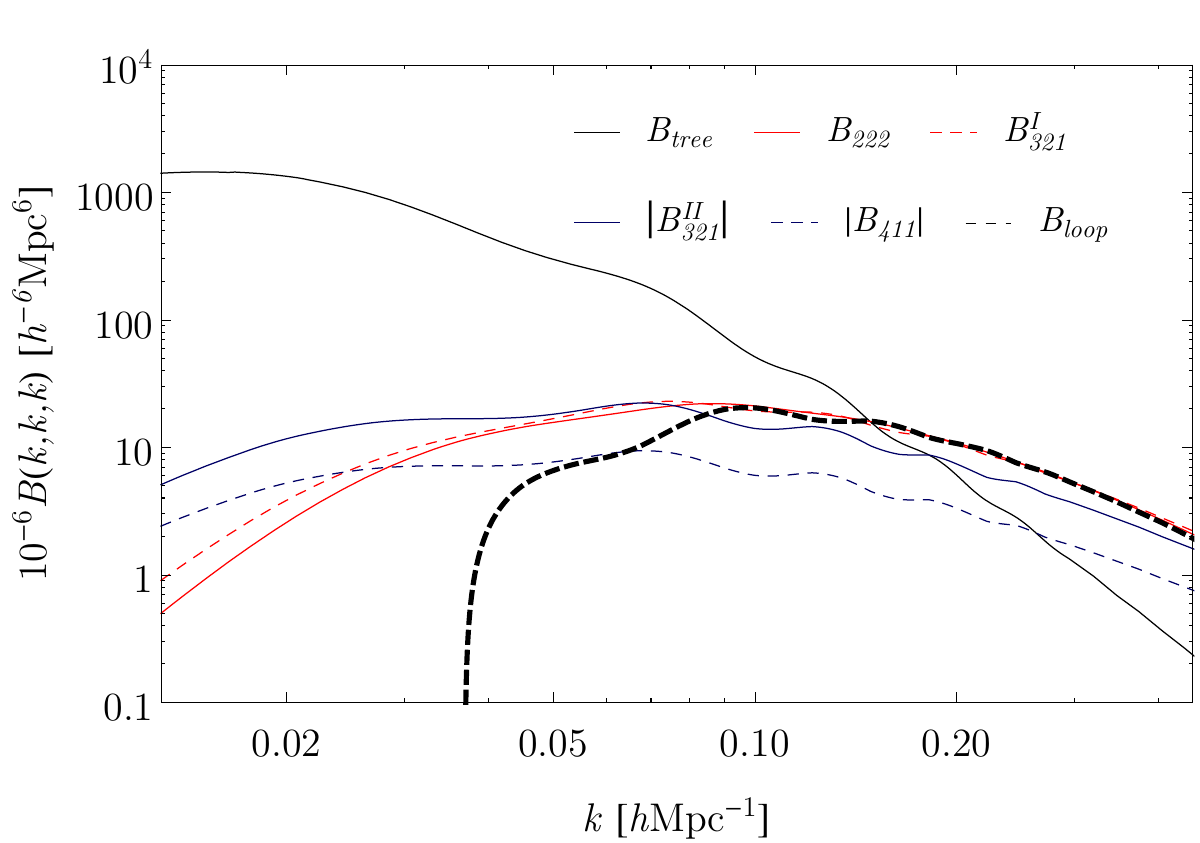} }
\hspace{10pt}
\subfloat{
\includegraphics[width=0.47\textwidth]{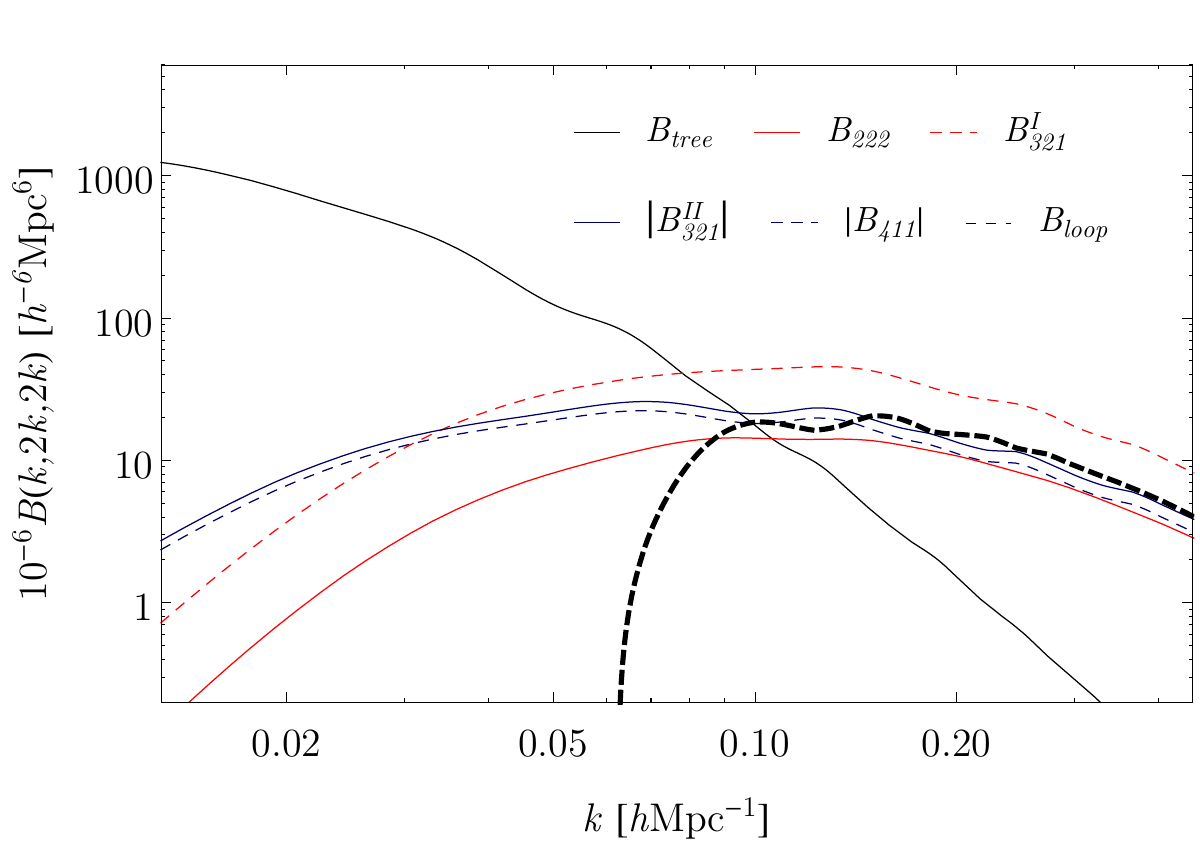} } \\
\subfloat{
\includegraphics[width=0.47\textwidth]{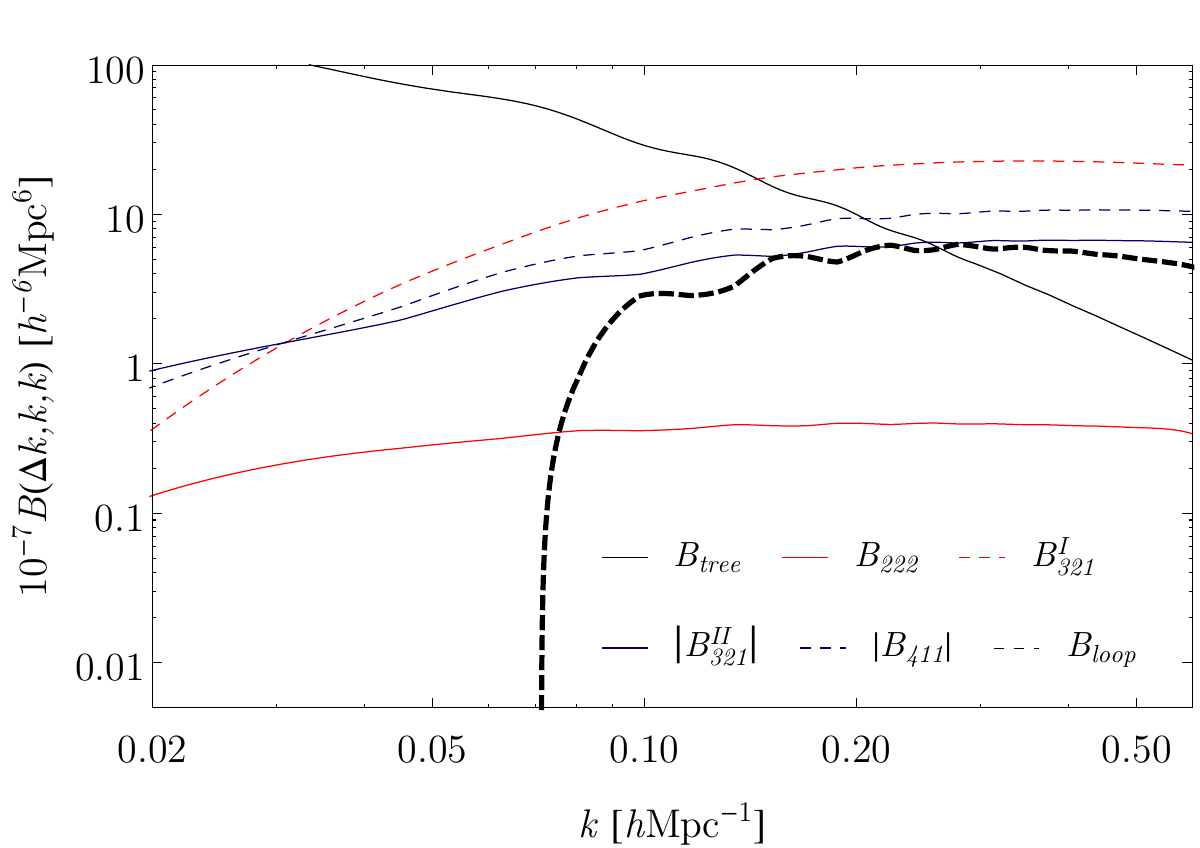} }

\caption{The one-loop diagramgs $B_{222}$ (solid red), $B_{321}^I$ (dashed red), $|B_{321}^{II}|$ (solid blue), $|B_{411}|$ (dashed blue) and the sum of all four (dashed black) is shown for three special configurations of $k_1$, $k_2$ and $k_3$ on a logarithmic scale. The solid black curve is the tree-level contribution. In the squeezed limit, we set $\Delta k= 0.013 \; h \mbox{Mpc}^{-1}$. We plot the absolute value of the diagrams $|B_{321}^{II}|$ and $|B_{411}|$ since they are negative. Our figures agree well with the ones in Ref.~\cite{Sefusatti2010} and, as can be seen, there are no large cancellations among the diagrams.  }\label{fig:bloop}
\end{figure}

\begin{figure}
\centering
\includegraphics[scale=0.5]{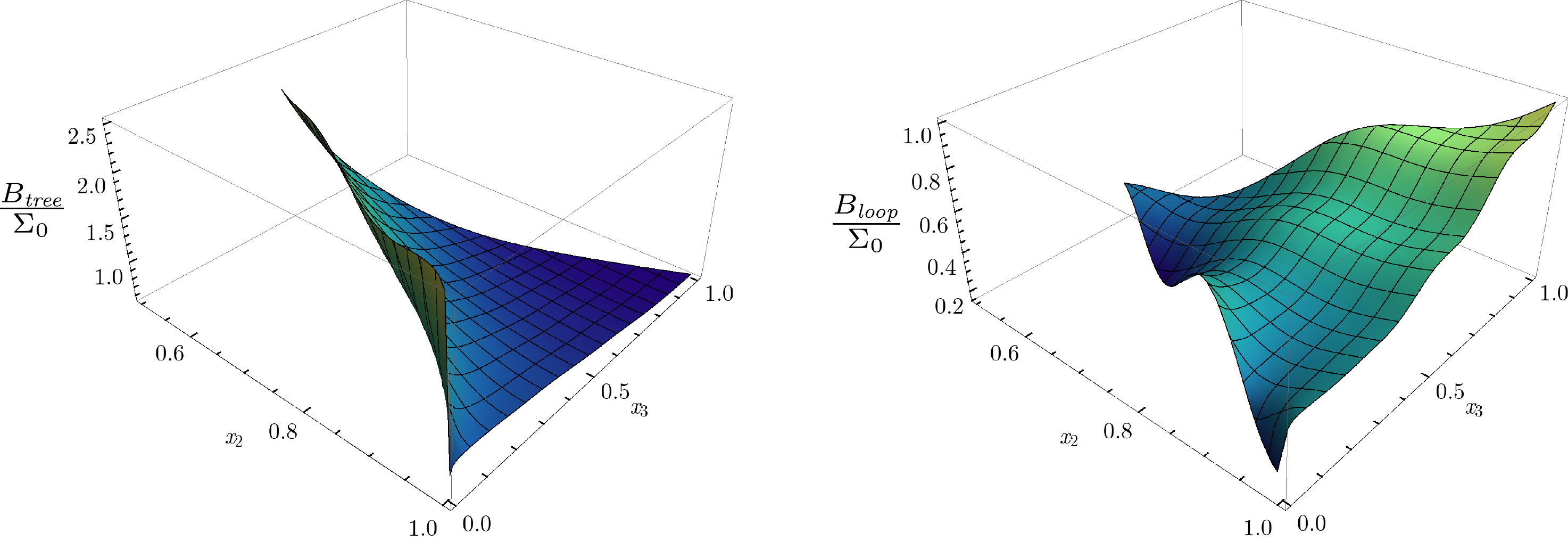} 
\caption{The shapes of the tree-level and one-loop bispectrum divided by $\Sigma_0$ are plotted as a function of $x_2 = k_2/k_1$ and $x_3=k_3/k_1$ for a fixed $k_1 = 0.2 \, \ihMpc$. The shape is restricted to the range of $x_2 \geq x_3 \geq 1-x_2$.}\label{fig:shapesSPT}
\end{figure}

We compute the one-loop integrals using two independent codes: once using the built-in routines of \emph{Mathematica} and a \texttt{C++} code which uses the \emph{CUBA} libraries \cite{CUBA}. The two calculations agree very well with each other and we can easily reproduce the results found in the literature, e.g.~in Ref.~\cite{Sefusatti2010}. To avoid numerically unstable situations, the one-loop contribution to the bispectrum is most conveniently computed using the IR-safe integrand discussed in Appx.~\ref{a:IRsafe} (see also Ref.~\cite{Blas2013a,Carrasco2013}). In Fig.~\ref{fig:bloop} we show the one-loop diagrams in three special configurations. As opposed to the one- and two-loop power spectrum, in the bispectrum there are not very large cancellation among the single diagrams. Nevertheless, the IR-safe integrand improved somewhat the precision of the numerical computation. Fig.~\ref{fig:shapesSPT} shows the two shapes of the tree-level and one-loop diagrams. In order to emphasize the shape dependence of the bispectrum, it is convenient to consider a quantity similar to the reduced bispectrum $Q$ (see Ref.~\citep{Bernardeau2002}). $Q$ is defined as the ratio of the bispectrum and the product of two power spectra (more precisely, the sum over permutations thereof). We choose to divide the bispectrum only by the combination $\Sigma_0$ made of linear power spectra

\begin{equation}
\Sigma_0 \eeq \Pl(k_1)\Pl(k_2) + \mbox{2 cycl. perm.}
\end{equation}
as opposed to $Q$ where the full power spectra are considered. Following Ref.~\cite{Babich2004}, we can then plot this quantity as a function of the two variables $x_2 = k_2/k_1$ and $x_3=k_3/k_1$ that satisfy the relation $x_2 \geq x_3 \geq 1-x_2$.

\subsection{UV-limit of the loop integrals}\label{s:uvlimits}

In the context of the EFTofLSS we are particularly interested in the UV-limit of the loop integrals. Therefore, we will have a closer look at this regime before discussing in detail the EFT contributions to the bispectrum. When the loop momentum becomes much larger than the external momenta, we can expand the kernels according to the scaling laws in Eq.~\eqref{e:Fscaling}. The resulting expressions give us a hint of how the contributions from the effective stress tensor should look like in order to cancel the possible divergences. 

Let us first consider the UV-limit of the one-loop power spectrum (see also Ref.~\cite{Pajer2013a}). Looking at the diagrams in Fig.~\ref{f:ps}, we can imagine that the momentum that runs inside the loop becomes much larger than the external momentum. Since the vertices scale as $\propto k^2/ q^2$ in this limit (see Eq.~\eqref{e:Fscaling}), we conclude that the two diagrams behave as $P_{22} \propto k^4$ and $P_{13} \propto k^2$. Including the correct numerical factors, in the UV-limit the two one-loop integrals in Eq.~\eqref{e:ploop} take the form

\begin{eqnarray}
&& P_{22} (k) \Big|_{q\rightarrow \infty } \eq \frac{9}{196 \pi^2} \, k^4 \int dq \;q^2\; \frac{\Pl^2(q)}{q^4} \;, \label{e:ploopuv22}\\[1.5ex]
&& P_{13} (k) \Big|_{q\rightarrow \infty } \eq - \frac{61}{630 \pi^2} \Pl(k) \, k^2 \int dq\;q^2\;  \frac{\Pl(q)}{q^2}  \label{e:ploopuv13}\\
&& \phantom{P_{13} (k) \Big|_{q\rightarrow \infty }} \eq - \frac{61}{105} \, \Pl(k) \, k^2 \sigma_v^2  \;, \nonumber
\end{eqnarray}
where we defined the quantity $\sigma_v^2 \equiv 1/3 \intq \Pl(q)/q^2$ as the UV-limit of this integral.

For the bispectrum, we follow exactly the same procedure as for the power spectrum in order to get the UV-limits of the integrals in Eqs.~\eqref{e:b222}, \eqref{e:b321i}, \eqref{e:b321ii} and \eqref{e:b411}. We can again look at the one-loop diagrams in Fig.~\ref{f:bs} and insert for every vertex a factor $k^2/ q^2$. Assuming that all three external momenta are of the same order $ k \sim k_1 \sim k_2 \sim k_3$ and $q \gg k$, we get the rough scaling of

\begin{equation}\label{e:Bscaling}
\begin{split}
& B_{222} \Big|_{q\rightarrow \infty } \; \propto \;  k^{6} \int dq \;q^2\; \frac{\Pl^{3}(q)}{q^{6}} \;, \\[1.5ex]
& B_{321}^{I} \Big|_{q\rightarrow \infty } \; \propto \; k^{4} \Pl(k) \int dq \;q^2\; \frac{\Pl^{2}(q)}{q^{4}} \;,\\[1.5ex]
& B_{321}^{II} \Big|_{q\rightarrow \infty } \; \propto \; k^{2} \Pl^{2}(k) \, \sigma_v^2 \;, \\[1.5ex]
& B_{411} \Big|_{q\rightarrow \infty } \; \propto \; k^{2} \Pl^{2}(k) \, \sigma_v^2 \;.
\end{split}
\end{equation}
Of course, the UV-limits of the loop integrals depend on a shape, i.e.~rather than having just some power of $k$ in front of a $k$-independent integral, we have to consider the triangular configuration of $\kv_1$, $\kv_2$ and $\kv_3$. The exact UV-limit of the diagram $B_{222}$ has the following shape 

\begin{equation}\label{e:b222div}
\begin{split}
B_{222} \Big|_{q\rightarrow \infty } \eq & - \frac{1}{4802 \pi ^2 } \, \Big\{ 30 k_1^6 -30 k_1^4 \left( k_2^2 + k_3^2 \right) + k_1^2 \left(-30 k_2^4+ k_2^2 k_3^2 - 30 k_3^4 \right) \\
&  +30  \left( k_2^2- k_3^2\right)^2  \left(k_2^2+ k_3^2\right) \Big\} \,    \int dq \;q^2\; \frac{\Pl (q)^3}{q^6} \;, 
\end{split}
\end{equation}  
where the $k^6$ scaling of Eq.~\eqref{e:Bscaling} is apparent. In the form of Eq.~\eqref{e:b222div}, it is clear that in the squeezed limit $k_1 \to 0$ the shape goes to zero. The same is true also for the other squeezed limits $k_2 \to 0$ and $k_3 \to 0$ although it is not explicitly apparent. Analogously, the shape of the divergence of $B_{321}^I$ is given by

\begin{equation}\label{e:b321Idiv}
\begin{split}
B_{321}^I \Big|_{q\rightarrow \infty } \eq & \frac{1}{35280 \pi^2 \, k_3^2} \Big\{ 170 {k_1}^6+ {k_1}^4 \left(83
   {k_2}^2+190 {k_3}^2\right) +2 {k_1}^2 \left(67 {k_2}^4 +256 {k_2}^2 {k_3}^2 \right. \\
& \left. -445 {k_3}^4\right)-\left(387 {k_2}^2-530 {k_3}^2\right) \left({k_2}^2-{k_3}^2\right)^2  \Big\} \; \Pl (k_3) \int dq \;q^2\; \frac{\Pl(q)^2}{q^4} \\
& + \mbox{5 perm.} 
\end{split}
\end{equation}
Notice that it is important that all possible permutations of the integral above are taken into account since otherwise, as we shall see in Sec.~\ref{s:eftbispectrum}, the renormalization does not work. The UV-limit of the $B_{321}^{II}$ diagram is rather simple. We can write $B_{321}^{II}$ in terms of $P_{13}$ (see Eq.~\eqref{e:b321ii}) and therefore, the large-$q$ limit reduces to the one of $P_{13}$

\begin{equation}
\begin{split}
B_{321}^{II} \Big|_{q\rightarrow \infty } \eq &  F_2(\kv_2, \kv_3) \Pl(k_2) \,   P_{13}(k_3) \Big|_{q\rightarrow \infty }  + \mbox{5 perm.}   \\[1.5ex]
\eq & - \frac{61}{105} \, F_2(\kv_2, \kv_3) \Pl(k_2)  \Pl(k_3) \, k_3^2 \sigma_v^2 + \mbox{5 perm.}  
\end{split}
\end{equation}
Although $B_{411}$ and $B_{321}^{II}$ have the same overall scaling with $k$, they have two different shapes, i.e. 

\begin{equation}
\begin{split}\label{e:b411div}
B_{411} \Big|_{q\rightarrow \infty } \eq & - \frac{1}{226380 }  \, \frac{1}{k_2^2 k_3^2} \Big\{  12409 {k_1}^6+20085   {k_1}^4 \left({k_2}^2+{k_3}^2\right) \\
&  +{k_1}^2   \left(-44518 {k_2}^4  + 76684 {k_2}^2 {k_3}^2-44518 {k_3}^4\right) \\
 &  +12024 \left({k_2}^2- {k_3}^2\right)^2  \left({k_2}^2+{k_3}^2\right) \Big\}  \,  \Pl(k_2) \Pl(k_3) \, \sigma_v^2 \\
 & + \mbox{2 perm.} 
\end{split}
\end{equation}

The aim of the EFTofLSS is to provide a consistent framework where the possible UV-divergences discussed in this section can be canceled by appropriate counterterms. This means that the counterterms that come from the effective stress tensor must not only scale with $k$ in the same way as the UV-limits of the loop integrals but also the shapes have to match. These counterterms come from the derivative expansion of the stress tensor, which we will introduce in the following sections.

\subsection{The scaling in EdS} \label{s:edspowercounting}

Before deriving the explicit form of the effective stress tensor, it is possible to get a feeling of the relative sizes of the various terms that contribute to the bispectrum by considering them in the case of EdS with power-law initial conditions. In Ref.~\cite{Pajer2013a} it was pointed out that the self-similarity of the equations of motion makes it possible to write the power-spectrum in the EFTofLSS as a polynomial in $k/\kNL$. Provided that the parameters in effective stress tensor transform accordingly, the equations of motion in EdS are invariant under a rescaling of $\xv \rightarrow \lambda_x \xv$ and $\tau \rightarrow \lambda_\tau \tau$. We can define the quantities $\Delta^2$ and $\mathcal{I}_B$ that correspond to the power- and bispectrum which are invariant under a self-similarity transformation

\begin{equation}\label{e:Delta}
\Delta^2 (k,\tau) \eeq \frac{k^3}{2\pi^2} \, P(k,\tau)  \eq \Delta^2(k/\lambda_x, \lambda_\tau \, \tau) \;, \\[1.5ex]
\end{equation}
and
\begin{equation}
\mathcal{I}_B (k_1, k_2, k_3,\tau) \eeq \left( \frac{k_1^3}{2\pi} \right) ^2 \, B(k_1, k_2, k_3,\tau) \eq \mathcal{I}_B (k_1/\lambda_x, k_2/\lambda_x, k_3/\lambda_x, \lambda_\tau \, \tau)\;. \quad 
\end{equation}
Note that by choosing to define $\mathcal{I}_B$ with a factor $k_1^6$ we broke the permutation symmetry of the momenta that we have in the bispectrum. Assuming power-law initial conditions 

\begin{equation}\label{e:plinpowerlaw}
\Pl(k) \eq a^2 \, A k^n \;,
\end{equation}
with some amplitude $A$, breaks self-similarity down to only one rescaling where $\lambda_x$ is fixed to $\lambda_x = \lambda_\tau ^{4/(3+n)}$. This ensures that the rescaled power spectra have still the same linear power spectrum. Since the two quantities $\Delta^2$ and $\mathcal{I}_B$ are invariant under self-similarity, they can only depend on ratios of physical scales. It turns out that it is convenient to measure all physical scales with respect to the non-linear scale $\kNL$. Following Ref.~\cite{Pajer2013a}, we define the non-linear scale $\kNL$ such that $\Delta^2$ is exactly one when evaluated at the linear level at $\kNL$, i.e.~$\Delta^2_\textit{lin}(\kNL) \equiv 1$.

Since the EFTofLSS is constructed as an expansion in derivatives, the two invariants $\Delta^2$ and $\mathcal{I}_B$ can be written as a polynomial in the ratio $k/\kNL$. For the power spectrum up to one-loop and including the LO EFT contributions, Ref.~\cite{Pajer2013a} found the remarkably simple form

\begin{equation}\label{e:generalD}
\begin{split}
\Delta^2(k) \eq & \left( \frac{k}{\kNL} \right)^{n+3} \left\{ 1 + \left( \frac{k}{\kNL} \right)^{n+3} \left[ a + \tilde{a} \ln \frac{k}{\kNL} \right] \right\}   \\[1.5ex]
& + b_c \, \left( \frac{k}{\kNL} \right)^{n+5} + b_J \, \left( \frac{k}{\kNL} \right)^{7} \;.
\end{split}
\end{equation}
The coefficients $a$ and $\tilde a$ come from the one-loop contribution and can be computed explicitly in SPT, while $b_c$ and $b_J$ are related to the LO viscosity and noise terms in the effective stress tensor. For the phenomenologically interesting value of $n \approx -3/2 $ we immediately realize that the viscosity terms are more important than possible two-loop contributions and that the noise terms are strongly suppressed. 

If we know the scaling in $k$, self-similarity tells us that every power of $k$ comes with the corresponding power of $\kNL$ and Eq.~\eqref{e:generalD} follows directly. The linear power spectrum, as given in Eq.~\eqref{e:plinpowerlaw}, gives the a contribution to $\Delta^2$ which is $\Delta^2 = (k/\kNL)^{(3+n)}$. The scaling of the finite part of $P_{22}$ and $P_{13}$ in Eq.~\eqref{e:ploop} can be obtained by dimensional analysis. From the explicit form in Eq.~\eqref{e:ploop}, we can infer that the finite part of the integrals must scale as 

\begin{equation}
P_{22} \; \simeq \; P_{13} \; \simeq \; k^3 \, \Pl(k)^2 \, C
\end{equation}
since the kernels are dimensionless quantities. From dimensional analysis we would infer that $C$ is just some computable constant (this argument does not tell us that there can be a $\ln k/\kNL$ in $C$). The first line of Eq.~\eqref{e:generalD} then follows directly. The above scalings can be easily generalized to multi-loop diagrams. Each loop adds another factor of $k^3\Pl(k) $ to the finite part of the loop contribution.

Next, we consider the EFT contribution to the power spectrum. Even without knowing the exact form, we want that the $k$-dependence of the contribution from the free parameters matches the one of the UV-limit of the loop integrals (see Ref.~\cite{Pajer2013a} for the details regarding the time dependence). The scaling of $P_{22}$ and $P_{13}$ is not the same for large loop momenta as we found in Eqs.~\eqref{e:ploopuv22} and \eqref{e:ploopuv13}. We can therefore infer that there are two different kinds of EFT contributions that scale as $\propto k^2 \Pl$ and $k^4$ at the level of power spectra, leading to $(k/\kNL)^{5+n}$ and $(k/\kNL)^7$ in Eq.~\eqref{e:generalD}. These two terms are nothing but the LO viscosity and noise terms as we shall discuss more carefully in Sec.~\ref{s:eftbispectrum}.

Also $\mathcal{I}_B$ can be expressed as a polynomial in powers of $k_1/\kNL$ and the derivation is analogous to the one of the power spectrum. The main difference is that while $a$, $\tilde{a}$, $b_c$ and $b_J$ are numbers that depend only on the spectral index $n$, in the case of the bispectrum they become shapes that depend on the specific configuration of $\kv_1$, $\kv_2$ and $\kv_3$. Let us start by considering the finite SPT part of the bispectrum. The tree-level bispectrum in Eq.~\eqref{e:blin} contains two linear power spectra and a dimensionless kernel. Hence, the tree-level part of $\mathcal{I}_B$ scales as $ (k/\kNL)^{2(n+3)}$. As in the case of power spectrum each loop adds a factor of $k^3\Pl(k)$, leading to

\begin{equation}\label{e:ISPT}
\mathcal{I}_B \; \supset \;  \left(\frac{k_1}{\kNL} \right)^{2(3+n)} \, \Bigg\{ \zeta_\textit{lin} (k_1, k_2, k_3)  + \sum_{i=1}^N \,  \left(\frac{k_1}{\kNL} \right)^{ (3+n)i}  \zeta_\textit{loop}^{(i)} (k_1, k_2, k_3)    \Bigg\} \;.
\end{equation}
For simplicity we did not explicitly write the possible logarithmic $k$-dependence of $\zeta_\textit{loop}$. It is important to note that all $\zeta$ now depend on the two ratios $k_1/k_3$ and $k_2/k_3$ as well as on the spectral index $n$. The form of Eq.~\eqref{e:ISPT} is somewhat arbitrary since it is always possible to trade $k_1/\kNL$ for $k_{2,3}/\kNL$, thereby absorbing dimensionless ratios $k_i / k_j$ into the coefficients $\zeta$.

Next, we determine the scaling of the effective stress tensor contributions, which are dictated by the ultraviolet behavior of SPT diagrams. We found in Eq.~\eqref{e:Bscaling} that we have three different UV-behaviours, i.e.~$B_{222}$, $B_{321}^{I}$, while $ B_{321}^{II}$ and $B_{411}$ scale the same way. The counterterms for the first two diagrams therefore scale as $\propto k^6$ and $\propto k^4 \Pl$ at the level of the bispectrum. For the remaining two, the scaling is just $k^2 \Pl^2$. The resulting powers of $k/\kNL$ are therefore 

\begin{equation}
\mathcal{I}_B \; \supset \; \left( \frac{k}{\kNL} \right)^p \quad \mbox{with} \quad \left\{ \begin{array}{ll} p = 12  & \quad \mbox{for } B_{222} \\
	p = n + 10 & \quad \mbox{for } B_{321}^{I} \\ p = 8 + 2n  & \quad \mbox{for } B_{321}^{II} \mbox{ and } B_{411} \end{array} \right.  \;.
\end{equation}  

Taking into account the shape dependence of the dimensionless factors and adding these contributions from the effective stress tensor to the SPT part, we obtain the general form of the bispectrum in the EFTofLSS in an EdS universe with power-law initial conditions

\begin{equation}\label{e:Igeneral}
\begin{split}
\mathcal{I}_B \eq & \left(\frac{k_1}{\kNL} \right)^{3+n} \, \Bigg\{ \zeta_\textit{lin}  \left(\frac{k_1}{\kNL} \right)^{3+n}  + \sum_{i=1}^N \,  \left(\frac{k_1}{\kNL} \right)^{ (3+n)(i+1)} \zeta_\textit{loop}^{(i)}  \\[1.5ex]
& +  \zeta_\textit{c}  \left(\frac{k_1}{\kNL} \right)^{ 5+n } + \zeta_J \left(\frac{k_1}{\kNL} \right)^{ 7 } \Bigg\} + \zeta_{JJJ}  \left(\frac{k_1}{\kNL} \right)^{ 12 } + \dots \;, 
\end{split}
\end{equation}
where we omitted in our notation the fact that all $\zeta$ are dimensionless shape functions. The ellipses denote higher order EFT terms. Although the functions $\zeta$ may take a rather complicated form, Eq.~\eqref{e:Igeneral} shows that the scaling of the various contribution is just as simple as in the case of two-point correlators. Apart from the overall factor $(k/\kNL)^{3+n}$ and the $(k/\kNL)^{12}$ noise term, the various contributions have exactly the same relative scaling as in Eq.~\eqref{e:generalD}. Ignoring the shape dependence of the coefficients $\zeta$, we can therefore draw the conclusion that noise terms in the bispectrum are negligible for $n\approx -3/2$. The viscosity terms, however, should be treated on an equal footing as the one-loop corrections. Hence, we will neglect the noise terms for the remainder of this paper, except when discussing the cancellation of the UV-divergences.

These simple scaling relations also allow us to estimate the size of the terms that we neglect. Throughout this paper, we neglect two-loop contributions, the next-to-next-to-leading order (NNLO) viscosity and the LO and NLO noise terms. The two-loop corrections would scale as $\sim (k_1/\kNL)^{4(3+n)}$ while, as seen in Eq.~\eqref{e:Igeneral}, the noise terms scale as $\sim (k_1/\kNL)^{10+n}$ and $(k_1/\kNL)^{12}$. The NNLO viscosity contribution, i.e.~terms that come with four derivatives and two fields in the effective stress tensor $\tau_\theta$, would scale as $\sim (k_1/\kNL)^{10+2n}$. $\tau_\theta$ may contain also terms with two derivatives and three fields. Such terms would scale as $\sim (k_1/\kNL)^{11+3n}$ in the bispectrum. For $n\approx -3/2$ it is the two-loop and the NNLO viscosity terms that give the largest contributions that we are neglecting

\begin{equation}
\left( \frac{k}{\kNL} \right)^p \quad \Longrightarrow \quad \left\{ \begin{array}{ll} p \approx 6 & \mbox{two-loop} \\ p \approx 7  & \mbox{NNLO viscosity} \\ p \approx 11/2 & \mbox{NNLO viscosity} \\ p \approx 21/2 \qquad &  \mbox{noise} \\ p = 12 &  \mbox{noise} \end{array} \right.
\label{eq:contribcount}
\end{equation}
These values should be compared to the one-loop term with $p=9/2$ and the NLO viscosity terms with $p=5$. Interestingley, even the leading noise contribution is suppressed compared to the two-loop for $n \approx -3/2$. This is, of course, a rather crude estimate of what would be the next larger corrections to the bispectrum beyond the one-loop computation and the NLO viscosity terms that we consider in this paper. In Sec.~\ref{s:simulations} we will estimate the two-loop contribution and add it as a theoretical uncertainty.


\section{The effective stress tensor at next-to-leading order}\label{s:tau}

Having discussed in detail the one-loop contribution to the bispectrum, we shall now investigate the effective stress tensor. First, we show that the effective stress tensor can be treated as being local in time, by which we mean that it can be written as a sum of operators which are evaluated at the same time. Finally, we derive its form up to NLO, i.e.~including terms with two derivatives and two fields.

\subsection{Local or non-local in time?}\label{local}

As we have seen in Sec.~\ref{s:SPT}, the equations of motion of the EFTofLSS in Eq.~\eqref{e:EOMrealspace} contain a stress tensor

\be
\label{dtau}
\frac{1}{a\,\rho}\, \d_j \tau^{ij} \;, 
\ee
which incorporates the effects of the short-distance physics. Most notably, as explained in Ref.~\cite{Baumann2012}, Eq.~\eqref{dtau} includes the response of the short modes to the tidal forces produced by the large scale perturbations

\be
\label{tide}
\tau^{ij} \; \supset \; \frac{\rho \expect{v_s^2}}{\cH^2}\d^i\d^j \phi \;.
\ee
However, this EFT has a peculiar feature: the characteristic time-scale of the short-scale modes is not necessarily shorter than the time-scales of interest, namely $\cH^{-1}$. Before getting virialized even the non-linear modes evolve quite slowly. Equivalently, in the fluid dynamics language, the issue is the very long free-streaming time $\cH^{-1}$ of Dark Matter particles, which are essentially collisionless. \comment{What allows this system to be described as a fluid is not that rapid collisions erase the short-scale memory, but rather the smallness of particle velocities, $v_{\rm max}\sim \cH/k_{NL}$ before virialization, and the finite age of the universe which results in a small mean-free path $1/k_{NL}$ (see Ref.~\cite{Baumann2012}).}

A more general version of the back reaction term \eqref{tide}, which satisfies the symmetries of the EFT (see Appx.~\ref{s:equiv}), and accounts for the evolution of the short-scale modes is an integral along the history of fluid elements as discussed in Ref.~\cite{Carrasco2013a}
\be
\label{tide2}
\frac{1}{a\,\rho}\d_i\d_j \tau^{ij} \; \supset \; \lap \int d\tau' K(\tau,\tau')\delta(\xv_{fl}[\xv,\tau;\tau'],\tau'),
\ee
where we have used the Poisson equation in Eq.~\eqref{e:EOMrealspace} to express $\phi$ in terms of $\delta$, and introduced $\xv_{fl}$ as the position of the fluid element $(\xv,\tau)$ at time $\tau'$. This is given recursively by 

\be
\xv_{fl}[\xv,\tau;\tau'] \eq \xv-\int^\tau_{\tau'} d\tau'' \vv(\xv_{fl}[\xv,\tau;\tau''],\tau'').
\ee
We shall often drop $\xv,\tau$ from the argument of $\xv_{fl}$ for brevity. The invariance of Eq. \eqref{tide2} under symmetries of EFTofLSS is verified in Appx. \ref{s:nonlocal}.

The long memory of the short modes implies that the kernel $K(\tau,\tau')$ has a temporal extent of order $\cH^{-1}$. This property of EFTofLSS has recently attracted a lot of attention (see Refs.~\cite{Carrasco2013a,Carroll2013}) and is often characterized as ``non-locality in time''. The purpose of this section is to argue that: (a) Locality in time of the EFT is guaranteed as long as perturbation theory is applicable. (b) The long memory effect induces spatially non-local terms in the EFT, but those are irrelevant for the one-loop bispectrum. We also explain from a Lagrangian perspective why in practice we can use local kernel $K(\tau,\tau')\propto \delta_D(\tau-\tau')$ rather than a naive perturbative expansion of $\xv_{fl}$ around $\xv$ in Eq.~\eqref{tide2}. This agrees with the observation made in \cite{Carrasco2013a}, based on matching with simulation results. 

Let us start by considering a toy model of two coupled harmonic oscillators
\begin{equation}
\begin{split} \label{X10}
& \ddot X_1+\omega_1^2 X_1 \eq \alpha X_2,\\[1.5ex]
& \ddot X_2+\omega_2^2 X_2 \eq \alpha X_1,
\end{split}
\end{equation}
and integrate out $X_2$. Ignoring the $X_1$ independent piece (which would correspond to the stochastic noise term in the EFTofLSS), we obtain
\be
\label{X1}
\ddot X_1+\omega_1^2 X_1 = \alpha^2 \int ^t_{t_0} dt' G_2(t,t') X_1(t'),
\ee
where $G_2(t,t')=\theta(t-t')\sin\omega_2(t-t')$ is the retarded Green's function of $X_2$, and the mixing $\alpha$ is turned on at $t_0$. This equation has the generic form of \eqref{tide2}. In the context of EFT, one is usually interested in the case $\omega_2^2\gg \omega_1^2,\alpha$ in which case $G_2(t,t')$ has a width of order $1/\omega_2$. The resulting effective equation of motion for $X_1$ can be expressed as an expansion in powers of $\d_t/\omega_2$ acting on $X_1$.

Now consider the more relevant case of $\omega_1\sim \omega_2$, where the above Taylor expansion in $\d_t/\omega_2$ breaks down. In this case there is still a chance of obtaining a local equation of motion for $X_1$: as long as $X_1$ is not violently perturbed by $X_2$, that is when $\alpha\ll \omega_1^2$, a local equation can be obtained by an expansion in powers of $\alpha/\omega_1^2$. More explicitly, at zeroth order in $\alpha$
\be
X_1^{(0)}(t')= X_1(t)\cos\omega_1(t'-t)+\frac{\dot X_1(t)}{\omega_1}\sin\omega_1(t'-t) \;,
\ee
which after plugging in the r.h.s. of \eqref{X1} gives
\begin{equation}
\begin{split}
\ddot X_1+\omega_1^2 X_1 \eq  &
\frac{\alpha^2}{2\omega_2} \left\{ \frac{1}{\omega_+}(1-\cos\omega_+T)+\frac{1}{\omega_-}(1-\cos\omega_-T)\right\} X_1 \\[1.5ex]
& +\frac{\alpha^2}{2\omega_2} \left\{ \frac{1}{\omega_+}\sin\omega_+T-\frac{1}{\omega_-}\sin\omega_-T \right\} 
\frac{\dot X_1}{\omega_1} +\O(\alpha^4) \;,
\end{split}
\end{equation}
where $\omega_\pm = \omega_2\pm \omega_1$ and $T=t-t_0$.\footnote{If we further assume $\omega_2\gg \omega_1$, and average the r.h.s. over time scales $\omega_2^{-1}\ll \Delta t \ll \omega_1^{-1}$ we obtain a conservative equation for $X_1$ as expected.} This procedure can be continued to yield a controlled expansion, essentially in powers $\delta X_1/ X_1$ with $\delta X_1$ characterizing perturbations due to the coupling to $X_2$. 

The same reasoning applies to the cosmological fluid. Now there are also self-interactions of the large scale modes which would be analogous to adding quadratic terms in $X_1$ to \eqref{X1}
\be
\ddot X_1+\omega_1^2 X_1 =\beta X_1^2+ \alpha^2 \int ^t_{t_0} dt' G_2(t,t') X_1(t'),
\ee
but those too are treated perturbatively. In terms of $X_1(t)$ and $\dot X_1(t)$ one can perturbatively solve the homogeneous equation in powers of $\beta/\omega_1^2$, substitute in the last term to get $\O(\alpha^2)$ local source terms, and iterate the procedure to obtain a double expansion in powers of $\alpha^2$ and $\beta$. 

In the case of EFTofLSS the same procedure can be carried out using the information on the final time slice $\tau$ to bring the effective Euler equation into the form of a double expansion. However, there are two differences with respect to the oscillator toy example. First, the expansion is not in small coupling constants such as $\alpha$ and $\beta$. Rather, what ensures the smallness of the back-reaction at large scales is the fact that modes are coupled through the tidal forces. The expansion parameter is therefore $k/k_{NL}$. 

Second, and more important, is the emergence of spatially non-local terms due to the long memory $T$ of the short-scale modes. Consider an EFT in which signals propagate at the speed $v_{\rm sig}$. To solve for the history of the long modes throughout the interval $t-t_0 = T$, one would need to know the information in a patch of size $R=v_{\rm sig}T$ on the final time slice. This leads to spatial non-locality of the EFT unless $k \, R\ll 1$. For the pressureless cosmic fluid this would not be a concern since $c_s^2=0$ at the leading order.\footnote{One can also check the signal-propagation speed induced by the back reaction of the short scale modes. This can be estimated by the velocity dispersion of the short modes which are not yet virialized, that is $k_s$ not much larger than $k_{NL}$. Using $v_i\simeq -\cH (\d_i/\lap)\delta$, which is valid for $k<k_{NL}$, and $\delta(k_{NL})\sim 1$, we get $v_{\rm max}\sim \cH/k_{NL}$. Assuming $v_{\rm sig}\sim v_{\rm max}$ and $T\sim \cH^{-1}$ we get $k R\sim k/k_{NL}\ll 1$ \cite{Baumann2012}.} However, in the Newtonian limit, the gravitational interaction is instantaneous (in other words, $c \,T$ is much larger than the length scales of interest). Hence, the long memory effect leads to appearance of spatially non-local terms in the EFTofLSS as we will explicitly see below.

\comment{In practice one just writes down the most general effective fluid equation as a double expansion in powers of $k$ and perturbation fields with unknown time-dependent coefficients which have to be fixed by simulations or observations. }

Let us have a closer look at the particular non-local term in Eq.~\eqref{tide2}. For a fixed $(\xv,\tau)$ we can Taylor-expand $\delta(\xv_{fl}[\tau'],\tau')$ in the integrand around the final time

\be
\label{Dtn}
\delta(\xv_{fl}[\tau],\tau') \eq \sum_n \frac{1}{n!}(\tau'-\tau)^n\frac{d^n}{{d\tau'}^n} \delta (\xv_{fl}[\tau],\tau)  \eq \sum_n \frac{1}{n!} (\tau'-\tau)^n D_\tau^n \delta (\xv,\tau) \;,
\ee
where the convective derivative is defined as
\be
D_\tau\equiv \d_\tau+\vv\cdot \nabla \;,
\ee
and we used $d \xv_{fl}/d\tau' ={\vv}(\xv_{fl}[\tau'],\tau')$. At this point, the integration over $\tau'$ just gives time-dependent coefficients for an expansion in convective derivatives: $(TD_\tau)^n \delta(x,\tau)$, where $T$ is the characteristic memory of the kernel $K(\tau,\tau')$. Using the fluid equations, we can rewrite $D_\tau^n\delta$ fully in terms of boost invariant quantities which contain only spatial derivatives. However, $D_\tau\delta = \O(\cH \delta)$ and for $T\sim \cH^{-1}$ all terms in the expansion of Eq.~\eqref{Dtn} contribute at the same order. A similar situation was encountered in the harmonic oscillator toy model, where the derivative expansion broke down for $\omega_2\sim \omega_1$. The right strategy, as before, is to find the full time dependence of $\delta(\xv_{fl}[\tau'],\tau')$ order by order in perturbation theory rather than Taylor expanding it around the final time. This problem is best formulated in Lagrangian perturbation theory.

\begin{figure}
\centering
\includegraphics[width=0.49\textwidth]{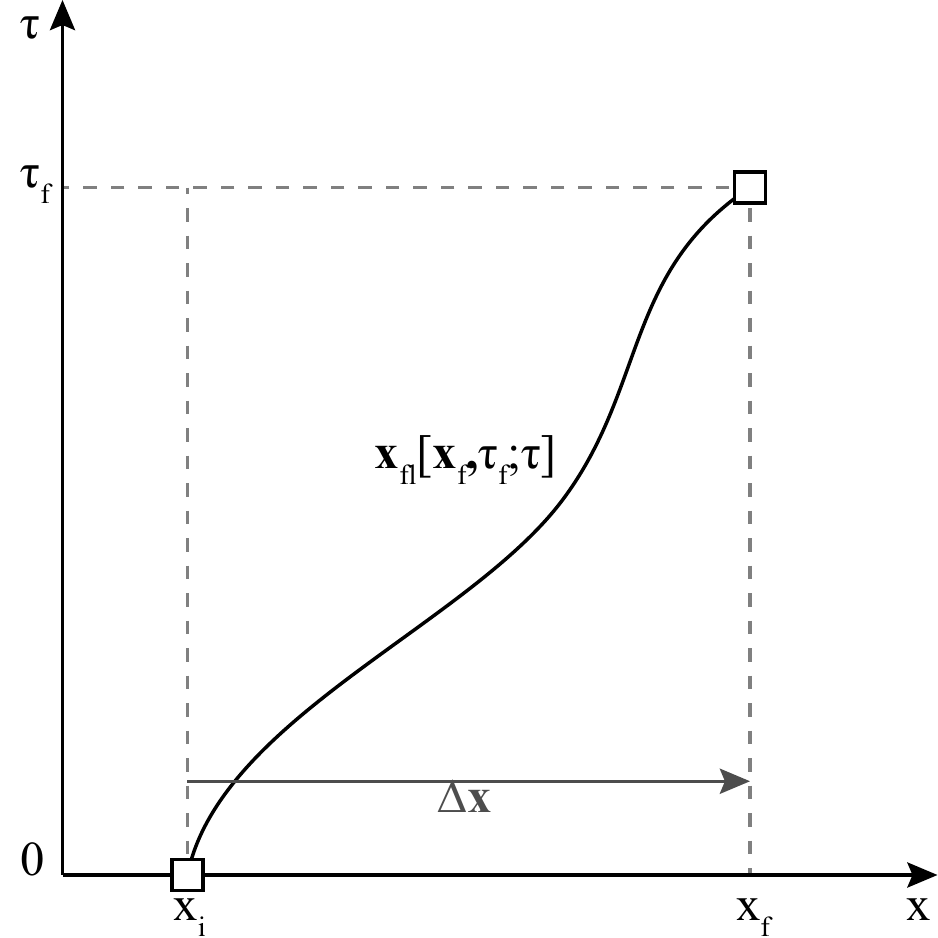}
\caption{Sketch of the fluid trajectory.}
\label{fig:localintime}
\end{figure}

To find $\delta(\xv_{fl}[\tau'],\tau')$ in terms of final data (such as $\delta(\xv,\tau)$ etc.), we can first use perturbation theory to find $\delta(\xv,\tau)$ in terms of the initial data $\delta_{1}$. Inverting this to find $\delta_{1}$ as a function of the final data allows us to express the fields at $\tau'$ in terms of the final data. Consider the solution of the fluid equations \eqref{e:EOMrealspace} with back reaction set to zero, i.e. $\d_j\tau^{ij}=0$. To second order, we find 
\be
\label{delta2}
\delta(\xv,\tau) \eq D_1(\tau) \, \delta_1 + D_1^2(\tau)\left[\frac{5}{7}\,\delta_1^2
+\frac{2}{7}\left(\frac{\d_i\d_j}{\lap} \,\delta_1\right)^2+\frac{\nabla}{\bigtriangleup} \delta_1 \cdot \nabla \delta_1\right]+\O(\delta_1^3) \;.
\ee
Let us calculate the first order displacement of a fluid element between the initial and final time
\be
\Delta \xv_{(1)} \eq \int _{0}^\tau d\tau' \vv_{(1)}(\xv,\tau') \eq  \frac{1}{\cH f} \, \vv_{(1)}(\xv,\tau) \;,
\ee
where we have used that the velocity field is given to first order by
\be
\vv_{(1)}(\xv,\tau) \eq -\cH f \frac{\nabla}{\bigtriangleup} \, \dn{1}(\xv,\tau) \;.
\ee
Equation \eqref{delta2} can now be brought into the Lagrangian form
\begin{equation}
\begin{split}
\delta(\xv,\tau) \eq  & D_1(\tau) \delta_1(\xv_{fl}[0]) \\
& + D_1^2(\tau) \Bigg\{ \frac{5}{7} \, \delta_1(\xv_{fl}[0])^2
+ \frac{2}{7} \left( \frac{\d_i\d_j}{\lap} \, \delta_1(\xv_{fl}[0])\right) ^2 \Bigg\} +\O(\delta_1^3) \;,
\end{split}
\label{eq:finxfl}
\end{equation}
where we have used the shorthand notation $\bm x_{fl}[0]=\bm x_{fl}[\bm x,\tau;0]$. Let us write the equivalent for the above equation for an intermediate point on the fluid trajectory
\begin{equation}
\begin{split}
\delta(\bm x_{fl}[\bm x,\tau;\tau'],\tau')= &D_1(\tau')\delta_1(\bm x_{fl}[0])\\
&+D_1^2(\tau')\Bigg\{ \frac{5}{7} \, \delta_1(\xv_{fl}[0])^2
+ \frac{2}{7} \left( \frac{\d_i\d_j}{\lap} \, \delta_1(\xv_{fl}[0])\right)^2 \Bigg\}+\O(\delta_1^3) \;.
\end{split}
\label{eq:intxfl}
\end{equation}
Here we have used that the initial position is the same for all points on the fluid trajectory, i.e., $\xv_{fl}[\xv_{fl}[x,\tau;\tau'];0]=\xv_{fl}[\xv,\tau;0]$.
To obtain $\delta(\xv_{fl}[\tau'],\tau')$ in terms of $\delta(\xv,\tau)$ we can perturbatively solve Eq.~\eqref{eq:finxfl} for $\delta_1(\xv_{fl}[0])$ and use the result in Eq.~\eqref{eq:intxfl} to obtain
\begin{equation}
\begin{split} \label{zeld}
\delta(\xv_{fl}[\xv,\tau;\tau'],\tau') \eq & \frac{D_1(\tau')}{D_1(\tau)}\delta(\xv,\tau)  +\left(\frac{D_1^2(\tau')}{D_1^2(\tau)}-\frac{D_1(\tau')}{D_1(\tau)}\right) \, \Bigg\{\frac{5}{7} \, \delta^2(\xv,\tau)  \\[1.5ex]
&  +\frac{2}{7} \left(\frac{\d_i\d_j}{\lap}\delta(\xv,\tau)\right)^2 \Bigg\}+\O(\delta^3) \;.
\end{split}
\end{equation}
Although we derived this equation up to second order, the fact that $\delta(\xv_{fl}[\tau'],\tau')$ can be expressed entirely in terms of boost symmetric quantities at the final time $\tau$ is a universal property of the LSS. While the above derivation employed the $\dn{n}=D_1^n \delta_1^n$ approximation it can be made exact for EdS by simply replacing $D_1\to a$. Also, we already see the appearance of spatial non-locality in the last term induced by the Poisson equation of Newtonian gravity. Working with the gravitational potential this reduces to the local term $(\d_i\d_j\phi)^2$. However, this will not continue to be the case at higher orders unless convective time derivatives are included \cite{Mirbabayi:2014}.\footnote{Consider for instance $[\d_i\d_j\phi(\xv_{fl}[\tau'],\tau')]^2$ when one or both $\phi$'s are expanded to second order in terms of the final $\phi(\xv,\tau)$.} These interactions with time derivatives start to appear at cubic order which is relevant for the one-loop four-point function or two-loop two-point function. 

Finally, let us note that the first term in Eq.~\eqref{zeld} is just the Zel'dovich approximation which after substitution in Eq.~\eqref{tide2} gives the usual sound speed term $c_s^2 \lap\delta$ in the Euler equation. Compared to an Eulerian expansion of $\delta(\xv_{fl}[\tau'],\tau')$ in powers of displacement $\Delta \xv  = \xv_{fl}-\xv$:
\be
\label{disp}
\delta(\xv_{fl}[\tau'],\tau')=\sum_n (\Delta \xv \cdot \nabla)^n \delta(\xv,\tau')
\ee
the Zel'dovich approximation $c_s^2 \lap\delta$ resums an infinite number of terms, namely the full displacement of the linearly evolving $\delta$. We believe this to be the reason why a local kernel $K(\tau,\tau')\propto \delta_D(\tau-\tau')$ which collapses Eq.~\eqref{tide2} to the sound-speed term was preferred by the simulations in Ref.~\cite{Carrasco2013a}. Higher order terms on the r.h.s. of Eq.~\eqref{zeld} correspond to higher orders in Lagrangian perturbation theory and contribute to the higher order counterterms in the EFT. Note that unlike in Eq.~\eqref{disp}, each term is manifestly boost invariant.

Let us summarize again the important result of this section. We showed that even if we start with a non-local kernel $K(\tau, \tau')$ in Eq.~\eqref{tide2}, in the perturbative regime we can always write the operators in $\tau^{ij}$ as operators which are evaluated at the same point in time. 
In practice this means that we can assume locality in time from the start: all memory effects are captured in the free, time-dependent  parameters of $\tau^{ij}$.


\subsection{The effective stress tensor}

Having found that we can write down effective stress tensor as a quantity that is local in time and space (at least at the order which is needed here), we can proceed and explicitly derive its form at the NLO. There are two different kinds of terms that can appear: viscosity terms, which are terms written in terms of the smoothed fields, and noise or stochastic terms which take into account the difference of the actual value of the stress tensor and its expectation value. We therefore rewrite the quantities  $- \partial_j \tau^{ij}  /(a \rho) $ and $\tau_\theta$ as

\begin{equation}
 -  \frac{1}{a \rho} \, \partial_j \tau^{ij}  \eeq \tau^i_\textit{vis} + \tau^i_\textit{noise}  \;, \qquad \quad \mbox{and} \quad \qquad \tau_\theta \eeq \tau_\theta^\textit{vis} + \tau_\theta^\textit{noise} \;. 
\end{equation}
The relevant difference between the noise and viscosity terms is that they scale differently with $k$. This is what at the end of the day will allow us to renormalize all one-loop integrals of the bispectrum despite their distinct scalings (see Eq.~\eqref{e:Bscaling}).

\paragraph{Viscosity terms}

In the spirit of the EFT philosophy, the aim is to write down all possible terms that are allowed by symmetry in a double expansion in derivatives and powers of smoothed fields. In the literature on EFTofLSS, mostly terms of the order $~\partial^i \delta$, i.e.~with one derivative and one field, have been considered at the level of $\tau^i_\textit{vis}$ (see Refs.~\cite{Carrasco2013a,Assassi2014} for exceptions). However, for the bispectrum we need to include terms that have the same number of derivatives but with one more field $\sim \partial^i \delta^2$.  It is convenient to work with the rescaled Newtonian potential $\Phi \equiv 2 \phi/ (3 \Omega_m \cH^ 2) $ and the velocity potential $u$ that satisfy

\begin{equation}\label{Phi}
\bigtriangleup \Phi \eq \delta\;, \qquad \quad \mbox{and} \quad \qquad \nabla u = \vv \;,
\end{equation}
rather than with the density and velocity fields themselves. In order to satisfy the equivalence principle and Galileo invariance, we can only allow for terms that have two derivatives acting $\Phi$ and $u$. 

At the LO the only possible terms with at least two derivatives on $\Phi$ and $u$ are

\begin{equation}\label{e:tauLO}
\tau^i_\textit{vis} \Big|_{LO} \eq - c_s^2 \partial^i \bigtriangleup \Phi + \frac{c_v^2}{\cH \, f} \, \partial^i \bigtriangleup u \;,
\end{equation}
where $c_s^2$ and $c_v^2$ are functions of time. Eq.~\eqref{e:tauLO} reduces to the usual $\tau_\theta^\textit{vis} = \partial_i \tau^i_\textit{vis} = - d^2 \bigtriangleup \dn{1}$ when we exchange $\Phi$ and $u$ for $\delta$ and $\theta$ and insert the linear solution $\dn{1} = \tn{1}$ in Eq.~\eqref{e:tauLO}. We shall often use the variable $d^2$ which is defined as

\begin{equation}\label{e:defd}
d^2 \eeq c_s^2 + c_v^2 \;.
\end{equation}

When considering the next higher order in powers of fields, we have more possibilities to combine derivatives and fields. 
The new NLO viscosity terms in the stress tensor are given by all possible combinations of products of two fields with at least two derivatives acting on them

\begin{equation}
\tau^{ij} \big|_\textit{NLO} \eq - d^2 \,  \delta^{ij}  \,  \dn{2}  +   \tilde{c}_1 \, \delta^{ij} (\bigtriangleup \Phi )^2+ \tilde{c}_2\, \partial^i \partial^j \Phi \, \bigtriangleup \Phi + \tilde{c}_3 \, \partial^i \partial_k \Phi \, \partial^j \partial^k \Phi\; .
\end{equation}
Note that we get a NLO contribution from the LO term in Eq.~\eqref{e:tauLO} when $\delta$ is replaced with $\dn{2}$. 
The right hand side of the Euler equation is obtained by taking one spatial derivative of the above equation and dividing by $a\rho$. We slightly reorder the terms and choose the following set of independent operators:

\begin{equation}\label{e:NLOoperators}
\partial^i (\Delta \Phi)^2 \;, \qquad \quad \partial^i s^2 \;, \qquad \mbox{and} \qquad s^{ij} \partial_j \bigtriangleup \Phi \;,
\end{equation}
where we defined the tidal tensor as
\begin{equation}\label{defs}
s_{ij} \eeq \partial_i \partial_j \Phi - \frac{1}{3} \delta_{ij} \, \Delta \Phi \;,
\end{equation}
and $s^2 \equiv s_{ij} s^{ij}$. Since the terms in Eq.~\eqref{e:NLOoperators} contain always two fields $\Phi$, we can have the same terms with either one or both $\Phi$ replaced by the velocity potential $u$. However, since we do not need any higher order terms, we can use the fact that at the linear order we have $\dn{1} = \tn{1}$ which means that at this order $\Phi = - \cH f \, u $. Therefore, adding terms with $u$ to the ones in Eq.~\eqref{e:NLOoperators} would not generate anything new at this order in perturbation theory. Note that expanding the factor $1/\rho$ that multiplies $\tau^{ij}$ in the Euler equation \eqref{e:EOMrealspace} does not lead to additional operators. It would generate terms that are simply the LO expression for $\tau^{ij}$ multiplied by $\delta$. Such terms can be reabsorbed through a redefinition of $\tilde{c}_{1,2}$. We should keep in mind that the terms we wrote down in Eq.~\eqref{e:tauLO} will generate two new terms at the NLO which are not contained in Eq.~\eqref{e:NLOoperators}. If we insert the second order solutions $\dn{2} $ and $\tn{2}$ in Eq.~\eqref{e:tauLO}, we get the following expression for the NLO $\tau^i_\textit{vis}$
\begin{equation}\label{e:tauNLOe}
\tau^i_\textit{vis} \Big|_{NLO} \eq  - c_s^2 \, \partial^i  \dn{2} + \frac{c_v^2}{\cH \, f} \, \partial^i \tn{2}  - c_1 \, \partial^i \dn{1}^2 - c_2 \, \partial^i s^2 - c_3 \, s^{ij} \partial_j\dn{1}  \;.
\end{equation}
Since the two kernels $F_2$ and $G_2$ are not the same, it looks as if at this order we can break the degeneracy between $c_s^2$ and $c_v^2$. As we shall show explicitly in Sec.~\ref{s:NLOsolution}, the difference between $\partial^i \dn{2}$ and $\partial^i \tn{2}$ can be absorbed in a redefinition of $c_1$ and $c_2$. This means that we will not be able to distinguish $c_s^2$ and $c_v^2$ after all. The first two terms in Eq.~\eqref{e:tauNLOe} collapse to one that we choose to coincide with what is found in the literature, i.e.~$- d^2 \, \partial^i \dn{2}$. 

In sum, the viscosity terms of the effective stress tensor that enter the equations of motion for $\theta$ are given by

\begin{eqnarray}
&& \tau_\theta^\textit{vis} \big|_{LO} \eq  - d^2 \, \bigtriangleup \dn{1}   \;, \label{e:tauvisLO} \\[1.5ex]
&& \tau_\theta^\textit{vis} \big|_\textit{NLO} \eq  - d^2 \, \bigtriangleup  \dn{2}  - e_1 \, \bigtriangleup \dn{1}^2 - e_2 \, \bigtriangleup s^2 - e_3 \, \partial_i \left[ s^{ij} \partial_j\dn{1}  \right] \,, \label{e:tauvisNLO}
\end{eqnarray}
and contain four free parameters that are not determined by the theory itself ($e_i$ are a function of $c_i$ and $c_v^2$) . 

\paragraph{Noise terms}

The noise terms in the effective stress tensor play a crucial role in the renormalization procedure of the EFTofLSS. Although they can be considered as suppressed compared to the viscosity terms in $\Lambda$CDM, we need to include them if we want to check that there is a bispectrum that is free of any UV-divergences. The LO noise term has been discussed already in the literature

\begin{equation}
\tau^i_\textit{noise} \big|_\textit{LO} \eq - J_0^i  \;,
\end{equation}
where the quantity $J_0 \equiv \partial_i J_0^i $ is only correlated with itself or other noise terms. In particular, it is uncorrelated with the smoothed density field $\EV{J_0}{\delta} = 0$. For the renormalization of the bispectrum we need to include terms that are of the order of $\sim J_0 \, \delta$. Since the noise terms can be scalars, vectors and tensors, we can build quite a few terms at the NLO

\begin{equation}
\begin{split}
\tau^i_\textit{noise} \big|_\textit{NLO} \eq & - \tilde{J}_1^{\,i} \,  \bigtriangleup \Phi -   \tilde{J}_2^{\,j} \,  \partial_j \partial^i  \Phi -  \tilde{J}_3^{\,ijk} \,  \partial_j \partial_k \Phi -  \tilde{J}_4 \,  \partial^i \bigtriangleup  \Phi \\[1.5ex]
&  -  \tilde{J}_5^{\,ij} \,  \partial_j \bigtriangleup  \Phi  -  \tilde{J}_6^{\,jk} \,  \partial^i \partial_j \partial_k  \Phi -  \tilde{J}_7^{\, ijkl} \,  \partial_j \partial_k \partial_l \Phi \;.
\end{split}
\end{equation}
As for the viscosity terms, we only add terms that contain two derivatives acting on $\Phi$. At the level of $\tau_\theta$ these expressions reduce to 

\begin{eqnarray}
&& \tau_\theta^\textit{noise} \big|_\textit{LO} \eq  - J_0  \;, \label{e:taunoiseLO} \\[1.5ex]
&& \tau_\theta^\textit{noise} \big|_\textit{NLO} \eq   - \bar{J}_1 \, \bigtriangleup \Phi  - \bar{J}_2^{ij} \, \partial_i \partial_j \Phi - \bar{J}_3^i \, \partial_i \bigtriangleup \Phi - \bar{J}_4^{ijk} \, \partial_i \partial_j \partial_k \Phi \nonumber \\
&& \phantom{\tau_\theta^\textit{noise} \big|_\textit{NLO} \eq} -   \bar{J}_5 \, \bigtriangleup^2 \Phi  - \bar{J}_6^{ij} \, \partial_i \partial_j \bigtriangleup \Phi - \bar{J}_7^{ijkl}\partial_i \partial_j \partial_k \partial_k \Phi \;, \label{e:taunoiseNLO}
\end{eqnarray}
where we absorbed derivatives acting on the $\tilde{J}$ into a redefinition of $\bar{J}$. Note that the correlators $\EV{\bar{J}_r}{\bar{J}_s}$ are all different and not related to $\EV{J_0}{J_0}$. Note that one could imagine to add terms with five or more derivatives acting on $\Phi$. However, since also in the viscosity terms we stopped at the level of four derivatives, we do the same here. As we shall see in Sec.~\ref{s:renorm}, the terms in Eq.~\eqref{e:taunoiseNLO} are sufficient to renormalize the one-loop bispectrum.

\section{EFT contributions to the bispectrum}\label{s:eftbispectrum}

In this section we discuss in more detail the EFT contributions to the bispectrum. From the NLO effective stress tensor we derive the corrections to the SPT solution and show that the resulting three-point correlators renormalize the one-loop integrals that were discussed in Secs.~\ref{s:sptbispectrum} and \ref{s:uvlimits}.


\subsection{NLO EFT solution} \label{s:NLOsolution}

\begin{figure}
\centering
\includegraphics[width=5cm]{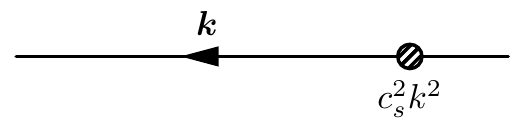}
\caption{LO counterterm.}
\label{fig:loctr}
\end{figure}

We want to compute the LO and NLO corrections to the SPT solution in Eq.~\eqref{e:SPTsol} that are generated by the effective stress tensor. Let us start by writing the perturbative solutions for $\delta$ and $\theta$ as

\begin{equation}
\delta=\delta^\textit{SPT}+\delta^c+\delta^J \;, \quad \qquad \theta=\theta^\textit{SPT}+\theta^c+\theta^J
\; .
\end{equation}

\begin{equation}\label{e:fulldeltatau}
\delta^\textit{SPT,c,J} \eq \sum_n   \dn{n}^\textit{SPT,c,J}  \;, \quad \qquad \theta^\textit{SPT,c,J} \eq - \cH \, f \sum_n   \tn{n}^\textit{SPT,c,J} \;,
\end{equation}
where we separate the SPT from the noise and viscosity contributions. For the SPT and viscosity part the index $n$ denotes the number of $\dn{1}$ that are convolved, i.e.~$\dn{n} \sim \dn{n}^c \sim \dn{1}^n$ (see also Eq.~\eqref{e:fngn}), while the noise part has one $\dn{1}$ less $\dn{n}^J \sim \dn{1}^{n-1}$ and likewise for $\theta$.  For the bispectrum, we will need the EFT terms up to second order $\dn{1,2}^{c,J}$. At the LO, we can easily compute $\dn{1}^{c,J}$ and $\tn{1}^{c,J}$ using the Green's function of Eq.~\eqref{e:greens}

\begin{equation}\label{e:deltatauc1}
\begin{split}
& \dn{1}^{c,J} \eq \int da' G_\delta (a, a') \tau_\theta^\textit{vis,noise} \Big|_{ LO} \;, \\[1.5ex]
& \tn{1}^{c,J} \eq  - \frac{1}{\cH \, f} \int da' G_\theta (a, a') \partial_{a'} \left\{ a'\, \tau_\theta^\textit{vis,noise} \Big|_{ LO} \right\} \;,
\end{split}
\end{equation}
where one just picks the corresponding LO viscosity or noise terms in the effective stress tensor $\tau_\theta$ that are found in Eqs.~\eqref{e:tauvisLO} and \eqref{e:taunoiseLO}. Note that for $\Lambda$CDM one should use the continuity equation $a\partial_a \dn{1}^{c,J} = - f \, \tn{1}^{c,J}$ to compute $\tn{1}^{c,J}$ rather than the Green's function $G_\theta$. 
The diagrammatic representation of $\dn{1}$ is fairly simple. It is just represented by a vertex with two legs as shown in Fig.~\ref{fig:loctr}.

In general we do not know the time dependence of free parameters in $\tau_\theta$ and it is not possible to evaluate the above integral explicitly. One possibility is to just redefine the LO coefficients $d$ and $J_0$ such that e.g.~for $\dn{1}^c$ we have 

\begin{equation}\label{e:defgamma}
\dn{1}^c (\kv,a) \eq - \gamma \, k^2 \dn{1}(\kv,a) \;, \qquad 
\end{equation}
where $\gamma$ is the time integral over the Green's function (the precise definition will follow). The time dependent parameter $\gamma$ can then be fitted directly at a certain point in time without having to assume a certain time dependence of $d^2$. This, however, has the disadvantage that the parameters that are determined through simulations or observations are not the parameters that enter the equations of motion. As we shall see, at NLO it is not possible to simply absorb all time integrals over the Green's function in the definition of new parameters. We therefore assume an explicit time dependence for the free parameters that allows us to perform the time integrals over the Green's function. In Sec.~\ref{s:simulations} we will use a set of parameters that have a minimal time dependence..  In EdS, the time dependence of the free parameters is fixed by the self-similarity of the equations of motion and the cancellation of the divergences (see Ref.~\cite{Pajer2013a}) and we shall use this fact to infer the time dependence of the parameters in $\Lambda$CDM. Let us discuss this more in detail.

\paragraph{Viscosity terms} 

\begin{figure}
\centering
\includegraphics[width=16cm]{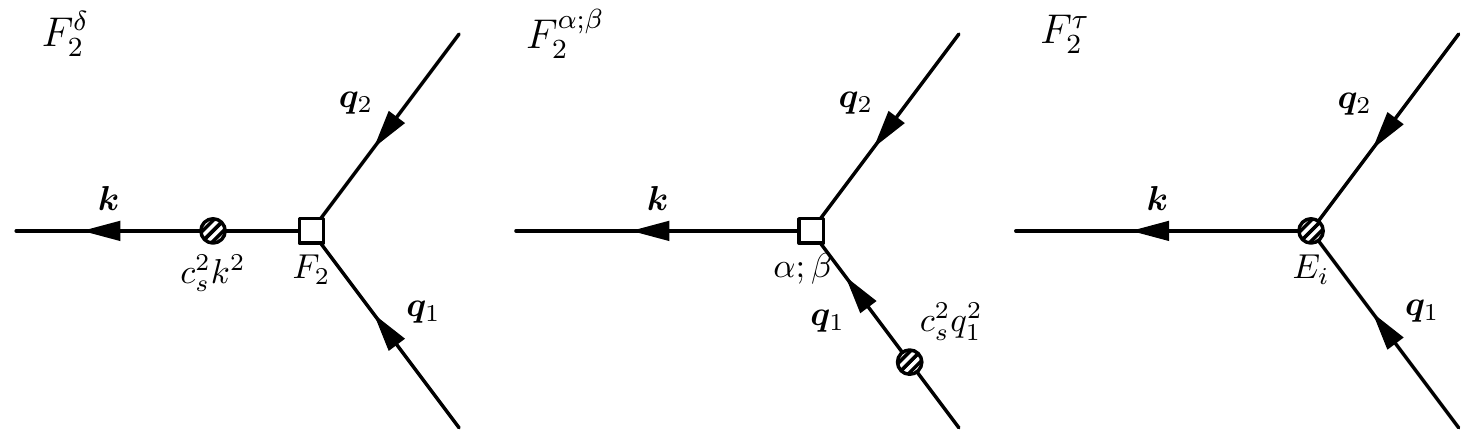}
\caption{NLO counterterms.}
\label{fig:nloctr}
\end{figure}

Before explicitly deriving $\dn{2}^c$, let us try to guess what we will eventually get.\footnote{We do not consider $\tn{2}^c$ since at the one-loop level for the bispectrum it is not needed. But its computation is analogous to the one for $\dn{2}^c$.} From the form of the NLO stress tensor in Eq.~\eqref{e:tauvisNLO}, we already know that in Fourier space $\dn{2}^c$ will be a convolution of of two $\dn{1}$ with some kernel that depends on time through the EFT coefficients of the effective stress tensor
\begin{equation}\label{e:deltac2}
\dn{2}^c (\kv , a) \eq \intq \, F_2^c (\qv, \kv-\qv, a)  \, \dn{1} (\qv,a) \dn{1}(\kv -\qv,a)  \;.
\end{equation}
The kernel $F_2^c$ receives three contributions that are quite different in their origin. These contributions can be represented diagrammatically as shown in Fig.~\ref{fig:nloctr}. The three diagrams represent the following contributions 
\begin{equation}\label{e:Fc2}
 F_2^c (\qv_1, \qv_2, a) \eeq F_2^\tau(\qv_1, \qv_2, a) + F_2^{\alpha\beta}(\qv_1, \qv_2, a) + F_2^{\delta}(\qv_1, \qv_2, a)\;.
\end{equation}
Note that, unlike the SPT kernels $F_n$ and $G_n$ the kernel $F_2^c$ has two more powers of momentum in the numerator. $F_2^\tau$ is generated by the NLO terms in Eq.~\eqref{e:tauvisNLO} that introduce new free parameters, i.e.~it contains $e_1$, $e_2$ and $e_3$. Therefore, in Fig.~\ref{fig:nloctr} it takes the form of a (point-like) three-leg vertex. The other two kernels contain only $d^2$. While $F_2^{\alpha\beta}$ stems from the non-linear corrections to the LO EFT solution $\dn{1}^c$, i.e. from inserting $\dn{1}^c$ and $\tn{1}^c$ in the source terms $\mathcal{S}_\alpha$ and $\mathcal{S}_\beta$, $F_2^\delta$ is merely $ \sim d^2 \, k^2 F_2$ with an extra factor that depends on the time integration. The latter kernel obviously captures the contribution from the first term in Eq.~\eqref{e:tauvisNLO}. Diagrammatically, the kernels $F_2^{\alpha\beta}$ and $F_2^\delta$ are therefore represented by the combination of a counterterm vertex with a three-point (SPT) vertex attached to it which then stands for either $F_2$ or a combination of $\alpha$ and $\beta$.  

Let us be a bit more specific. To compute the kernels $F_2^\tau$, $F_2^{\alpha \beta}$ and $F_2^\delta$ we need to discuss the time dependence of the free parameters. This has been considered in detail in Ref.~\cite{Pajer2013a} in the case of a scale-free universe. It was found that the time dependence of the counterterm part of the parameters has to be such that it matches exactly the time dependence of the divergent one-loop integrals. In practice this means that the divergent (counterterm) part scales with time as 
\begin{equation}
d^2 \big|_\textit{ctr} \; \sim \;  e_{1,2,3} \big|_\textit{ctr} \; \sim \; a \;. 
\end{equation}
On the other hand, the time dependence of the finite (renormalized) part of the free parameters is fully determined by the self-similarity of the equations of motion and it was shown that 
\begin{equation}\label{e:EdSscalingren}
d^2 \big|_\textit{ren} \; \sim \;  e_{1,2,3} \big|_\textit{ren} \; \sim \; a^{(1-n)/(n+3)} \;.
\end{equation}
where $n$ is the spectral index of the linear power spectrum $\Pl (k) \propto k^n$. The time dependence of the NLO parameters $e_{1,2,3}$ is same as the one of the LO coefficients since at the level of the equations of motion both have to satisfy the same self-similarity condition (both are of the order $\sim k^2$ in the derivative expansion). Knowing how $d^2$ and $e_{1,2,3}$ depend on time, it is possible to explicitly compute the time integrals in Eq.~\eqref{e:deltatauc1} (as well as in the analogous NLO expressions).

In $\Lambda$CDM, however, we do not have a symmetry that fixes the time dependence of the parameters. Even though at the end of the day we measure the parameters through simulations at a fixed moment in time, we still need to make some assumptions about the time dependence of the free parameters. The reason is that $\dn{2}^c$ also depends on the LO coefficient $d^2$ and we need to know how exactly. Let us therefore assume that the following scaling holds in a $\Lambda$CDM universe
\begin{equation}\label{e:mscaling}
d^2 \; \sim \; e_{1,2,3} \; \sim \; D_1^{m}(a)
\end{equation}
for some number $m$ (in $\Lambda$CDM there are no divergences, so we do not need to distinguish between the counterterm and renormalized part of the free parameters). We factorize the time dependence and a factor of $\cH_0^{2} \Omega_m^0$ in order to work with the constant parameters $\bar{d}^{\, 2}$ and $\bar{e}_{1,2,3}$ that are defined through 
\begin{equation}\label{e:constpar}
d^2 \eeq D_1^{m}(a) \, \cH_0^2 \Omega_m^0 \;  \bar{d}^{\,2}  \;, \qquad \qquad e_{1,2,3} \eeq D_1^{m}(a) \,\cH_0^2 \Omega_m^0 \; \bar{e}_{1,2,3}   \;,
\end{equation} 
One possible choice for $m$ is to use the EdS scaling of $m=(1-n)/(n+3)$ of Eq.~\eqref{e:EdSscalingren} and apply it also in $\Lambda$CDM for the phenomenologically viable value of $n \approx -3/2$. Or we can choose $m=1$ in order to match the time dependence of the loop integrals. When fitting for the free parameters in Sec.~\ref{s:simulations}, we will consider only these two cases
\begin{equation}\label{e:mvalues}
m \eq \left\{  \begin{array}{c} \frac{1-n}{n+3} = \frac{5}{3} \\[1.5ex] 1 \end{array} \right.  \;.
\end{equation} 
Assuming the scaling of Eq.~\eqref{e:mscaling} for the free parameters, we can write the LO EFT solutions $\dn{1}^c$ and $\tn{1}^c$ as 
\begin{eqnarray}
&& \dn{1}^c (\kv, a) \eq  g_1(a,m) \bar{d}^{\,2} \,  k^2 \dn{1}(\kv,a)  \;, \label{e:deltac1} \\[1.5ex]
&&  \tn{1}^c (\kv, a) \eq  h_1(a,m) \bar{d}^{\,2} \,  k^2 \dn{1}(\kv,a) \;, \label{e:thetac1}
\end{eqnarray}
where the two dimensionless functions $g_1$ and $h_1$ contain the time integrals of Eq.~\eqref{e:deltatauc1}
\begin{eqnarray}
&& g_1(a,m) \eeq   \frac{\cH_0^2\Omega_m^0}{D_1(a)} \int da'\; G_\delta (a,a')  D_1^{m+1}(a') \;, \\[1.5ex]
&& h_1(a,m) \eeq - \frac{\cH_0^2\Omega_m^0}{\cH f(a) \, D_1(a)} \int da'\; G_\theta (a,a') \partial_{a'} \left\{ a'  D_1^{m+1}(a') \right\} \;.
\end{eqnarray}
Of course, once $g_1$ is computed, it is easier to use the continuity equation to get $\theta$, i.e.~$a \partial_a \, (g_1\,D_1) = f\, h_1 D_1$. 
The fitting parameter $\gamma$ defined in Eq.~\eqref{e:defgamma} is then related to the constant parameter $\bar{d}^{\,2}$ as 
\begin{equation}\label{dtogamma}
\gamma \eq - g_1(a,m) \, \bar{d}^{\,2} \;.
\end{equation}
Note that we factored one power of $D_1$ in order to restore the correct time dependence of $\dn{1}$ in Eqs.~\eqref{e:deltac1} and \eqref{e:thetac1}. In EdS, the integrals in $g_1$ and $h_1$ can be computed analytically and give rather simple results for the dependence on $a$ and $m$. We can then extrapolate the EdS result to $\Lambda$CDM by replacing $a$ with $D_1$ and by adopting $\cH_0$ accordingly (remember that when matching $\cH$ in EdS and $\Lambda$CDM at early times one has to keep track of the factor $\sqrt{\Omega_m^0}$). The results are

\begin{eqnarray}\label{g1}
&& g_1(a,m)  \eq  - \frac{2\, }{(m+1)(2m+7)} \, D_1^{m+1}(a) \;, \\[1.5ex]
&& h_1(a,m)  \eq (m+2) \, g_1(a,m) \;.
\end{eqnarray}
%
For convenience, we will always use the approximated forms of $g_1$ and $h_1$. Note that to link the parameters in the equations of motion to the parameters in $\dn{1}^c$ and $\tn{1}^c$ one has to rely on the values of $g_1$ and $h_1$.

Next, let us compute the kernel $F_2^c$. $F_2^\tau$ can be computed in complete analogy to Eq.~\eqref{e:deltatauc1} by evaluating integral over the Green's function and the $d^2$-independent NLO terms in the effective stress tensor $\tau_\theta^\textit{vis}\big|_{NLO}$ (see Eq.~\eqref{e:tauvisNLO}). It takes the form
\begin{equation}\label{e:F2tau}
F_2^\tau (\qv_1, \qv_2, a) \eq  g_2^e(a,m) \, \sum_{i=1}^3  \bar{e}_i \,  E_i(\qv_1 , \qv_2) \;,
\end{equation}
where the (time independent) shape functions $E_i$ are given by 
\begin{equation}\label{e:Ei}
\begin{split}
E_1 (\qv_1, \qv_2) \eq &  (\qv_1 +\qv_2)^2   \;,\\[1.5ex]
E_2 (\qv_1, \qv_2) \eq & (\qv_1 +\qv_2)^2 \left[ -\frac{1}{3} + \frac{(\qv_1\cdot \qv_2)^2}{\qv_1^2 \qv_2^2 } \right]   \;, \\[1.5ex]
E_3 (\qv_1, \qv_2) \eq &  \left[- \frac{(\qv_1 +\qv_2)^2}{6}   +\frac{\qv_1\cdot \qv_2}{2} \left[\frac{(\qv_1 +\qv_2) \cdot \qv_2}{\qv_2^2} + \frac{\qv_1 \cdot (\qv_1 +\qv_2) }{\qv_1^2}\right] \right]  \;.
\end{split}
\end{equation}
The function $g_2^{e}$ is the analogue of $g_1$ 

\begin{equation}
g_2^e (a,m) \eeq \frac{\cH_0^2 \Omega_m^0 }{D_1^2(a)} \int da' \; G_\delta (a, a') \, D_1^{m+2}(a') \;.
\end{equation}
Computing the time integral in EdS and then generalizing the result to $\Lambda$CDM gives 

\begin{equation}\label{g2e}
g_2^e(a,m)  \eq  -  \frac{2}{(m+2)(2m+9)} \, D_1^{m+1}(a) \;.
\end{equation}
As in the case of $g_1$ and $h_1$ we factored out $D_1^2$ in order to restore the time dependence of the two $\dn{1}$ in Eq.~\eqref{e:deltac2}. Similar to the parameter $\gamma$, it will be useful for the fitting procedure in Sec.~\ref{s:simulations} to define three parameters $\epsilon_{1,2,3}$ that relate to $e_{1,2,3}$ as

\begin{equation}\label{e2e}
\epsilon_i \eeq - g_2^e(a,m) \, \bar{e}_i \;.
\end{equation}
Again, we effectively absorbed the time integral into the definition of $\epsilon_i$.

The second kernel $F_2^{\alpha\beta}$ is obtained by inserting $\dn{1} + \dn{1}^c$ and $\tn{1} + \tn{1}^c$ into the source terms $\mathcal{S}_\alpha$ and $\mathcal{S}_\beta$ in Eq.~\eqref{e:Salphabeta} and collecting all terms that give something of the order $\sim d^2 \bigtriangleup \delta^2$. Taking the integral over the Green's function of these new source terms gives

\begin{equation}\label{e:Falphabeta}
\begin{split}
F_2^{\alpha\beta}(\qv_1,\qv_2,a) \eeq & - \gamma \, \bar{F}_2^{\alpha \beta}(\qv_1,\qv_2,a) \\[1.5ex]
\eq & - \gamma \Bigg\{ \qv_1^2 \left[ \frac{g_2^\alpha(a,m)}{g_1(a,m)}  \, \alpha (\qv_2, \qv_1) + \frac{ \tilde{g}_2^\alpha (a,m) }{g_1(a,m)} \, \alpha (\qv_1, \qv_2) + \frac{ g_2^\beta(a,m) }{g_1(a,m)} \, \beta (\qv_1, \qv_2) \right]   \\[1.5ex]
& +  \qv_2^2 \left[ \frac{g_2^\alpha (a,m)}{g_1(a,m)} \, \alpha (\qv_1, \qv_2) + \frac{\tilde{g}_2^\alpha (a,m) }{g_1(a,m)} \, \alpha (\qv_2, \qv_1) + \frac{ g_2^\beta  (a,m) }{g_1(a,m)} \, \beta (\qv_1, \qv_2) \right] \Bigg\} \;,
\end{split}
\end{equation}
where we directly wrote $F_2^{\alpha\beta}$ in terms of $\gamma$ rather than with $\bar{d}^{\,2}$. As before, the functions $g_2^\alpha$, $\tilde{g}_2^\alpha$ and $g_2^\beta$ encode the time dependence and are given by

\begin{eqnarray}
 g_2^\alpha (a,m) &\eeq &- \frac{1 }{2 \, D_1^2(a)}\int da' \; G_\delta(a, a') \,\cH(a') \, \partial_{a'} \Big\{ a'\, \cH(a') f(a') \, g_1(a',m) D_1^2(a') \Big\} \,, \\[1.5ex]
 \tilde{g}_2^\alpha (a,m) & \eeq &  - \frac{1}{2 \, D_1^2(a)} \int da' \; G_\delta(a, a') \,\cH(a')\, \partial_{a'} \Big\{ a' \, \cH(a') f(a') \, h_1(a',m) D_1^2(a') \Big\} \,,\\[1.5ex]
 g_2^\beta (a,m) & \eeq & -  \frac{1}{D_1^2(a)} \int da' \; G_\delta(a, a') \, \Big[ \cH(a') f(a') \Big]^2 \, h_1(a',m) D_1^2(a') \,.
\end{eqnarray}
It now becomes obvious why we need to know explicitly the time dependence of the LO solutions $\dn{1}^c$ and $\tn{1}^c$. While we can absorb all time dependence of the LO into $\gamma$, the time dependence of $\bar{F}^{\alpha\beta}$, shown in Eq.~\eqref{e:Falphabeta} by the ratios of the various functions, cannot be avoided because $\alpha$ and $\beta$ come with a different time dependence. Also, it should be noted that it is \emph{not} possible to fully rewrite the terms in the $F_2^{\alpha\beta}$ kernel as a linear combination of the terms in $F_2^{\tau}$ and it will turn out to be crucial to have these terms included in order to cancel all UV-divergences of the bispectrum. We will use the functions $g_2^\alpha$, $\tilde{g}_2^\alpha$ and $g_2^\beta$ as computed in an EdS universe

\begin{eqnarray}\label{g2alpha}
&& g_2^\alpha(a,m)  \eq -  \frac{1}{(m+1)(m+2)(2m+9)}  \, D_1^{m+1}(a)\;, \\[1.5ex]
&& \tilde{g}_2^\alpha(a,m)  \eq (m+2) \, g_2^\alpha(a,m) \;, \\[1.5ex]
&& g_2^\beta(a,m)   \eq -  \frac{4}{(m+1)(2m+7)(2m+9)} \, D_1^{m+1}(a) \;.
\end{eqnarray}
%
As for $g_1$ and $h_1$ we will use the approximated, i.e.~EdS expressions for $g_2^\alpha$, $\tilde{g}_2^\alpha$ and $g_2^\beta$. Let us stress again that while we may fit for the parameters $\gamma$ and $\epsilon_i$, the knowledge of $g_2^\alpha$, $\tilde{g}_2^\alpha$ and $g_2^\beta$ is crucial if we want to measure the same parameters in the power- and bispectrum simultaneously.

Finally, let us consider $F_2^\delta$. As we showed before, we would naively expect that at the NLO it is possible to break the degeneracy between $c_s^2$ and $c_v^2$ since at the NLO the two terms come with two different kernels, i.e.

\begin{equation}
 \sim \; k^2 \intq \, \left\{ c_s^2 \, F_2(\qv, \kv-\qv) + c_v^2 \, G_2(\qv, \kv-\qv) \right\} \dn{1}(\qv,a) \dn{1}(\kv - \qv,a)
\end{equation}
This would give a total number of five free parameters of which the combination $d^2 \equiv c_s^2 + c_v^2$ can be fixed through the power spectrum (at the one-loop level). However, these five parameters are degenerate. In particular, even by measuring the bispectrum we can only determine one combination of $c_s^2$ and $c_v^2$: the LO quantity $d^2$. The SPT kernels $F_2$ and $G_2$ are both made up of three terms, i.e. $\vv \cdot \nabla \delta$, $\delta^2$ and $s^2$ (in real space), that come with different numerical coefficients. But the $\delta^2$ and $s^2$ terms can be absorbed through a redefinition of $c_1$ and $c_2$. The term $\vv \cdot \nabla \delta$, however, appears with exactly the same numerical coefficient in $F_2$ as in $G_2$. This means that the coefficient of $\vv \cdot \nabla \delta$ is proportional to $d^2 \equiv c_s^2 + c_v^2$. The total number of free parameters is therefore only four: $d^2$ and $e_{1,2,3}$ where the latter are a combination of $c_{1,2,3}$ and $c_{s,v}^2$.\footnote{This statement could be made at a more fundamental level \cite{Mirbabayi:2014}. Namely the difference between the $F_i$ and $G_i$ kernels can always be expressed by counterterms that one would have to introduce anyways at the corresponding order. A proof of this statement is beyond the scope of this study.} Choosing the relation between the new parameters $e_{1,2,3}$ and the parameters in Eq.~\eqref{e:tauNLOe} as

\begin{equation}\label{eq:credef}
e_1 \eeq c_1 -  \frac{8}{21} c_v^2 \;, \qquad \quad e_2 \eeq c_2 + \frac{2}{7} \, c_v^2 \;, \qquad \quad e_3 \eeq c_3 \;,
\end{equation}
we can write the kernel $F_2^\delta$ as
\begin{equation}\label{e:Fdelta}
F_2^\delta (\qv_1, \qv_2, a) \eeq  - \gamma \bar{F}_2^\delta(\qv_1, \qv_2, a)  \eq - \gamma \, \frac{g_2^e(a,m)}{g_1(a,m)} \, (\qv_1 +\qv_2)^2 \, F_2(\qv_1, \qv_2) \;. 
\end{equation}
Since we fixed the time dependence of $d^2$ to be the same as the one of $e_{1,2,3}$, the time dependence both $F_2^\tau$ and $F_2^\delta$ is given by the function $g_2^e$.

To summarize, let us rewrite the second order EFT solution for the density contrast
\begin{equation}\label{e:deltac2epsilongamma}
\begin{split}
\dn{2}^c (\kv,a) \eq &  \intq \, F_2^c (\qv, \kv-\qv, a)  \, \dn{1} (\qv,a) \dn{1}(\kv -\qv,a)  \\[1.5ex]
\eq & \intq \, \Big\{ - \sum_{i=1}^3 \, \epsilon_i E_i (\qv, \kv-\qv) - \gamma \, \Big[  \bar{F}_2^{\alpha \beta} (\qv, \kv-\qv,a) + \bar{ F}_2^\delta (\qv, \kv-\qv,a)  \Big] \Big\} \\[1.5ex]
& \phantom{intq } \cdot \dn{1} (\qv,a) \dn{1}(\kv -\qv,a) \;.
\end{split}
\end{equation}
For clarity, let us rephrase the set of parameters that we consider. In the expansion of the effective stress tensor we found in Eqs.~\eqref{e:tauvisLO} and \eqref{e:tauvisNLO} that there are four time dependent viscosity parameters $d^2$ and $e_{1,2,3}$. We then factorize the time dependence and define the constant (and cosmology independent) parameters $\bar{d}^{\,2}$, $\bar{e}_{1,2,3}$. For the fitting procedure, however, it is most convenient to use the (time dependent) fitting parameters $\gamma$ and $\epsilon_{1,2,3}$. They are essentially the time integrals over the Green's function and the parameters from the equations of motion. The relation between $d^2$, $e_{1,2,3}$ and $\gamma$, $\epsilon_{1,2,3}$ is given by the various $g$-functions. Despite the definition of $\gamma$, $\epsilon_{1,2,3}$, there is an implicit time dependence in $\bar{F}_2^{\alpha\beta}$ and $\bar{F}_2^{\delta}$, meaning that they depend on the parameter $m$ (all powers of $D_1$ have been factorized) . Schematically, we have

\begin{equation}
\begin{array}{ccccc}
\textit{stress tensor} & & \textit{stress tensor} & & \textit{density field}\\
\bar e &\xleftarrow{D^m\cH_0^2\Omega_m^0} &e &\xrightarrow{\int G} & \epsilon_i\\
\bar d &\xleftarrow{D^m\cH_0^2\Omega_m^0} &d &\xrightarrow{\int G} & \gamma
\end{array}
\end{equation}

\paragraph{Noise terms}

According to the power counting in EdS (see discussion in Sec.~\ref{s:edspowercounting}), the noise terms are suppressed compared to the viscosity terms for phenomenologically viable spectral indices. We assume that this also holds in a $\Lambda$CDM universe, which is why we will neglect the noise terms in general through out this paper. However, in order to cure all the UV-divergences of the bispectrum we cannot do without the noise terms. We therefore simplify the following discussion by considering only the EdS case and taking only the counterterm part of the various noise terms in Eqs.~\eqref{e:taunoiseLO} and \eqref{e:taunoiseNLO}. 

For $J_0^\textit{ctr} \sim a$ and an EdS universe the LO noise terms are simply given by the integral over $\tau_\theta^\textit{noise}\big|_{LO}$ in Eq.~\eqref{e:taunoiseLO} 

\begin{eqnarray}
&& \dn{1}^J (\kv, a) \Big|_\textit{EdS} \eq - \int da' \, G_\delta(a, a') J_0^\textit{ctr}(\kv, a') \eq \frac{2}{7} \frac{1}{\cH^2} \, J_0^\textit{ctr} (\kv, a)\;, \\[1.5ex]
&& \tn{1}^J (\kv, a) \Big|_\textit{EdS} \eq 2 \dn{1}^J(\kv, a) \Big|_\textit{EdS}
\end{eqnarray}
At the NLO, we have again two contributions to $\dn{2}^J$: from the non-linear corrections to $\dn{1}^J$ and from $\tau_\theta^\textit{noise}\big|_{NLO}$. Both can be written in the same form

\begin{equation}\label{e:deltaJ2}
\dn{2}^J (\kv, a) \Big|_\textit{EdS} \eq \frac{1}{\cH^2} \, \intq \, \sum_{n=0}^8 F_2^{J_n}  (\qv, \kv-\qv) \,  \dn{1}(\qv)  J^\textit{ctr}_n(\kv-\qv,a) \;, 
\end{equation}
The first kernel $F_2^{J_0}$ is generated analogously to $F_2^{\alpha\beta}$ by plugging in $\dn{1} + \dn{1}^J$ and $\dn{1} + \tn{1}^J$ in the sources $\mathcal{S}_\alpha$ and $\mathcal{S}_\beta$ of Eq.~\eqref{e:Salphabeta} and collecting everything that contains one power of $J_0^\textit{ctr}$. The result is

\begin{equation}
F_2^{J_0} (\qv_1 , \qv_2 ) \eq \frac{1}{9} \left\{ \alpha(\qv_1 , \qv_2) + 2 \alpha (\qv_2, \qv_1) + \frac{8}{7} \beta(\qv_1, \qv_2) \right\}  \;. 
\end{equation}
In order to compute the rest of the kernels, we need to clarify what we mean by the (scalar) noise terms $J_n^\textit{ctr}$ of Eq.~\eqref{e:deltaJ2}. We want all $J_n^\textit{ctr}$ to have the same dimension as $J_0^\textit{ctr}$ and to be scalar quantities. This means that we have to decompose the noise terms in Eq.~\eqref{e:taunoiseNLO} and write their tensor structure in terms of wave vectors. Take for example $\bar{J}^{ij}_2(\kv,a)$. In Fourier space it depends only on the vector $\kv$. Since we already know that the noise will be correlated with other noise terms, we can make use of statistical isotropy and homogeneity. Hence we can write $\bar{J}^{ij}_2 $ as 

\begin{equation}
\bar{J}^{ij}_2 (\kv,a) \; \overset{\langle .. \rangle}{=} \; \frac{\kv^i \kv^j }{k^2} \, J_2^\textit{ctr}(\kv,a) + \delta^{ij} K_2(\kv,a) \;,
\end{equation}
where brackets indicate that this equation is only valid when $\bar{J}^{ij}_2 $ is correlated with some other noise term. All other possible terms in the decomposition of $\bar{J}^{ij}_2 $ would vanish inside an expectation value.
The noise term $K_2$ can then be absorbed in a redefinition of $J_1^\textit{ctr}$. We can apply this procedure to all other terms in Eq.~\eqref{e:tauvisLO} and, upon some reshuffling of terms, the expressions for the kernels $F_2^{J_n}$ read

\begin{eqnarray}
&& F_2^{J_1} (\qv_1 , \qv_2)  \eq \frac{1}{9} \phantom{q_1}  \;,  \qquad \qquad \qquad F_2^{J_2} (\qv_1 , \qv_2)  \eq \frac{1}{9} \frac{(\qv_1 \cdot \qv_2)^2}{q_1^2 \, q_2^2}  \;,  \\[1.5ex]
&&  F_2^{J_3} (\qv_1 , \qv_2)  \eq \frac{1}{9} \frac{\qv_1 \cdot \qv_2}{q_2^2}   \;,  \qquad \qquad  F_2^{J_4} (\qv_1 , \qv_2)  \eq \frac{1}{9}  \frac{(\qv_1 \cdot \qv_2)^3}{q_1^2 \, q_2^4} \;, \\[1.5ex]
&&  F_2^{J_5} (\qv_1 , \qv_2)  \eq \frac{1}{9}  \frac{q_1^2}{q_2^2}  \;,  \qquad \qquad \qquad  F_2^{J_6} (\qv_1 , \qv_2)  \eq \frac{1}{9}  \frac{(\qv_1 \cdot \qv_2)^2}{ q_2^4} \;, \\[1.5ex]
&& F_2^{J_7} (\qv_1 , \qv_2)  \eq \frac{1}{9}  \frac{(\qv_1 \cdot \qv_2)^4}{q_1^2 \, q_2^6}  \;.   \qquad \qquad  
\end{eqnarray}
As we shall discuss in the next section, these are all terms that we need for cancelling the UV-divergences of the one-loop integrals of the bispectrum.


\subsection{EFT contributions to the bispectrum and renormalization} \label{s:renorm}

\begin{figure}
\centering
\includegraphics[width=10.5cm]{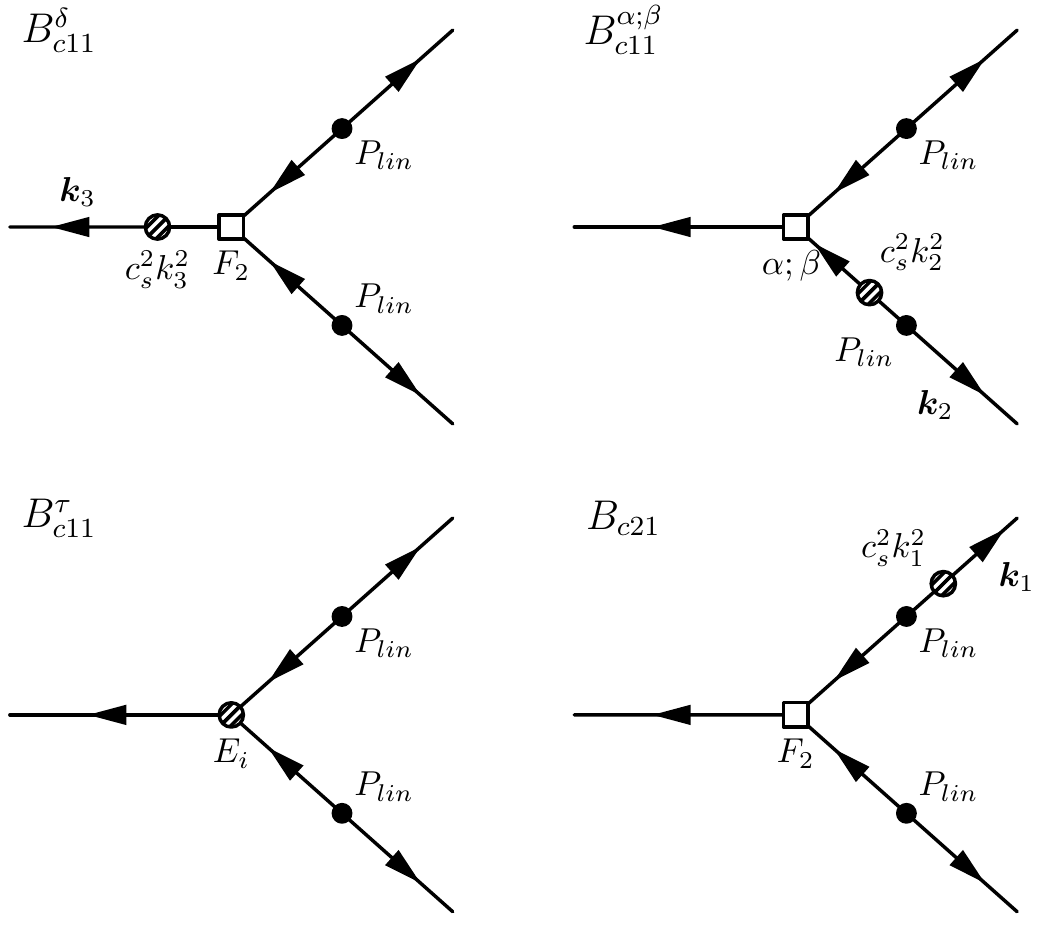}
\caption{Bispectrum contribution from the viscosity terms.}
\label{fig:bispecctr}
\end{figure}

\begin{figure}
\centering
\includegraphics[scale=0.5]{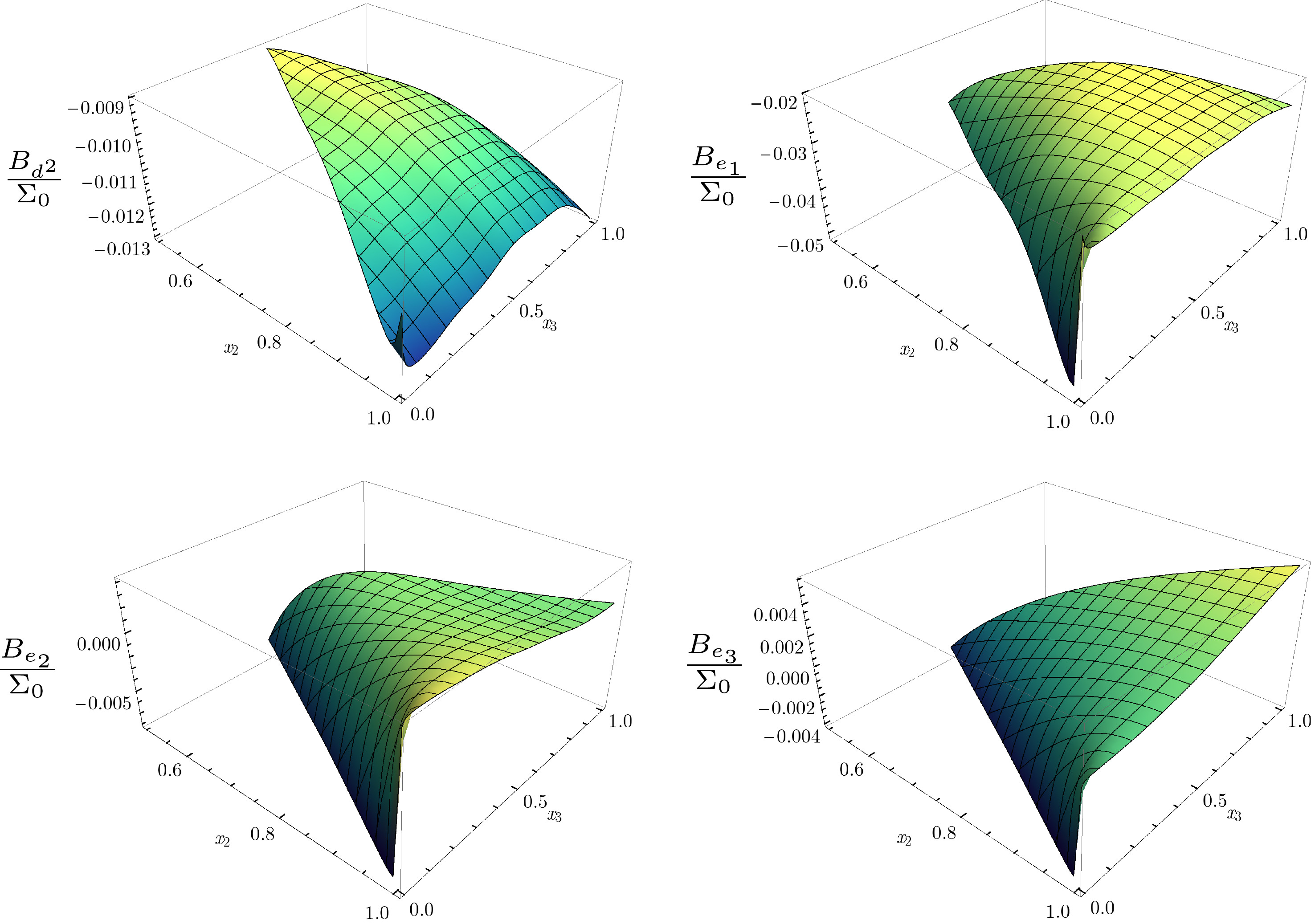} 
\caption{In this figure we show the shapes of the contributions of the four LECs to the bispectrum as a function of $x_1$ and $x_3$ (see also Fig.~\ref{fig:shapesSPT}). While $e_{1,2,3}$ only contribute to $B_{c11}$, the shape of $d^2$ contributes to both $B_{c21}$ and $B_{c11}$. We used $m=5/3$ (see Eq.~\eqref{e:mvalues}) and fixed $k_1 = 0.2 \, \ihMpc$.}\label{fig:shapesEFT}
\end{figure}

We can now compute the contributions to the bispectrum that stem from the effective stress tensor. For the bispectrum we need to consider all contributions that can be constructed using $\dn{1}^{c,J}$ and $\dn{2}^{c,J}$. Starting with the viscosity terms $\dn{1,2}^c$ we can build two kinds of three-point correlators e.g.

\begin{equation}
\EVb{\dn{1}^c(\kv_1)}{\dn{2}(\kv_2)}{\dn{1}(\kv_3)} \;, \qquad \quad \mbox{and} \qquad \quad  \EVb{\dn{2}^c(\kv_1)}{\dn{1}(\kv_2)}{\dn{1}(\kv_3)}  \;,
\end{equation}
The contribution to the bispectrum that comes from such correlators can be represented as in Fig.~\ref{fig:bispecctr} and is given by
\begin{eqnarray}
&& B_{c21}(k_1, k_2, k_3) \eq  F_2(\kv_2, \kv_3) \, \Pl(k_2) P_c(k_3) + \mbox{5 perm.} \;, \label{e:bc21} \\[1.5ex]
&& B_{c11}(k_1, k_2, k_3) \eq 2 F_2^c(\kv_2, \kv_3, a)\, \Pl(k_2) \Pl(k_3) + \mbox{2 cycl. perm.} \;, \label{e:bc11}
\end{eqnarray}
where the EFT contribution to the power spectrum $P_c$ is given by correlating $\dn{1}$ with $\dn{1}^c$ and vice versa, giving 
\begin{equation}
 P_c(k,a) \eq 2 \bar{d}^{\, 2 }g_1(a,m) k^2 \Pl(k) \eq -2 \gamma \,  k^2 \Pl(k)  \;.
\end{equation}
The function $g_1$ stems from the time integral over the Green's function. In Fig.~\ref{fig:shapesEFT} we take the four contributions of $d^2$ and $e_{1,2,3}$ that enter $B_{c21}$ and $B_{c11}$ and plot them separately as a function of $x_2$ and $x_3$ at fixed $k_1$ as we did in the case of the SPT contributions in Fig.~\ref{fig:shapesSPT}. These are nothing but the shapes of the kernels $E_i$, $\bar{F}_2^{\alpha\beta}$ and $F_2^\delta$ from Eq.~\eqref{e:deltac2epsilongamma}. It is clearly seen that the shapes of $e_{1,2,3}$ share some similarity since they all contain the same term $(\qv_1 + \qv_2)^2$ (see Eq.~\eqref{e:Ei}).

The noise terms also give two contributions to the bispectrum which stem from correlating either three $\dn{1}^J$ or one $\dn{2}^J$ with $\dn{1}^J $ and $\dn{1}$ 
\begin{eqnarray}
&& (2\pi)^3 \dirac(\kv_1 + \kv_2 +\kv_3) \, B_{JJJ}(k_1,k_2,k_3) \eeq  \EVb{\dn{1}^J(\kv_1)}{\dn{1}^J(\kv_2)}{\dn{1}^J(\kv_3)} \;, \label{e:bjjj} \\[1.5ex]
&& (2\pi)^3 \dirac(\kv_1 + \kv_2 +\kv_3) \, B_{JJ1} (k_1,k_2,k_3) \eeq \EVb{\dn{2}^J(\kv_1)}{\dn{1}^J(\kv_2)}{\dn{1}(\kv_3)} + \mbox{5 perm.} \qquad 
\end{eqnarray}
As opposed to $B_{c21}$ and $B_{c11}$, it is not obvious how these contributions actually look like. Note that the only information we have on $J_0$, and all other noise terms, is the overall scaling with $k$: $J_n \sim k^2$. As has been shown long time ago in Refs.~\cite{Zeldovich1965, Peebles1980}, the correlator of two noise terms scales as $\sim k^4$. Therefore we can infer that $B_{JJJ}$ has an overall scaling of $B_{JJJ} \sim k^6$. But we do not have any information about the shape of $B_{JJJ}$. For $B_{JJ1}$ the situation is somewhat better since we can reduce it to two-point correlators. From Eq.~\eqref{e:deltaJ2} we get
\begin{equation}\label{e:bjj1}
B_{JJ1} (k_1,k_2,k_3) \eq \frac{2}{7\, \cH^4} \sum_{n=0}^8 \, F_2^{J_n}(\kv_3 , \kv_1) P_{n0}^J(k_1) \Pl(k_3) \;, 
\end{equation}
where we defined the correlator of two $J_0^\textit{ctr}$ with a $J_i^\textit{ctr}$ as 

\begin{equation}\label{e:PJJ}
\EV{J_0^\textit{ctr}(\kv)}{J_n^\textit{ctr}(\kv')} \eeq (2\pi)^3 \dirac(\kv + \kv') \, P_{0n}^J(k) \;.
\end{equation}

Summing up all contributions, the bispectrum in the EFTofLSS is given by

\begin{equation}\label{e:Ball}
\begin{split}
B(k_1, k_2,k_3) \eq & B_\textit{112} + B_{222} + B_{321}^I + B_{321}^{II} + B_{411} + B_{c21} + B_{c11} + B_{JJJ} + B_{JJ1} \\[1.5ex]
&+ \mbox{higher orders} \;,
\end{split}
\end{equation}
where by higher orders we mean two- or higher loop terms and EFT terms with three fields or more than two derivatives in the effective stress tensor $\tau_\theta$. The tree-level and one-loop expressions have been defined in Eqs.~\eqref{e:blin}, \eqref{e:b222}, \eqref{e:b321i}, \eqref{e:b321ii} and \eqref{e:b411}, while the EFT contributions can be found in Eqs.~\eqref{e:bc21}, \eqref{e:bc11}, \eqref{e:bjjj} and \eqref{e:bjjj}. From Eq.~\eqref{e:Igeneral} we can infer that the noise terms are suppressed also in a $\Lambda$CDM universe. We will therefore neglect the noise terms and only consider them when discussing the renormalization procedure.

\begin{figure}
\centering
\includegraphics[scale=1]{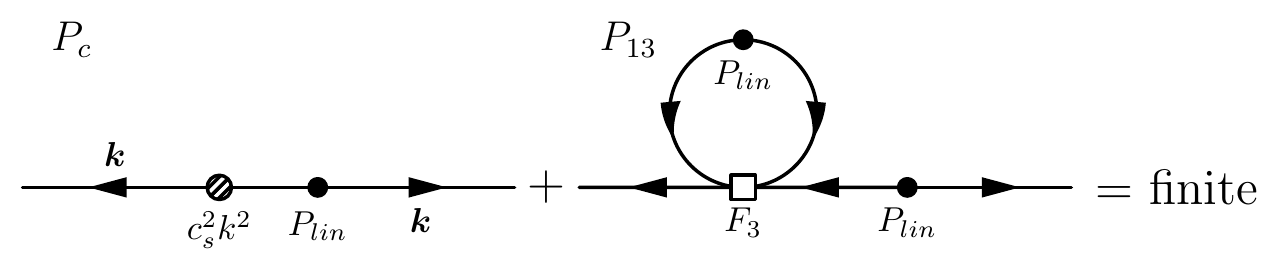}
\caption{Renormalization of the $P_{13}$ diagram.}\label{fig:p13ren}
\end{figure}

Finally, we can check whether the possible divergences discussed in Sec.~\ref{s:uvlimits} are really cancelled by the four EFT contributions  $B_{c21}$, $ B_{c11}$, $ B_{JJJ}$ and $B_{JJ1} $. In the case of the power spectrum, we know form the literature that the following combinations are finite, i.e.~that the UV-divergent part of $P_{13}$ and $P_{22}$ shown in Eqs.~\eqref{e:ploopuv13} and \eqref{e:ploopuv22} can be cancelled by the counterterm part of $d^2$ and $J_0$

\begin{eqnarray}\label{e:ploopren} 
P_{22} (k) + \frac{4}{49 \, \cH^4} P_{00}^J(k) \;, \qquad \mbox{and} \qquad P_{13}(k) + P_c(k) \;.
\end{eqnarray}
A diagramatic representation of the above relation is given in Fig.~\ref{fig:p13ren}.
Note that $P_{00}^J$ scales as $\propto k^4$ and $P_c \propto k^2 \Pl$. The counterterm part of $d^{\,2}$, i.e. the component that scales as $\sim a$ in time, is fixed by the relation in Eq.~\eqref{e:ploopren} to be 

\begin{equation}\label{ddiv}
\bar{d}^{\,2} \big|_\textit{ctr} \eq - \frac{183}{70 } \, \sigma_v^2 \;,
\end{equation}
where we just give the constant part of $d^2$.

In Sec.~\ref{s:uvlimits} we discussed the UV-limits of the one-loop integrals that contribute to the bispectrum. We found that, ignoring the triangular shape of the external momenta, the possible divergences have a $k$-dependence of $B_{222} \sim k^6 $, $B_{321}^I \sim k^4 \Pl $ and $B_{321}^{II} \sim B_{411} \sim k^2 \Pl^2$. From this scaling, we can already anticipate which counterterms are needed to renormalize the four diagrams. While $B_{222}$ and $B_{321}^I$ clearly require a noise term to cancel their divergences, the other two diagrams $B_{321}^{II}$ and $B_{411}$ will be renormalized through viscosity terms. Of course, we have to consider the specific shapes of the UV-divergences and the counterterms in order to be sure that the full one-loop bispectrum in Eq.~\eqref{e:Ball} is free of divergences. Actually, it is this shape dependence that provides a crucial check for the correctness of the form of the effective stress tensor at NLO.

As mentioned before, we do not know the shape dependence of $B_{JJJ}$. The only certain thing is that $B_{JJJ}$ has exactly the same scaling with $k$ as $B_{222}$. Unfortunately, we cannot do more than to assume that among every possible shape that $B_{JJJ}$ can have, there is one that matches the divergence of $B_{222}$ leaving us with  

\begin{equation}
B_{222} + B_{JJJ} \eq \mbox{finite} \;. 
\end{equation}

Next, we notice from Eq.~\eqref{e:Bscaling} that $B_{321}^I$ scales the same way as a correlator that is composed of $\dn{1} $, $\dn{1}^J$ and $\dn{2}^J$. This means that we expect the divergence of $B_{321}^I$ to be cancelled by $B_{JJ1}$ of Eq.~\eqref{e:bjj1}. When the shape of the $B_{321}^I$ in Eq.~\eqref{e:b321Idiv} is compared with what we get from the counterterm part of $B_{JJ1}$ (using $J_n^\textit{ctr} \sim a$ for the time dependence) it turns out that we can exactly match the two shapes. This means that 

\begin{equation}
B_{321}^I  + B_{JJ1} \eq \mbox{finite} \;. 
\end{equation}
Note that the renormalization of $B_{321}^{II}$ does not work at the level of the single integral. The sum of all possible permutations has to be taken into account since otherwise one may be misled to think that also counterterms involving only one derivative on the gravitational potential $\partial \Phi$ (thus violating the equivalence principle) are needed for the renormalization. 

The renormalization of the $B_{321}^{II}$ diagram is rather simple. We can write $B_{321}^{II}$ in terms of $P_{13}$ (see Eq.~\eqref{e:b321ii}) the same was as we can write $B_{c21}$ in Eq.~\eqref{e:bc21} using $P_c$. The sum of $B_{321}^{II}$ and $B_{c21}$ can therefore be reduced to the sum of $P_{13}$ and $P_c$

\begin{equation}
B_{321}^{II} + B_{c21} \eq F_2(\kv_2, \kv_3) \Pl(k_2) \,  \Big\{ P_{13}(k_3) + P_c(k_3) \Big\} + \mbox{5 perm.} \eq \mbox{finite} \;.
\end{equation}
From Eq.~\eqref{e:ploopren} we know that $P_{13} + P_c$ is finite, meaning that also the divergence in $B_{321}^{II}$ is cured. 

Finally, we get to the most interesting divergence, namely the one of $B_{411}$. We checked that the counterterm part of $B_{c11}$ in Eq.~\eqref{e:bc11} is exactly such that it cancels the UV-divergence of $B_{411}$ when taking into account all permutations

\begin{equation}
B_{411} + B_{c11} \eq \mbox{finite} \;. 
\end{equation}
The counterterm part of the NLO parameters must be such that 
\begin{equation}
\bar{e}_1  \big|_\textit{ctr}  \eq - \frac{1733}{1470} \, \sigma_v^2 \;, \qquad \bar{e}_2  \big|_\textit{ctr}  \eq  - \frac{1457}{1372 } \, \sigma_v^2 \;, \qquad  \bar{e}_3  \big|_\textit{ctr}  \eq - \frac{20991}{3430 } \, \sigma_v^2 \;.
\label{e:e123values} 
\end{equation}
There are two important things that have to be noted. Firstly, the counterterm part of $d^2$ is already fixed through the renormalization of the power spectrum in Eq.~\eqref{e:ploopren}. As expected, when renormalizing $B_{411}$ we need the contribution of $d^2$ in $B_{c11}$ (it enters through the kernels $F_2^{\alpha \beta}$ and $F_2^\delta$) and the counterterm part of $d^2$ exactly matches the value which is obtained from the renormalization of $P_{13}$ (see Eq.~\eqref{e:ploopren}). Secondly, the renormalization of $B_{411}$ works when locality in time is assumed, i.e.~starting from a local form of the effective stress tensor we get all necessary counterterms to cancel all possible divergences in the bispectrum, as expected from our discussion in Sec.~\ref{local}.

\begin{figure}
\centering
\includegraphics[scale=0.7]{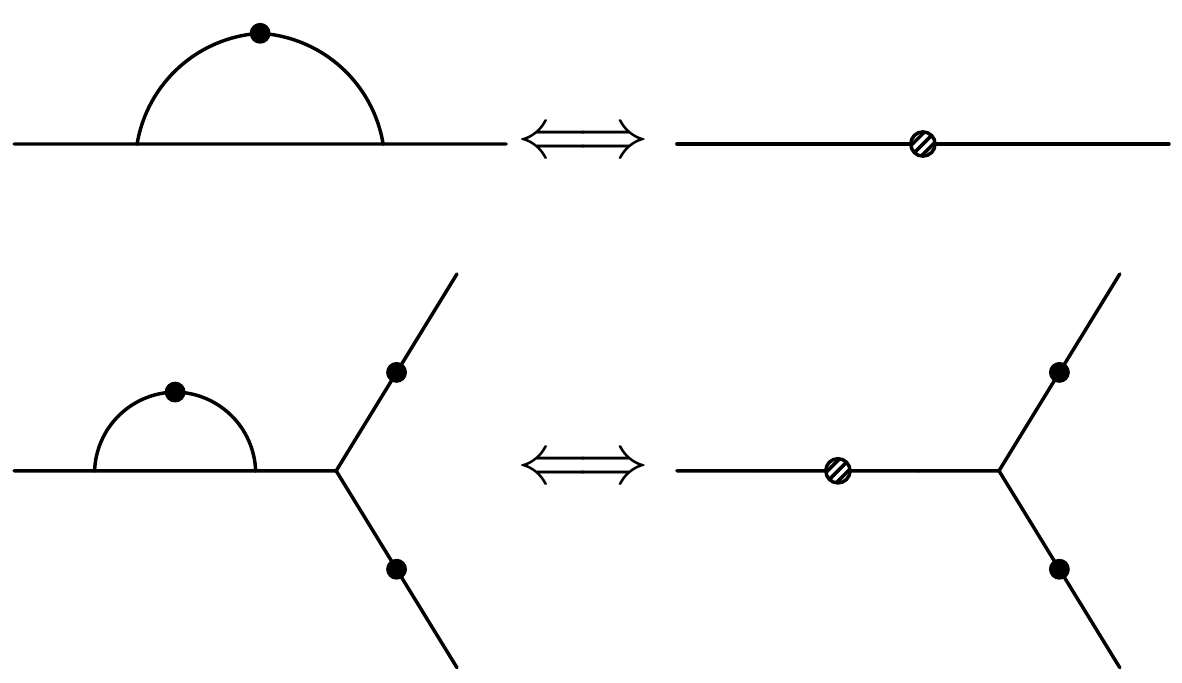}
\caption{A possible diagrammatic representation of the renormalization procedure. The 1PI diagram of the one-loop power spectrum and the corresponding (tree-level) counterterm diagram is shown on the top. 
.}\label{fig:counterterms}
\end{figure}

We just saw that the counterterm part of the free parameters in the effective stress tensor allow us to fully renormalize the one-loop bispectrum. So far we followed closely the SPT way of organizing the one-loop diagrams and everything works as expected. However, we find it somewhat intriguing that for the renormalization of $B_{321}^{II}$ and $B_{411}$ we need the counterterm part of $d^2$ which is already fixed by the renormalization of the one-loop power spectrum. Already in the context of RPT a diagrammatic language was introduced that, as opposed to SPT, allows to distinguish one-particle reducible (1PR) and one-particle irreducible (1PI) diagrams (see also Ref.~\cite{Assassi2014}). The former type of diagrams can be split into two disjoint parts by cutting a single internal line, while the same is not possible with the latter. We believe that the organization of the counterterms is related to the distinction of 1PR and 1PI diagrams. The idea is to set up the Feynman rules for the interactions in the effective stress tensor in such a way that the divergences of the 1PR loop diagrams in a $N$-point function are automatically cancelled by the renormalization of the $N'$-point function with $N'<N$. Only for the 1PI diagrams we would have to introduce new counterterms. However, one has to be somewhat careful. In the internal lines of the diagrams there is an implicit sum over the both $\delta$ and $\theta$. This means that for the renormalization of the 1PR diagrams, we need to renormalize all lower order $N$-point functions, also those involving the velocity field. At the end, in the final result for the density correlators the velocity counterterms drop out.
For the power- and bispectrum this would mean that the counterterm part of $d^2$ together with the counterterms for the two-point velocity correlator cancel all divergences in 1PR diagrams of the bispectrum, while the NLO counterterms fix the 1PI divergences. 
A diagrammatic sketch of this procedure is shown in Fig.~\ref{fig:counterterms}. To be able to distinguish the different counterterms and the corresponding divergences more clearly would allow for a better understanding of the renormalization in the EFTofLSS and possibly give us a procedure to compute all counterterms at any given order. We leave this for future work.

\section{Simulations and determination of parameters} \label{s:simulations}

In this section we describe the simulations we use to test the validity of the EFTofLSS approach and to measure its parameters. After describing our numerical data, we show the results of our fitting procedure. We always consider the simulations at the present epoch, i.e.~at z=0.



 
\subsection{Simulations}
\begin{figure}
\centering
\includegraphics[width=0.8\textwidth]{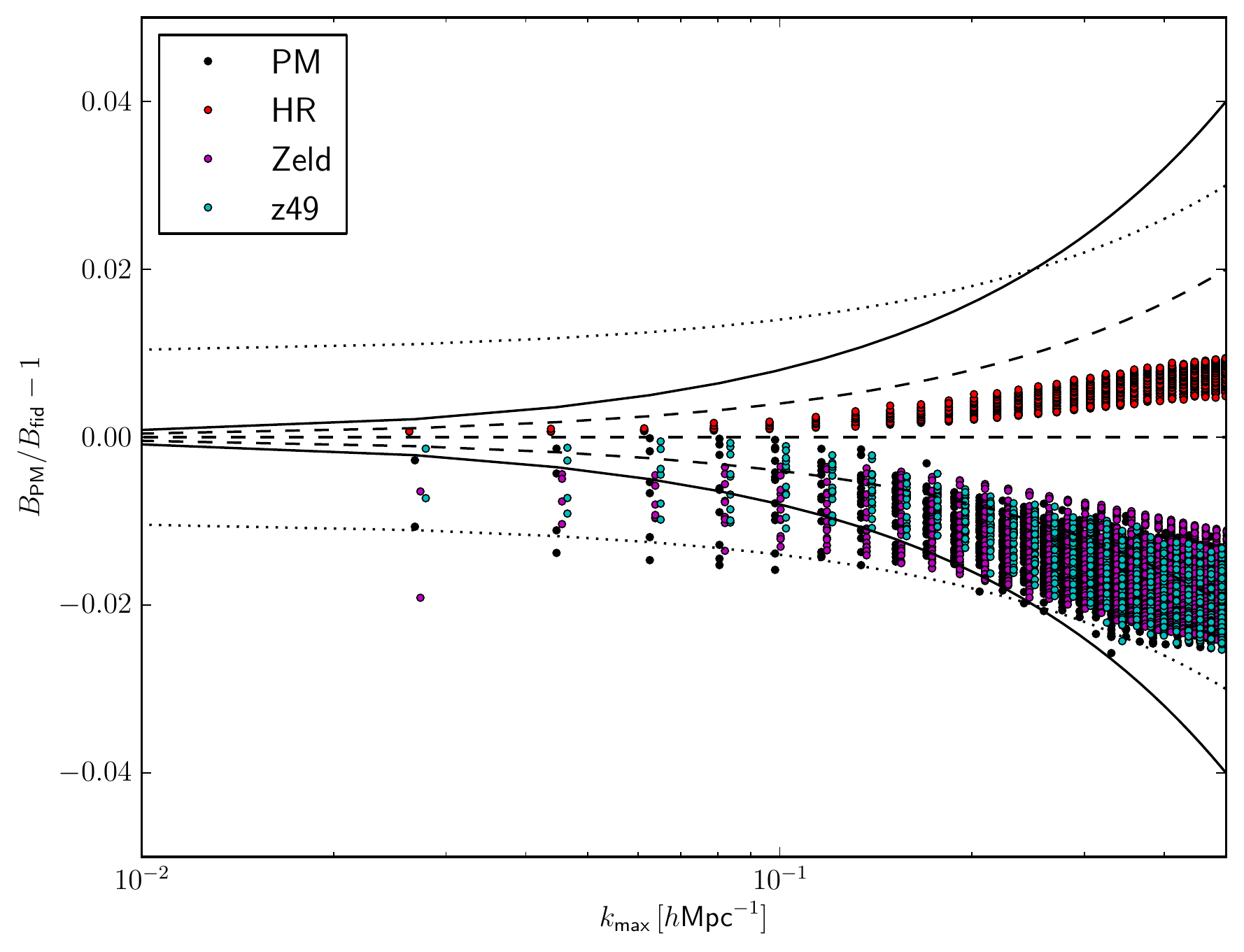}
\caption{Systematic deviations of the bispectra of $N$-body simulations with varying simulation parameters with respect to our fiducial simulation. We are varying \emph{i}) the starting redshift ($z_i=99 \to z_i=49$), \emph{ii}) the IC code (2LPT $\to$ 1LPT/Zeldovich), \emph{iii}) the size of the FFT mesh used to calculate long distance forces (PM) and \emph{iv}) the numerical precision of the $N$-body code (HR). The dashed lines show our most conservative ansatz for the systematic error, which will be employed in the simulation fits. For clarity the points have been slightly offset from their true $k_\text{max}$.}
\label{fig:bispect_sys}
\end{figure}

While the EFTofLSS provides a consistent tool for solving the fluid equations perturbatively, it has a number of free parameters that are related to the intrinsically non-linear small scale dynamics. These parameters can be either obtained from detailed measurements of the source terms of the equation of motion or from a fit to the data of the EFTofLSS predictions for distinct statistics. We will follow the latter approach. Furthermore we would like to understand to what extend the one-loop EFT can provide a model for the matter bispectrum.
For this purpose we employ a suite of collisionless $N$-body simulations. These simulations sample the cosmological Dark Matter fluid by $N_p=1024^3$ particles in a cubic box with dimension $L=1500 \hMpc$. 
The cosmology is based on the WMAP 7-year data analysis \cite{WMAP7} and is summarized in the following set of cosmological parameters: $\Omega_m^0=0.272$, $\Omega_\Lambda^0=0.728$, $h=0.704$, 
$\sigma_8=0.81$, $n_s=0.967$.
The initial conditions are set up at redshift $z=99$ using a second order Lagrangian perturbation theory code \texttt{2LPT} \cite{Crocce2006tr} and the subsequent gravitational evolution is computed using the Tree-PM code \texttt{Gadget}-II \cite{Springel2005}. 
We are assigning the particles to a cartesian grid of dimension $N_c=512^3$ using Cloud in Cell (CIC) interpolation, Fourier transform and correct for the CIC assignment window.
We are estimating the bispectrum in linearly binned shells in $k$-space using \cite{Scoccimarro1998}
\begin{align}
\hat B(k_i,k_j,k_l)=\frac{V_\textit{f}}{V_{ijl}}\int_{[\qv_1]_i}\int_{[\qv_2]_j}\int_{[\qv_3]_l}\delta(\qv_1)\delta(\qv_2)\delta(\qv_3)(2\pi)^3\dirac(\qv_1+\qv_2+\qv_3)\, ,
\end{align}
where $V_f$ is the volume of the fundamental cell $V_f=(2\pi/L)^3$.
The square brackets describe a linear bin around $k_i$ and 
\be
V_{ijl}=\int_{[\qv_1]_i}\int_{[\qv_2]_j}\int_{[\qv_3]_l}(2\pi)^3\dirac(\qv_1+\qv_2+\qv_3)\approx \frac{8\pi^2}{(2\pi)^6}k_i k_j k_l \Delta k^3
\ee
is the volume of shells.
A naive implementation of the above estimator is very slow, since one would have to check $N_c^2$ cells for the triangle condition. Rewriting the Dirac delta as an integral we have (see Refs.~\cite{Scoccimarro2000a,Sefusatti:2005})
\begin{align}
\hat B(k_i,k_j,k_l)=\frac{V_\textit{f}}{V_{ijl}} \int_{\bm x} \prod_{\kappa=i,j,l} \int_{[\qv]_\kappa} \text{exp}[i \qv \cdot \bm x] \delta(\qv)
\end{align}
Thus, we are selecting Fourier modes from a shell in $k$-space, Fourier transform these shells to real space and sum over the product of the Fourier transforms of the three shells. It is sufficient to follow this procedure for an ordered set $k_3\leq k_2\leq k_1$ since the algorithm implicitly accounts for the permutations of the momentum magnitudes.
The leading contribution to the cosmic variance of the bispectrum is given by 
\begin{align}
\Delta \hat B^2(k_i,k_j,k_l)=s_{123}\frac{V_f}{V_{ijl}}P(k_i)P(k_j)P(k_l),
\end{align}
where $s_{123}$ accounts for the number of permutations contributing to a certain bispectrum configuration and is 6, 2 and 1 for equilateral, isosceles and general triangles. 
We have compared the theoretical variance to the variance between our 16 simulation runs and found reasonable agreement on large scales.\\
Besides the cosmic variance errors, we would also like to quantify the systematic errors in our simulations. Due to finite numerical resources any simulation will eventually be limited by spatial and temporal resolution effects. While an extensive convergence study of simulation bispectra would be far beyond the scope of this paper, we would still like to obtain some intuition about the magnitude of residual systematic deviations that might plague our fiducial set of simulations and the bispectra extracted from these simulations. We thus consider some isolated modifications around our fiducial $N$-body runs starting from the same realization of the initial Gaussian random field. We \emph{i}) vary the starting redshift from $z=99$ to $z=49$, \emph{ii}) use the Zeldovich approximation instead of the 2LPT initial conditions, \emph{iii}) increase the size of the grid used for the particle mesh part of the $N$-body code, and \emph{iv}) increase the numerical accuracy of the $N$-body code. Fig.~\ref{fig:bispect_sys} shows how these changes affect the bispectrum.
While we believe that our choice of simulation parameters is favorable to cases \emph{i})-\emph{iii}), we still decide to be conservative and add the following systematic error to the cosmic variance error discussed above
\be
\frac{\Delta \hat B_{sys}}{\hat B}=0.01+0.02\left(\frac{k_\text{max}}{0.5 \ihMpc}\right),
\ee
where we defined $k_\text{max} \equiv \mbox{max}(k_1,k_2,k_3)$.
Both the measurement of the bispectrum in the simulations and the calculation of the loop integrals are time consuming tasks. In particular, the simulation measurement is necessarily an average of the true bispectrum over a finite bin, who's value is assigned to a certain set of wavenumbers appropriate for this bin. On the other hand the loop corrections are evaluated at a specific wavenumber and should in principle be averaged over the same bin employed in the simulation estimator. Since the parallelized computation of the bispectrum for the 2600 data points employed in our comparison takes order of days we refrain from doing this and rather choose an effective wavenumber for the simulation bins and evaluate the loop integrals at this wavenumber following the approach in \cite{Sefusatti2010}
\be
k_{i,\textit{eff}}=\frac{1}{V_{ijl}}\int_{[\qv_1]_i}\int_{[\qv_2]_j}\int_{[\qv_3]_l}q_i (2\pi)^3\dirac(\qv_1+\qv_2+\qv_3)\, .
\ee

 
\subsection{Fits}

In this section, we describe how the EFTofLSS bispectrum compares with the $N$-body simulation data discussed previously. 
Let us first remind the reader of the structure of the density field in the EFT
\begin{equation}
\delta=\delta^\textit{SPT}+\delta^c+\delta^J\; .
\end{equation}
Based on the power counting presented in Eq.~\eqref{eq:contribcount}, we will ignore the stochastic term $\delta_J$. Thus we are left with the three new viscosity parameters, who's counterterm parts are needed to cancel the divergences in the one-loop bispectrum, namely $ e_{1} $, $ e_{2} $ and $ e_{3} $, in addition to the parameter that one encounters in the one-loop power spectrum, namely $ d^2 $, propagated to second order
\begin{equation}
\begin{split}
\delta^c(\bm k,a) \eq & -\gamma k^2 \dn{1}(\bm k)-\int_{\bm q}\Biggl\{\sum_{i=1}^3 \epsilon_i E_i(\bm q,\bm k-\bm q)\\[1.5ex]
&+\gamma\bigl[ \bar F_2^{\alpha\beta}(\bm q,\bm k-\bm q,a)+\bar F_2^\delta(\bm q,\bm k-\bm q,a)\bigr]\Biggr\}\dn{1}(\bm q,a)\dn{1}(\bm k-\bm q,a) \;,
\end{split}
\end{equation}
where we write $\delta^c$ in terms of the fitting parameters $\gamma$ and $\epsilon_{1,2,3}$ (see Eqs.~\eqref{dtogamma} and \eqref{e2e}).
The first term correlated with another linear density field enters in the EFT power spectrum at the one-loop level and can thus be determined through a measurement of the power spectrum. It also enters in the bispectrum when replacing one of the legs in the $B_{112}$ term and forming $B_{c21}$ (see lower left panel of Fig.~\ref{fig:bispecctr}).
The terms quadratic in the linear density field contribute to the bispectrum when correlated with two external linear density fields giving $B_{c11}$. At the order we are interested in, the bispectrum in the EFTofLSS is then given by

\begin{equation}
\begin{split}
& B_{EFT} \eq B_{SPT}+B_{EFT}\,,\\[1.5ex]
& B_{EFT}(\kv_{1},\kv_{2},\kv_{3}) \eq  -\left\{  \sum_{i=1}^3 \epsilon_i E_i(\bm k_{2},\bm k_{3})
+\gamma \, \Bigl[ k_{1}^2 \, F_2(\kv_2, \kv_3) +\bar F_2^{\alpha\beta}(\bm k_{2},\bm k_{3},a) \right. \\[1.5ex]
& \hspace{3.5cm} \left. \phantom{ \sum_{i=1}^3} +\bar F_2^\delta(\bm k_{2},\bm k_{3},a)\Bigr]\right\}  \cdot  P_{lin}(k_{2})P_{lin}(k_{3})+\mathrm{5\, perm.}\;,   \label{BEFT}
\end{split}
\end{equation}
where the terms in brackets depend on the value of $ m $.

The goal of this section will be to assess to what extend the additional flexibility in the shape of the EFT terms can help to improve the agreement between theoretical and simulated bispectra. The agreement will be measured by the standard $\chi^2$
\be
\chi^2_B=\sum_{k_i=k_\textit{min}}^{k_\textit{max}}\left[\frac{B(k_1,k_2,k_3)-B_{SPT}(k_1,k_2,k_3)-B_{c11}(k_1,k_2,k_3)-B_{c21}(k_1,k_2,k_3)}{\Delta B(k_1,k_2,k_3)}\right]^2
\ee
We will minimize the above sum to obtain the parameters of the EFT and their errors as a function of the largest wavenumber $\km = \mathrm{max}(k_1,k_2,k_3)$. Subsequently we compute the reduced $\chi^2$ up to $k_{max}$ using the parameters obtained for the same $k_\textit{max}$. Once this function considerably deviates from unity we consider the theory to be a bad fit and thus select our fiducial $k_\textit{max}$ and the corresponding set of fiducial parameters. To check how well this set of parameters performs, we then compute the reduced $\chi^2$ again as a function of $k_{max}'$ (different from the fiducial $\km$), but with the fixed set of parameters. This will allow us to check the performance of our fiducial parameter set on scales different from the fiducial $k_{max}$. As a quantitative measure for the failure of the theoretical description, we can compute also the p-value 

\begin{equation}
\mbox{p-value} \eq 1- \mathrm{CDF}_{\chi^2}(\mbox{DOF},\chi^2_B) \;,
\end{equation}
where $\mathrm{CDF}$ is the cumulative distribution function of the $\chi^2$-distribution with the corresponding number of degrees of freedom. This gives us the probability of having a $\chi^2$ which is even larger than the one we get from the fitting procedure. Hence, we only need to find the scale where the p-value becomes very small.

\begin{figure}
\includegraphics[width=.49\textwidth]{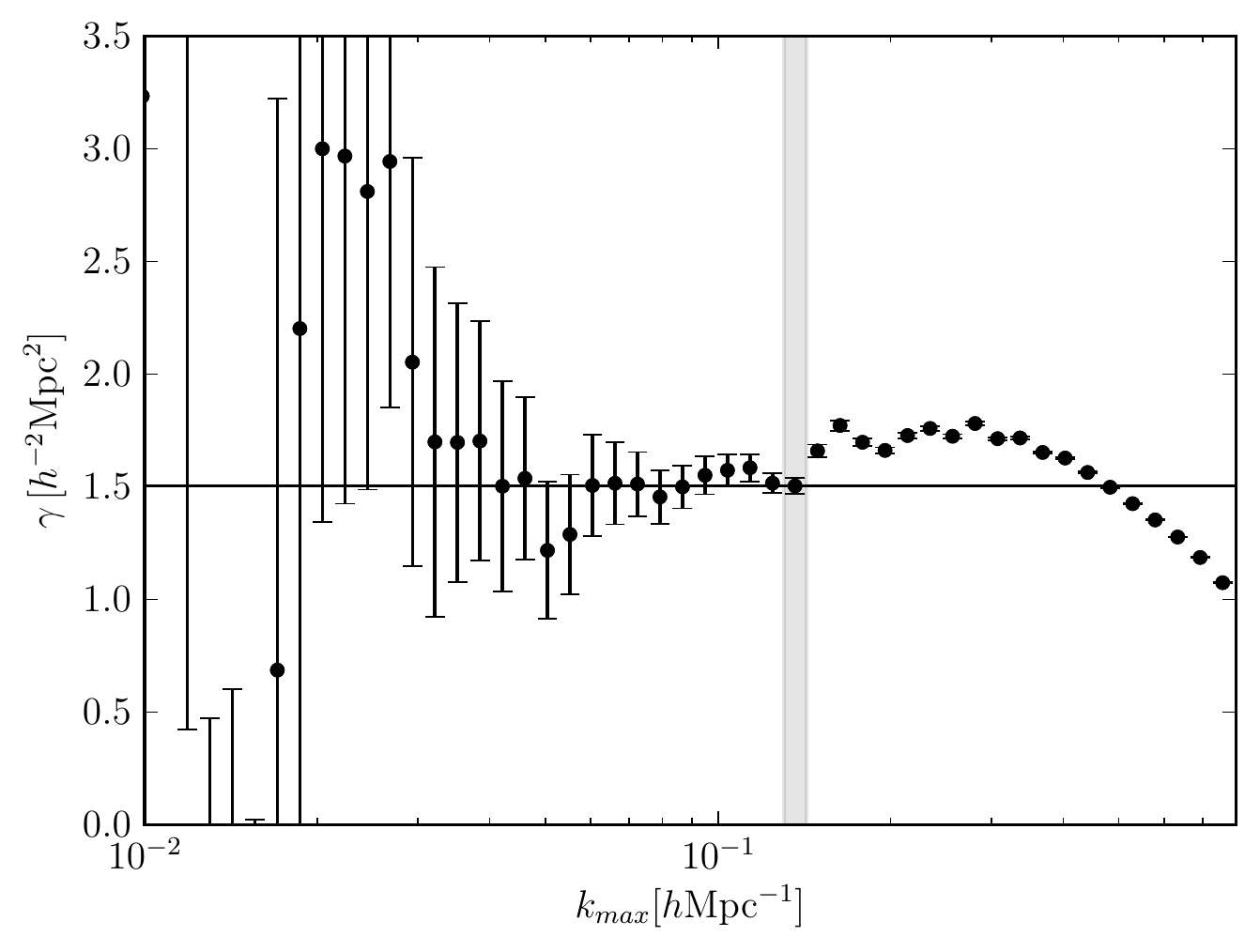}
\includegraphics[width=.49\textwidth]{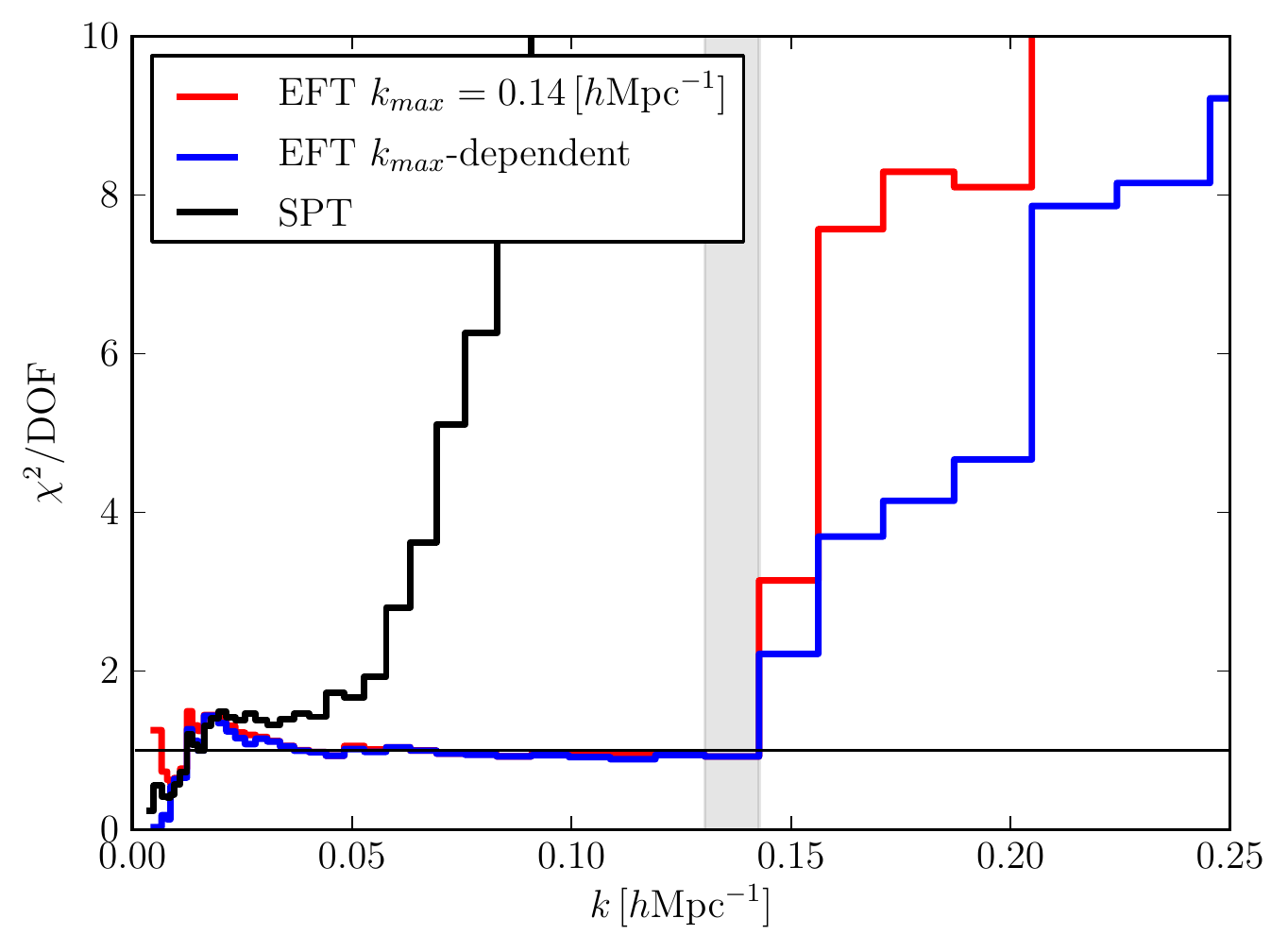}
\caption{Constraint on $\gamma$ from the power spectrum. \emph{Left panel: }Convergence of the constraint on $\gamma$ with increasing $k_{max}$ of the fit. Except for some scatter on very large scales, where the EFT correction is irrelevant, the fit converges to the value obtained at $k_{max}\approx0.14\, \ihMpc$ and then becomes incompatible with the constraints obtained from larger scales. \emph{Right panel: }Reduced $\chi^2$ according to Eq.~\eqref{eq:Pchi2}. We clearly observe a steep rise of the $\chi^2$ at $k=0.15 \ihMpc$, which we take to be our fiducial $k_{max}$ for the power spectrum (vertical gray band). The red line shows the $\chi^2$ with $\gamma$ fixed once and for all by the data up to $0.15 \ihMpc$, whereas for the blue line we fit $ \gamma $ at each $ k_{max} $.}
\label{fig:powerfit}
\end{figure}

\subsubsection{Re-fitting $\gamma$}
In a first step, we re-fit the free parameter for the power spectrum in our simulations. We measure the matter power spectrum using the same CIC mass assignment scheme described above for the bispectrum. The error on the power spectrum is estimated from the standard deviation between the 16 simulation boxes.
While we could decide to fit the speed of sound from the final power spectrum of the simulations, we decide to cancel the leading cosmic variance contribution by fitting the ratio of the final and initial power spectra in the simulations, i.e., we consider
\begin{equation}
\chi^2_P=\sum_{k=k_\textit{min}}^{k_\textit{max}}\frac{1}{\Delta (P_\textit{ff}/P_\textit{ii})^2}\left[\frac{P_\textit{ff}}{P_\textit{ii}}-1-\frac{P_{13}}{P_{11}}-\frac{P_{22}}{P_{11}}+2k^2 \gamma\right]^2\, ,
\label{eq:Pchi2}
\end{equation}
where $P_\textit{ff}$ is the final power spectrum and $P_{ii}$ the initial power spectrum linearly rescaled to redshift $z=0$.
The reduced $\chi^2$ shown in the right panel of Fig.~\ref{fig:powerfit} forces us to limit the fit to $k<0.14\, \ihMpc$. We are aware that IR resummation schemes might reduce the amplitude of the BAO wiggles in the residuals and thus extend the range of the fit (see Ref.~\cite{Senatore2014} for an EFT approach and Refs.~\cite{Crocce2006,Crocce2006a,Crocce2008} for resummation effects in RPT). We will not consider these extensions here and rather content ourselves with the fact that even the standard EFT formulation can improve the $k$-range by a factor of three. We note that the measurement of $\gamma$ from the final power spectrum alone is compatible with the one inferred from the above method, but that it has significantly larger uncertainties.
The obtained value of $\gamma=(1.5\pm0.03)\hMpcsq$ is slightly smaller than the corresponding value of $\gamma=(1.62\pm0.03)\hMpcsq$ reported in Ref.~\cite{Carrasco2013a}. This is likely due to their fitting range $0.15 \ihMpc \leq k \leq 0.25 \ihMpc$, due to the slightly different cosmology and their usage of the Cosmic Emulator  \cite{Heitmann:2014em} instead of $N$-body data.
Performing the above fit for a number of redshifts indicates that the integrated EFT coefficient scales as $D^{m+1}$, with $m \approx 1.5\pm0.5$.
This statement supports our assumption of $m=5/3$ but is somewhat dependent on the actual fitting range.
The details of this time dependence are beyond the focus of this study and will be discussed in more detail elsewhere.


\subsubsection{Zero parameters}

We have seen that, in addition to $ \gamma $, there are three new parameters in the EFTofLSS to be included in the bispectrum. However, the structure of the loop divergences suggests how to fix them \textit{all} using \textit{only} data from the power spectrum, i.e.~without ever looking at the bispectrum data. The form of the EFTofLSS bispectrum was given in \eqref{BEFT}, and we will use the value $ m=5/3 $ as suggested by the self-similarity. As we saw in Sec.~\ref{s:renorm}, the cancellation of divergences in $ B_{411} $ requires a certain relation among $ \bar d^{2} $ and $ \bar{e}_{i} $ (see \eqref{ddiv} and \eqref{e:e123values}). Rewriting those relations in terms of $ \gamma $ and $ \epsilon_{i} $ using \eqref{dtogamma} and \eqref{e2e}, one finds
\be\label{magic}
\epsilon_{1}\eq \frac{3466}{14091}\gamma\,, \qquad\epsilon_{2}\eq \frac{7285}{32879}\gamma\,,\qquad \epsilon_{3}\eq \frac{41982}{32879}\gamma\,.
\ee
This relation reflects the fact that $\sigma_v^2$ is the same for the power- and bispectrum. Therefore, we can expect that a single fitting parameter should suffice to capture the leading EFT correction in both the power- and the bispectrum. The equation above allows us to express all EFT parameters that appear in the bispectrum in terms of $\gamma$ which can be determined by measuring the power spectrum. Using Eq.~\eqref{magic} and the value $ \gamma=(1.5\pm 0.03 )\hMpcsq$ which we found in the previous section, we can predict the bispectrum of Eq.~\eqref{BEFT} without considering the bispectrum data from the simulations. The comparison of this zero parameter prediction with the measured bispectrum from the simulations is shown as the red line in Fig.~\ref{fig:chi2fit}. We show the reduced $\chi^2$ of the various scenarios (discussed later on) and the corresponding p-value. To determine at which scale our theoretical prediction fails to agree with the data, we simply look for the $k_\textit{max}$ where the p-value drops to values close zero or, equivalently, the reduced $\chi^2$ starts deviating significantly from $1$. Interestingly, the zero parameter prediction for the bispectrum performs as good as the other EFT fitting scenarios. One can see that even without any fitting parameters in the bispectrum (only fitting $\gamma$ from the power spectrum) the largest $ k $ up to which one can accurately predict the bispectrum has increased from $ k_\textit{max} \approx 0.13 \ihMpc $ in the case of SPT to $ k_\textit{max} \approx 0.22 \ihMpc$ when the EFT contributions are included. It should be noticed that Fig.~\ref{fig:chi2fit} does not change significantly if we set $m=1$. It is remarkable that without any free parameter left our prediction for the bispectrum agrees well with the simulations to such a high $k_\textit{max}$ and that, as we shall see, this scale does not change by adding more fitting parameters. This is a nice confirmation that are capturing the correct physics.

 
\subsubsection{One parameter}

\begin{figure}
\centering
\includegraphics[width=0.49\textwidth]{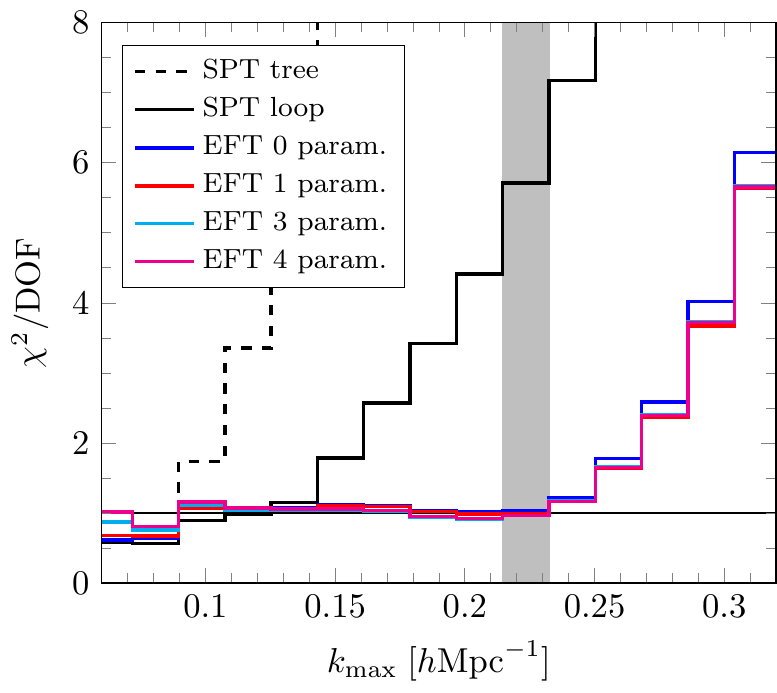} 
\includegraphics[width=0.49\textwidth]{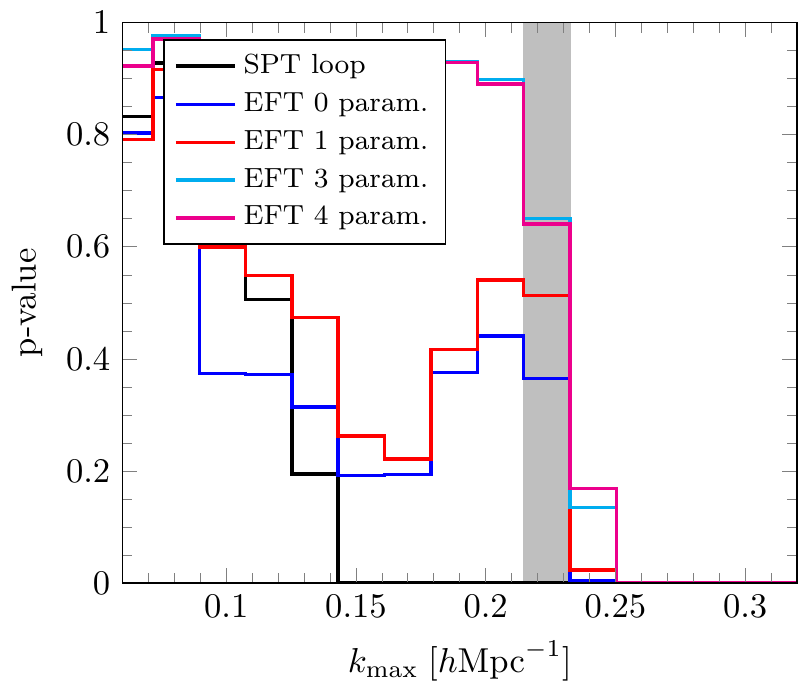}
\caption{\emph{Left panel: }Reduced $\chi^2$ for the fitting procedures described in the text. The fit is performed at every $\km$ and for $m=5/3$. Irrespective of the scenario considered, the reduced $\chi^2$ for the EFT stars exceeding 1 at $\km \approx 0.22 \ihMpc$ (gray area). Comparing to the one loop SPT this means an improvement by nearly a factor of two in $k_{max}$. The three- and four-parameter fits are overfitting the data below $\km \approx 0.22 \ihMpc$. \emph{Right panel: } The p-values that correspond to the $\chi^2$ in the left panel. Above $\km \approx 0.22 \ihMpc$ all fitting scenarios show a steep drop, indicating that the theory fails to describe the data.}
\label{fig:chi2fit}
\end{figure}

\begin{figure}
\centering
\includegraphics[width=0.49\textwidth]{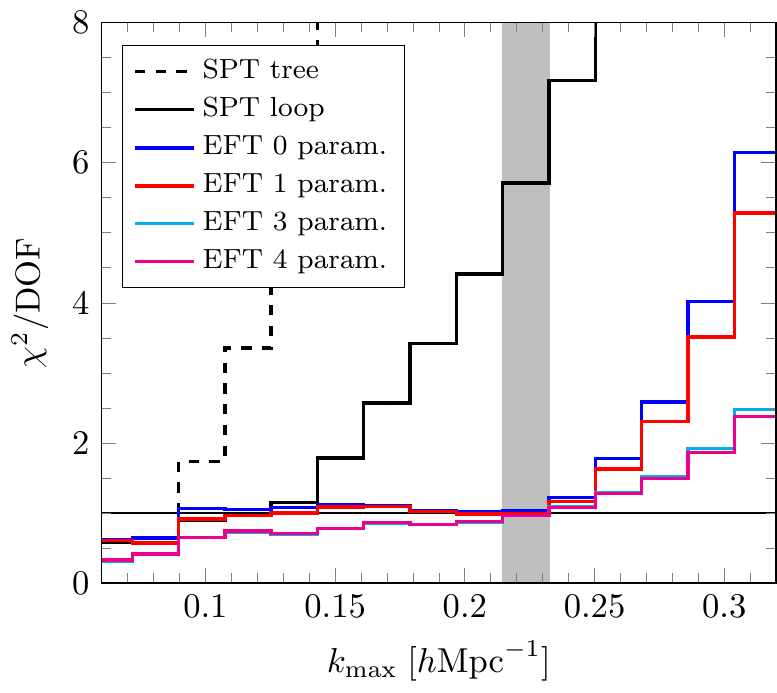} 
\includegraphics[width=0.49\textwidth]{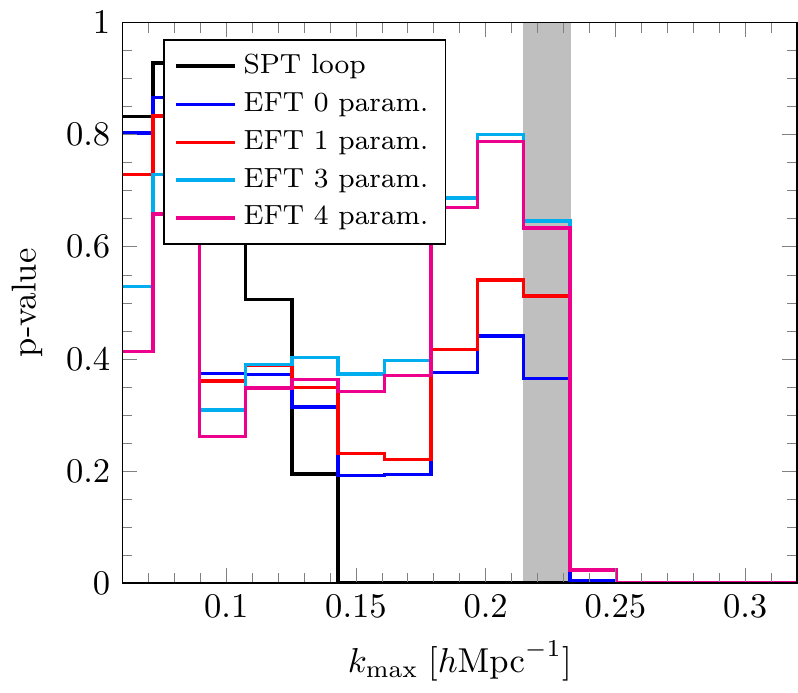}
\caption{\emph{Left panel: } Same as Fig.~\ref{fig:chi2fit}, but with the fit performed at the fiducial $\km \approx 0.22 \ihMpc$ (gray band) and for $m=5/3$. As before, the reduced $\chi^2$ for the EFT raises fairly steeply above the fiducial $\km$. The four EFT scenarios are almost indistinguishable, which is remarkable in particular for the zero-parameter fit. \emph{Right panel: } The p-values that correspond to the $\chi^2$ in the left panel. Again, the p-values fall below $1\%$ at $\km \approx 0.22 \ihMpc$. This shows that the fit does not improve compared to Fig.~\ref{fig:chi2fit} when a fixed $\km$ is chosen. }
\label{fig:chi2}
\end{figure}

We have just seen that the zero-parameter prediction for the bispectrum works very well, increasing the range of $ k_{max} $ by roughly a factor of two compared to SPT. It is natural to ask whether one can do even better by varying one or more parameters. When fitting parameters from the bispectrum data, we have two options. Either we fit at every $k_\textit{max}$, or we choose a fiducial $k_\textit{max}$ (at $\km \approx 0.22 \ihMpc$). The first situation is shown in Fig.~\ref{fig:chi2fit}, while the second one is represented in Fig.~\ref{fig:chi2}.

Let us start with keeping the relations among $\epsilon_{1,2,3}$ fixed as in \eqref{magic}, but fitting for $ \gamma $ instead of taking it from the power spectrum analysis. This is shown as the blue line in Figs.~\ref{fig:chi2} and \ref{fig:chi2fit}. As one can see from those figures, the $ \chi^{2} $ and relative p-value obtained from fitting $ \gamma $ are pretty close to those for the zero-parameter formula discussed in the previous section.\footnote{One should be cautious in fitting $ \gamma $ to the bispectrum with small $ k_{max} $ because at large scales the EFT correction is expected to be small and one ends up fitting the large scale ``noise'' in the simulation.} In particular, the range of $ \km $ covered is the same, i.e. there is a sharp drop after $\km \approx 0.22 \ihMpc$. This remains true as we freely fit more and more parameters. The best fit value for $\gamma$ from the bispectrum at our fiducial $\km \approx 0.22 \ihMpc$ is $\gamma = (1.36 \pm 0.04) \hMpcsq$ for $m=5/3$ (see also Tab.~\ref{tab:solconst}). Compared to the value of $\gamma=(1.5\pm0.03)\hMpcsq$ that we get from the power spectrum fit, this is somewhat off. Since the bispectrum contains precisely the same UV-divergence as the power spectrum one might think that there would be a better agreement between the power- and bispectrum values for $\gamma$.


\begin{figure}
\centering
\includegraphics[width=.99\textwidth]{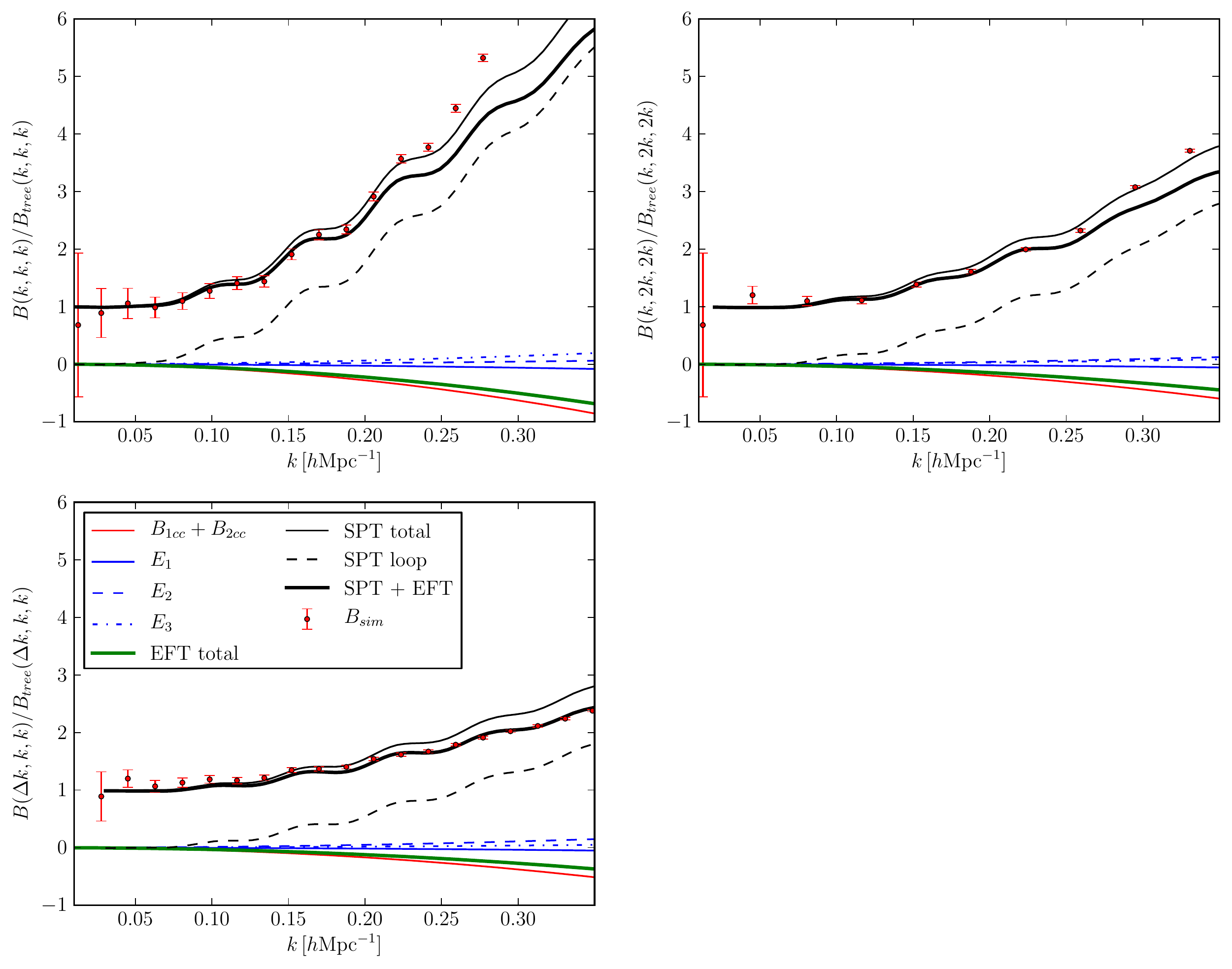}
\caption{Ratio of the various contributions of the one loop SPT + EFT bispectrum to the tree level bispectrum. The parameters are chosen according to the three parameter fitting procedure for $m=1$. As an overall observation we note that the loop corrections are always larger than the EFT terms and that the latter are dominated by the $B_{1cc}$ and $B_{2cc}$ parameters, whose amplitude is fixed by the fit to the power spectrum. 
\emph{Upper left panel: }Equilateral configuration.
\emph{Upper right panel: }Isosceles configuration with $k_3=k_2=2k_1$.
\emph{Lower left panel: }One of the modes is fixed to the second linear bin centred around $\Delta k=0.027 \ihMpc$. With increasing $k$ the configuration evolves from a equilateral to a squeezed shape.}
\label{fig:ratios}
\end{figure}

\subsubsection{Three parameters}

From the effective stress tensor in Eq.~\eqref{e:tauvisNLO}, we expect that the preferred scenario for modelling the bispectrum is to take $\gamma$ from the power spectrum fit and to use the bispectrum only for fitting the new intrinsically second order parameters $\epsilon_1,\epsilon_2$ and $\epsilon_3$. The reduced $\chi^2$ and the corresponding p-values are found in Figs.~\ref{fig:chi2fit} and \ref{fig:chi2} (green curves). 
Due to the explicit dependence of the shape of $B_{c11}$ on the EFT time exponent $m$, we will consider both time dependencies in Eq.~\eqref{e:mvalues}, the one suggested by the cancellation of the divergences, i.e.~$m=1$, and $m=5/3$ as suggested by the approximate self-similarity.
We quote our parameter constraints for the fiducial $\km \approx 0.22 \ihMpc$ in Tab.~\ref{tab:solconst} and remark that the errors on the new NLO stress tensor coefficients are fairly large. Furthermore, the constraints on $\epsilon_2$ and $\epsilon_3$ are quite degenerate with cross correlation coefficients of $r_{\epsilon_2,\epsilon_3}\approx -0.55$, $r_{\epsilon_1,\epsilon_3}\approx 0.55$ and $r_{\epsilon_1,\epsilon_2}\approx 0.21$. 
We checked the convergence of the constraint on $\epsilon_1$, $\epsilon_2$ and $\epsilon_3$ with varying $k_{max}$. While $\epsilon_1$ is relatively stable and seems to be converged, the constraints for $\epsilon_2$ and $\epsilon_3$ vary rapidly with $k_{max}$. This, together with the fact that the reduced $\chi^2$ is smaller than 1 up to $\km \approx 0.22 \ihMpc$ (see Fig.~\ref{fig:chi2fit}) suggests that we are overfitting the data. Let us stress that even with three free parameters the improvement of the bispectrum fit up to $\km \approx 0.22 \ihMpc$ is the same that is obtained through the zero-parameter fit discussed earlier. Also, in the zero- and one-parameter scenarios the problem of overfitting is either absent or less severe than in the case of three fitting parameters.

For the three-parameter case, we show the performance of the best fit free parameters with respect to the measured bispectra in Fig.~\ref{fig:ratios} for three different configurations. Remarkably, the SPT prediction for the bispectrum performs fairly well even beyond $\km \approx 0.1\; \ihMpc$. This leaves only very little room for the EFT contribution. One can clearly see that the latter is dominated by the part whose amplitude is fixed by the power spectrum, namely $B_{c21}$ and the contributions from $\bar{F}_2^{\alpha \beta} $ and $\bar{F}_2^\delta$. As in the zero- and one-parameter scenarios, this shows us again that the improvement in the fitting range compared to SPT is driven by the LO parameter $\gamma$. 

Finally, we note that changing the time dependence of the stress tensor coefficients hardly affects the goodness of fit. Both the time dependence of the divergence $m=1$ and the one inspired by self similarity $m=5/3$ lead to equally good results. Thus is due to the flatness of the ratios of the $g$ functions in Eq.~\eqref{e:Falphabeta}.

\begin{table}
\centering
\begin{tabular}{c|ccccc}
scenario & $\epsilon_1$ & $\epsilon_2$ & $\epsilon_3$ & $\gamma$ & $m$\\
\hline
\hline
1 parameter &---&---&---&$1.40\pm0.04$& $1$\\
1 parameter &---&---&---&$1.36\pm0.04$& $5/3$\\
\hline
3 parameters &  $0.31\pm 0.09$ & $1.89\pm 0.48$ & $1.29\pm 0.43$ & $1.50 \pm 0.03$ & $1$\\ 
4 parameters &  $0.95\pm 2.32$ & $2.21\pm 1.23$ & $1.59\pm 1.18$ & $1.09\pm 1.46$ & $1$\\ 
\hline 
3 parameters &  $0.23\pm 0.09$ & $1.87\pm 0.48$ & $1.16\pm 0.43$ & $1.50 \pm 0.03$ & $5/3$\\ 
4 parameters &  $0.89\pm 2.40$ & $2.19\pm 1.26$ & $1.49\pm 1.30$ & $1.09\pm 1.46$ & $5/3$\\ 
\hline
\hline
\end{tabular}
\caption{Parameter constraints for the fitting scenarios considered in the text. All the parameters are quoted in units of length squared ($\hMpcsq$) and contain all the time integration factors. The parameters at the level of the equation of motion can be computed using the conversion factors $e_i=\epsilon_i (m+2)(2m+9)\cH_0^2\Omega_m^0/2D_1 $ and $d^2=\gamma(m+1)(2m+7)\cH_0^2\Omega_m^0/2D_1 $.}
\label{tab:solconst}
\end{table}

\subsubsection{Four parameters}
For our most conservative fit we neglect the connection between the power spectrum and bispectrum and let $\gamma$ float in addition to the new second order parameters, thus we use the bispectrum to fit for $\gamma$ and $\epsilon_1,\epsilon_2,\epsilon_3$. The resulting reduced $\chi^2$ in Figs.~\ref{fig:chi2fit} and \ref{fig:chi2} does not show any considerable  improvement over the previously discussed scenarios. The constraints on the parameters in Tab.~\ref{tab:solconst} are extremely indecisive showing marginal detections of $\epsilon_2$ and $\gamma$ but no detection of non-zero $\epsilon_1$ and $\epsilon_3$. Except for $\epsilon_2$ and $\epsilon_3$ all the parameters are perfectly anti-correlated or correlated (cross correlation coefficient $r>0.9$). After noting that the constraints are consistent with the previously discussed constraints, in particular the quite tight constraint on $\gamma$ from the power spectrum, we conclude that the bispectrum does not provide sufficient leverage to constrain this amount of freedom in the EFT shape. That is to say that the more realistic scenarios considered above are sufficient to describe the bispectrum up to $\km\approx 0.22\, \ihMpc$.

\subsection{Higher order SPT contributions}
\begin{figure}
\centering
\includegraphics[height=7.5cm]{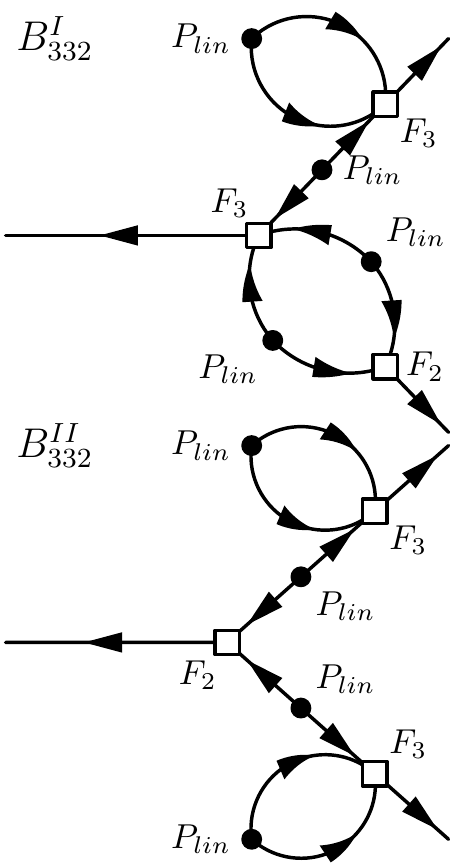}
\includegraphics[height=7.5cm]{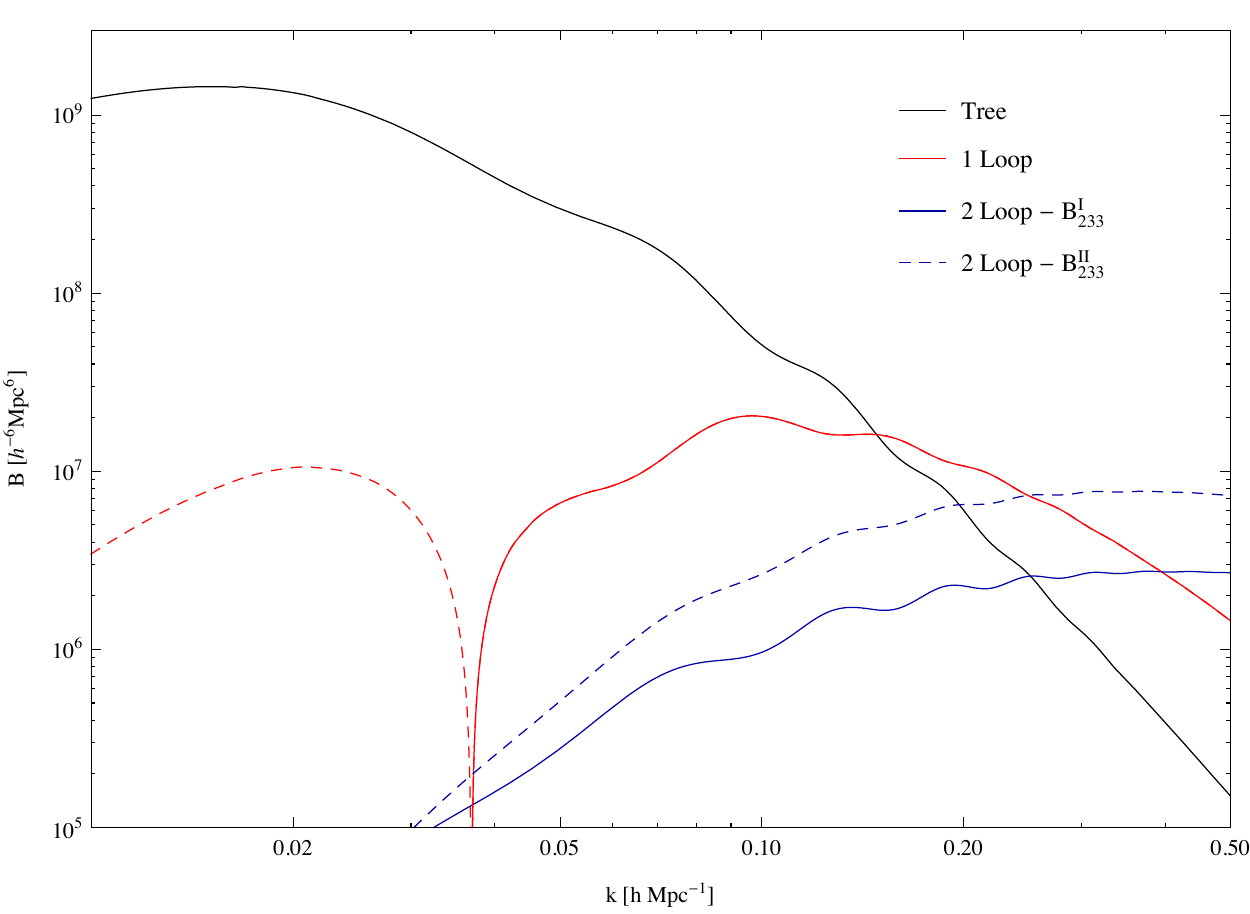}
\caption{Two-loop contributions to the equilateral bispectrum: \emph{Left panels: }Diagrams for the reducible diagrams considered in Eqs.~\eqref{eq:b332i} and \eqref{eq:b332ii}. \emph{Right panel:} Scale dependence of the reducible two-loop bispectra in comparison to the tree level and one-loop bispectrum. }
\label{fig:2loop}
\end{figure}

As discussed in Eq.~\eqref{eq:contribcount} we expect two-loop SPT bispectra to scale as $(k/k_\text{NL})^6$ and to be the next relevant contribution once the one-loop SPT and EFT terms have been considered. Without an explicit calculation of the two loop terms we are not able to give much more than this estimate on their scaling with $k$. Further considerations, such as the counting of free factors of $2\pi$ could be misleading since there are very subtle cancellations between the diagrams contributing to the two-loop bispectrum. With this warning in mind, let us still discuss a simple, i.e. reducible two-loop SPT bispectrum that can be obtained by replacing the lower leg in $B_{321}^{II}$ in Fig.~\ref{fig:biloopdiag} by the upper one
\begin{equation}
B_{332}^{I}=2 F_2(\bm k_1,\bm k_2) \frac{P_{13}(k_1)}{2}\frac{P_{13}(k_2)}{2}+2\, \text{cyc.}
\label{eq:b332i}
\end{equation}
as shown in the upper left panel of Fig.~\ref{fig:2loop}.
This terms scales as $k_1^2 k_2^2 \sigma_v^4 P(k_1)P(k_2) F_2(\bm k_1,\bm k_2)$ both in the IR and the UV, i.e., it has the same shape as the next term in the EFT expansion, which would be $k^4 P^2$.  Another easily computable bispectrum shape is given by replacing the free $P(k)$ in $B_{321}^{I}$ by a $P_{13}$, i.e., we calculate
\begin{equation}
B_{332}^{II}  = 6 \frac{P_{13}(k_3)}{2} \intq F_3(-\qv, \qv - \kv_2, -\kv_3 ) F_2(\qv, \kv_2 - \qv) \, \Pl(q) \Pl(|\qv -\kv_2|) + 5\, \text{perm.}
\label{eq:b332ii}
\end{equation}
as shown in the lower left panel of Fig.~\ref{fig:2loop}.
We show the scale dependence of these terms in the left panel of the same Figure. While they might be cancelled by other two-loop terms in a configuration dependent way, we note that their amplitude roughly equals the amplitude of the linear power spectrum at $k\approx 0.2\; \ihMpc$ for the equilateral bispectrum. This provides us with another indication that $k_{max}\approx 0.2\; \ihMpc$ is a reasonable scale up to which the one-loop SPT+EFT should describe the bispectrum.

%

\section{Conclusions}\label{concl}

In this paper we considered the bispectrum in the EFTofLSS. We first revisited the one-loop SPT contributions to the bispectrum with a particular emphasis on the UV-limit of the loop integrals. At this order in perturbation theory, we need to go beyond the leading order expansion of the effective stress tensor and include also terms that contain the product of two fields. As has been discussed in the literature, at this order the question of non-locality in time becomes important. We showed that the EFTofLSS can be treated as being local in time as long as perturbation theory is applicable. The effective stress tensor can therefore be expanded in derivatives and powers of fields evaluated at the same space-time position (spacial non-localities arise at higher orders). This means that there are only four free parameters in the theory at this order of which one can be determined through the one-loop power spectrum. 

We computed the one-loop SPT as well as the the EFT contributions to the bispectrum. The one-loop integrals are evaluated using an IR-safe integrand which we derived in Appx.~\ref{a:IRsafe}. The EFT part of the bispectrum receives contributions from the three new operators in the effective stress tensor, from the expansion of the LO term $\bigtriangleup \delta$, and from the non-linear corrections to the leading order EFT solution. While the former come with new fitting parameters, the latter two are functions of the leading order parameter. It is important to note that even at NLO, we cannot break the degeneracy of the two parameters $c_s^2$ and $c_v^2$. As shown, this is due to the fact that one can absorb the difference between the SPT kernels $F_2$ and $G_2$ into the next-to-leading order operators. We discussed the time dependence of the free parameters and showed that there is a residual time dependence in the EFT contribution (through the parameter $m$) from the time integrals over the Green's function that cannot be absorbed in a redefinition of the fitting parameters. 

In Sec.~\ref{s:eftbispectrum} we discussed the renormalization procedure. We checked that indeed the NLO counterterms allow for a full cancellation of the all possible UV-divergences of one-loop integrals. The renormalizations of the bispectrum and of the one-loop power spectrum are consistent, meaning that the renormalization of the power- and bispectrum yield the same numerical value for the leading order counterterm. We want to stress that without including the leading order counterterm, there is no cancellation of the UV-divergences. This is a non-trivial check. The renormalization of the one-loop bispectrum is fully consistent assuming an effective stress tensor that is local in time.

We measured the bispectrum in numerical simulations at $z=0$ and compared it with the theoretical prediction for the bispectrum in the EFTofLSS. In Figs.~\ref{fig:chi2fit} and \ref{fig:chi2} we showed the reduced $\chi^2$ that results form fitting the free parameters of the EFTofLSS to the simulation data. The EFTofLSS contains four fitting parameters and one can either fit all of them only using the bispectrum or use the one-loop power spectrum to determine the leading-order parameter. We also considered two scenarios where we used the relations among the parameters that are induced by the cancellation of the UV-divergences to reduce the number of fitting parameters. It is possible to fix the three next-to-leading order parameters in terms of the leading order parameter (see Eqs.~\eqref{e:e123values} and \eqref{magic}). In particular, if we determine the leading-order parameter through the power spectrum, we get a prediction for the bispectrum that has no free parameters. It is remarkable that this zero-parameter prediction has a $\chi^2$ that is very similar to the other scenarios where we fit one or more parameters. While the one-loop SPT bispectrum prediction fails to agree with the simulations at a scale of $k \approx 0.13 \, h \mbox{Mpc}^{-1}$, the $\chi^2$ of the EFTofLSS prediction becomes large only at a scale of $k \approx 0.22 \, h \mbox{Mpc}^{-1}$.  

In this paper we compared the results from the EFTofLSS with SPT. However, there are other methods on the market which one could consider. In Refs.~\cite{Bernardeau2008,Bernardeau2010,Bernardeau2012}, the bispectrum has been studied within the context of RPT. The authors of Ref.~\cite{Bernardeau2008} provide a comparison with data from numerical simulations showing the reduced bispectrum as a function of the angle between $k_1$ and $k_2$ while fixing them at  $k_1 = 0.068 \, \ihMpc$ and $k_2 = 0.136 \, \ihMpc$. Unfortunately, the authors of Ref.~\cite{Bernardeau2008} do not compute the $\chi^2$ as in our Figs.~\ref{fig:chi2fit} and \ref{fig:chi2} and it is hard to tell at which wave number RPT stops giving a good prediction of the numerical data. This makes a thorough comparison with between the results from the EFTofLSS and those from RPT (as well as other methods that go beyond SPT) difficult but not less interesting.

As an outlook for future work, we sketched a diagrammatic formulation of the EFTofLSS that allows for a better organization and understanding of the renormalization procedure in the EFTofLSS. It would be good to find a procedure that allows for the determination of the counterterms at any given order. We think that a more thorough understanding of how the LO counterterms are propagated to the higher order solutions is a key ingredient for such a construction. Also, it would be interesting to include the power spectrum at two loops and to perform a joint fit.

\section*{Acknowledgements}
We would like to thank Guido D'Amico, Roman Scoccimarro and Matias Zaldarriaga for useful discussions and comments on the manuscript. We would like to thank the authors of \cite{Sefusatti2010} M. Crocce, V. Desjacques and E. Sefusatti  for providing their bispectrum measurements, which we used to validate our code. We thank also L.~Senatore for coordinating the simultaneous submission our papers on the bispectrum in the EFTofLSS.
T.B. gratefully acknowledges support from the Institute for Advanced Study through the W.~M.~Keck Foundation Fund. L.M. is supported by a grant from the Swiss National Science Foundation. M.M. acknowledges support by NSF Grant PHY-1314311. E.P. is supported in part by the Department of Energy grant DE-FG02-91ER-40671.

\appendix

\section{Equivalence principle and the structure of EFTofLSS\label{s:equiv}}

The effective theory of large scale structure is obtained as a Newtonian fluid approximation to general relativity plus dark matter. As such it inherits a subset of symmetries from general relativity. The goal here is to identify them and see their implications for the structure of renormalized theory. The residual symmetry can be seen by inspecting the fluid and Poisson equations \eqref{Dtn} and \eqref{e:EOMrealspace}. The equations of motion are invariant under time-dependent boosts \cite{Kehagias,Peloso}
\be
\label{n}
\xv \to \xv' \eeq \xv+\n(\tau),
\ee
if the fields are shifted according to
\be\label{boost}
\delta_{n}(\xv,\tau)&\eeq& \delta(\xv',\tau) \,,\\
\vv_{n}(\xv,\tau)&\eeq&\vv(\xv',\tau)-\dot{\n}(\tau)\,,\\
\phi_{n}(\xv,\tau)&\eeq &\phi(\xv',\tau)+\xv\cdot \left(  \cH \dot{\n}(\tau)+\ddot{\n}(\tau)\right)\,,
\ee
where the label ``$n$'' stands for ``new'' and we introduced primed spatial coordinates for convenience. 
%
$\n(\tau)$ is an arbitrary function of time and over-dot denotes $\d/\d\tau$. This is of course the equivalence principle which relates the acceleration $\ddot{\n}$ of the observer to the gravitational force $\nabla \phi$; a residual gauge symmetry of the Newtonian gauge-fixing in general relativity. Despite the gauge freedom the initial value problem in our fluid-gravity system is well-defined because the coordinate transformation does not vanish at infinity. Applying an arbitrary boost to a SPT solution would generate an unphysical solution. However, a subclass of boosts do generate new solutions: those for which the non-linear shifts of $\vv$ and $\phi$ can be mimicked locally by a long-wavelength physical perturbation. 

To find this subclass in EdS, notice that the finite-$k$ solutions with zero vorticity satisfy at linear order
\be
2\phi=-3\cH u,
\ee
where $u$ is the velocity potential $\vv = \nabla u$. Taking $u =-\xv \cdot \dot{\n}$ and $\phi = \xv \cdot(\ddot{\n}+\cH \dot{\n})$ we find
\be
\dot{\n}(\tau)= \left(\frac{a}{a_{in}}\right)^{1/2} \dot{\n}_{in}.
\ee
These physical solutions that are locally equivalent to a coordinate transformation are the adiabatic modes in the Newtonian approximation, Ref.~\cite{Weinberg2008}. \footnote{The non-trivial transformation with non-zero $\dot{\n}_{in}\eq \qv$ corresponds to the gauge transformation
\begin{equation}
\begin{split}
&\tau\to (1+2 \qv \cdot \xv)\tau\\[1.5ex]
&\xv \to (1-10 \qv \cdot \xv )\xv + 5  \qv x^2 + \qv \tau^2,
\end{split}
\end{equation}
first studied in \cite{Creminelli}.}

\subsection{Coarse graining}

Cut-off regularization (namely cutting momentum integrals off at $ k=\Lambda $) preserves this symmetry. This is obvious at least for the subclass of Galilean boosts ($\dot{\n} = \rm const.$) on physical terms since basically every description and physical theory of the world around us follows a coarse graining procedure, the real-space equivalent to momentum cut-off. If this invalidated the symmetry, we would have needed to account for the non-invariance in our theories. 

Let us define the smoothed fields using a window function $W_\Lambda(\yv)$ with a characteristic size of $1/\Lambda$ and such that $\int d^3y\; W_\Lambda(\yv) = 1$
\be
\label{long}
X_l(\xv) \equiv [X(\xv)]_\Lambda = \int d^3y\;  W_\Lambda(\yv) X(\xv + \yv),
\ee
and the short-scale field as
\be
X_s(\xv)\equiv X(\xv)-X_l(\xv).
\ee
From \eqref{long} we see that smoothed fields inherit the boost symmetry property of the original field, simply because the boost transformation is at most linear in fields and the boost parameter $\n$ is spatially uniform. Suppose the original system is invariant under boosts. This implies that given a solution $\{\delta(\xv,\tau),\vv(\xv,\tau)\}  $ there is another solution of the same equation given by  $  \{\delta_n(\xv,\tau),\vv_n(\xv,\tau)\} =\{\delta(\xv',\tau),\vv(\xv',\tau)-\dot{\n}\}$ for the above mentioned subclass of boosts. Smoothing the two solutions implies that given a smoothed solution  $\{\delta_{l}(\xv,\tau),\vv_{l}(\xv,\tau)\}  $ of the coarse grained system, there is a new solution $  \{\delta_{l,n}(\xv,\tau),\vv_{l,n}(\xv,\tau)\} =\{\delta_{l}(\xv',\tau), \vv_{l}(\xv',\tau)-\dot{\n}\}$. 

One can also directly verify that the fluid equations remain boost symmetric under smoothing. Let us explicitly apply it ot the expression $(\d_\tau+\vv \cdot \nabla)\delta$:
\be
[(\d_\tau+\vv \cdot \nabla)\delta]_\Lambda=\d_\tau \delta_l +[\vv \cdot \nabla \delta]_\Lambda.
\ee
In the second term decompose $\vv = \vv_l+\vv_s$
\begin{equation}
\begin{split}
[\vv_l\cdot \nabla \delta]_\Lambda =& \vv_l \cdot \nabla \delta_l + \nabla_j v_l^i \nabla_i\nabla_j \delta_l
+\cdots\\
[\vv_s\cdot \nabla \delta_l]_\Lambda =& [\vv_s\cdot \nabla \delta_s]_\Lambda+\O\left(\frac{1}{\Lambda^2}\nabla^2 \vv_l \cdot \nabla \delta_l\right),
\end{split}
\end{equation}
where we used the general property of smoothing $[X_l Y_s]_\Lambda \sim X_l (\nabla^2/\Lambda^2) Y_l+\cdots$ with all terms containing at least two derivatives acting on $Y_l$. Summed with $\d_\tau \delta_l$ we get an expression which is invariant under boost transformation of the smoothed quantities. What remains is $[\vv_s\cdot \nabla \delta_s]_\Lambda$ which depends on the long modes only through tidal forces, if there is no initial correlation between the long and short modes (i.e. no primordial non-Gaussianity). So it is by itself invariant under boosts except for the common shift of coordinates \eqref{n}.


\subsection{Boost invariance of non-local-in-time terms\label{s:nonlocal}}

Consider a non-local-in-time source term like
\be\label{nl}
\int_{0}^{\tau} d\tau' \,K(\tau,\tau')\,\delta(\xv_{fl}[\xv,\tau;\tau'],\tau')\,,
\ee
where $ \xv_{fl}[\xv,\tau;\tau']$ is the position occupied at time $ \tau' $ by the element of fluid that at time $ \tau $ occupies the position $ \xv $. The implicit expression is 
\be
\xv_{fl}[\xv,\tau;\tau'] \eq \xv -\int_{\tau'}^{\tau}d\tau''\, \vv (\xv_{fl}[\xv,\tau;\tau''],\tau'')\,,
\ee
where one verifies that $ \xv_{fl}[\xv,\tau;\tau] =\xv $ as expected. We want to check that this term is invariant under a boost transformation of Eqs.~\eqref{n} and \eqref{boost}. More specifically, invariant means that the new fields evaluated at some point in the old coordinates are equal to the old fields at the same point written in terms of new coordinates. For example
\be
\left(  \partial_{\tau}+\vv_{n}(\xv,\tau)\cdot \nabla \right)\delta_{n}(\xv,\tau) \eq \left(  \partial_{\tau}+\xv(\xv',\tau)\cdot \nabla' \right)\delta(\xv',\tau)\,,
\ee
which is usually referred to as the covariance of the equations of motion. Rewriting \eqref{nl} in terms of the new fields one gets
\be
\eqref{nl}&\rightarrow& \int_{0}^{\tau} d\tau' \,K(\tau,\tau')\,\delta_{n}(\xv_{fl,n}[\xv,\tau;\tau'],\tau')\\
&=& \int_{0}^{\tau} d\tau' \,K(\tau,\tau')\,\delta(\xv_{fl,n}[\xv,\tau;\tau']+n(\tau'),\tau')\,,\label{nl2}
\ee
where $ \xv_{fl,n} $ is given by the same expression as $\xv_{fl}$ but with all fields replaced with the ``new'' fields. One can then compute
\be
\xv_{fl,n}[\xv,\tau;\tau']& \eq & \xv-\int_{\tau'}^{\tau}d\tau''\, \vv_{n}(\xv_{fl,n}[\xv,\tau;\tau''],\tau'')\\
&\eq & \xv-\int_{\tau'}^{\tau}d\tau''\, \vv(\xv_{fl,n}[\xv,\tau;\tau'']+\n(\tau''),\tau'')-\dot{\n}(\tau'')\\
&\eq&\xv+\n(\tau)-\n(\tau')-\int_{\tau'}^{\tau}d\tau''\, \vv(\xv_{fl,n}[\xv,\tau;\tau'']+n(\tau''),\tau'')\,,
\ee
which is solved by
\be
\xv_{fl,n}[\xv,\tau;\tau'] \eq \xv_{fl}[\xv+\n(\tau),\tau;\tau']-\n(\tau').
\ee
Plugging this into \eqref{nl2}, one finds
\be
\int_{0}^{\tau} d\tau' \,K(\tau,\tau')\,\delta(\xv_{fl}[\xv+\n(\tau),\tau;\tau'],\tau')\,,
\ee
which is the same as our starting expression \eqref{nl}, but at the new position $\xv' = \xv+\n(\tau)$.

\section{IR-safe integrand} \label{a:IRsafe}

Let us discuss the IR regime of the one-loop integrals, i.e.~the situation when the loop momentum $\qv$ is smaller or equal to the external momenta. In the case of power-law initial conditions, $\Pl \propto k^n$, all four one-loop integrals of Eqs.~\eqref{e:b222}, \eqref{e:b321i}, \eqref{e:b321ii} and \eqref{e:b411} are affected by IR-divergences for $n<-1$. However, all these IR-divergences cancel exactly in equal time correlators. In Refs.~\cite{Jain1996,Scoccimarro1996} it was shown that this cancellation is a consequence of the Galileo invariance of the equations of motion and holds for  $-3< n < -1$ to all orders in perturbation theory. This means that when summing all one-loop integrals, all their IR-divergences cancel. Despite having concluded that we do not have to worry about IR-divergences in the full bispectrum, there is still a technical point that deserves further considerations. 

When evaluating the one-loop integrals numerically, it is unfavourable to compute the integrals separately and summing them up only afterwards to get the final result. The problem is that we may run into the situation of having to subtract large numbers of similar size in order to get a much smaller final result. The absence of IR-divergences induces large cancellations which might severely affect the precision of the numerical computation. The authors of Refs.~\cite{Blas2013a,Carrasco2013} therefore derived an explicitly IR-safe integrand for the one- and two-loop power spectrum where the IR-divergences cancel at the integrand level. Although the cancellations among the single diagrams seem to be less severe for the bispectrum than in the case of the power spectrum (see Fig.~\ref{fig:bloop}), we shall derive the analogous IR-safe integrand for the one-loop bisepctrum which we have used in the computations of our theoretical predictions.

The one-loop integrals in Eqs.~\eqref{e:b222}, \eqref{e:b321i}, \eqref{e:b321ii} and \eqref{e:b411} all diverge for $\qv \rightarrow 0$, but there are also divergences as the loop momentum takes the value of an external momentum, i.e.~in the limit $\qv \rightarrow \kv_2$ in $B_{222}$ and $B_{321}^I$ as well as for $\qv \rightarrow -\kv_1$ in $B_{222}$. Roughly speaking, the IR behaviour of the integrands is given by

\begin{equation}
\begin{split}
& \qv \rightarrow 0 \; : \qquad  q^2 b_{222}  \; {\sim} \; q^2 b_{321}^I  \; {\sim} \; q^2b_{321}^{II} \; {\sim} \; q^2b_{411} \; {\sim} \; q^n  \;, \\[1.5ex]
& \qv \rightarrow \kv_2 \; : \qquad  q^2 b_{222} \; {\sim} \; q^2 b_{321}^I \; \sim \; |\qv - \kv_2|^n \;, \\[1.5ex]
& \qv \rightarrow -\kv_1 \; : \qquad  q^2 b_{222}  \; {\sim} \;  |\qv +\kv_1|^n \;,
\end{split}
\end{equation}
where we always assumed a linear power spectrum of the form $\Pl \propto k^n$ and explicitly included the $q^2$ term that comes from the integral measure. Lower case letters denote the integrands of the corresponding diagram but without the permutations. Apart from this leading IR-divergences, $b_{222}$ and $b_{321}^I$ are also affected by subleading divergences that scale as $\sim q^{n+1}$. 

As a first step, let us check that all IR-divergences really cancel when the full one-loop bispectrum is considered. To this end, we must be a bit more specific about the divergences, in particular about their shape. One can check that all IR-divergences have same form

\begin{equation}
\mathcal{D}(\qv, \kv_i, \kv_j; \kv_l, \kv_m ) \eq  \frac{\qv \cdot \kv_l \, \qv \cdot \kv_m}{\qv^2} \, F_2(\kv_i, \kv_j) \Pl(q) \Pl(k_i) \Pl(k_j) 
\end{equation}
%
%
In the limit $\qv \rightarrow 0$, the integrands have the following shapes

\begin{eqnarray}
b_{222}& = & -2 \mathcal{D}_\textit{IR} (\qv, \kv_1, \kv_2; \kv_1, \kv_2 ) +\cO{q^{n+1}} \;, \label{e:b2220} \\[1.5ex]
b_{321}^I  & = &  \mathcal{D}_\textit{IR} (\qv, \kv_2, \kv_3; \kv_2, \kv_2 ) +  \mathcal{D}_\textit{IR} (\qv,\kv_2, \kv_3; \kv_2, \kv_3 ) +\cO{q^{n+1}} \;, \\[1.5ex]
b_{321}^{II}  & = &  - \mathcal{D}_\textit{IR} (\qv,\kv_2, \kv_3; \kv_3, \kv_3 ) +\cO{q^{n+2}} \;, \\[1.5ex]
b_{411}  & = &  -\mathcal{D}_\textit{IR} (\qv,\kv_2, \kv_3; \kv_2, \kv_2 ) - 2  \mathcal{D}_\textit{IR} (\qv,\kv_2, \kv_3; \kv_2, \kv_3 ) - \mathcal{D}_\textit{IR} (\qv,\kv_2, \kv_3; \kv_3, \kv_3 ) \nonumber 	\\
&&  +\cO{q^{n+2}} \;.
\end{eqnarray}
The other three IR-divergences in the limits $\qv \rightarrow -\kv_1$ and $\qv \rightarrow \kv_2$ give a contribution of

\begin{eqnarray}
b_{222} & = & -2 \mathcal{D}_\textit{IR} (-\kv_1 -\qv, \kv_1, \kv_3; \kv_1, \kv_3 ) + \cO{|\kv_1 + \qv|^{n+1}} \;, \label{e:b222k1} \\[1.5ex]
b_{222} & = & -2 \mathcal{D}_\textit{IR} (\kv_2-\qv, \kv_2, \kv_3; \kv_2, \kv_3 ) + \cO{|\kv_2 - \qv|^{n+1}} \;, \label{e:b222k2} \\[1.5ex]
b_{321}^I & = & \mathcal{D}_\textit{IR} (\kv_2-\qv, \kv_2, \kv_3; \kv_2, \kv_2 ) +  \mathcal{D}_\textit{IR} (\kv_2-\qv,\kv_2, \kv_3; \kv_2, \kv_3 )  \nonumber \\
&& + \cO{|\kv_2 - \qv|^{n+1}}\;.
\end{eqnarray}
Including all permutations, it is then easy to check that all the IR-divergence add up to zero as expected. The subleading divergences in $b_{222}$ and $b_{321}^I$ that scale as $\sim q^{n+1}$ are canceled through the angular integration, leaving us with a IR-safe bispectrum for $n>-3$.

Next, let us find the IR-safe integrand by following the procedure outlined in Ref.~\cite{Carrasco2013}. The idea is fairly simple: first, one maps the $\qv \rightarrow \pm \kv_i$ divergences in $b_{222} $ and $b_{321}^I$ onto a $\qv \rightarrow 0$ divergence and reduce the integration limits in order to exclude this kind of divergences (using $\Theta$-functions). This will leave us with integrals that diverge only for $\qv \rightarrow0$. Then, we must sum all integrands before performing the integration in order to achieve the cancellation of the leading IR-divergences at the integrand level. To cancel the subleading divergences, the integrand has to be symmetrized.

The $b_{321}^{II}$ and $b_{411}$ integrands do not need any specific treatment. However, the integrands $b_{222}$ and $b_{321}^I$ need to be put in a more convenient form. Let us start with $b_{321}^I $ which is somewhat simpler. We can simply split the integration region into two parts and shift the integration variable in order to map the $\qv \rightarrow \kv_2$ divergence of $b_{321}^I $ into $\qv \rightarrow 0$

\begin{equation}
\begin{split}
\intq \tilde{b}_{321}^I  \defi & \int_{q<|\kv_2 - \qv|} b_{321}^I(\qv,\kv_2,\kv_3) + \int_{q \geq |\kv_2 - \qv|} b_{321}^I (\qv,\kv_2,\kv_3) + \mbox{5 perm.} \\[1.5ex]
 \eq &  \intq \Big\{  b_{321}^I (\qv,\kv_2,\kv_3)\,  \Theta(|\kv_2 - \qv| -q) +  b_{321}^I(-\qv,\kv_2,\kv_3) \, \Theta(|\kv_2 + \qv| -q) \Big\} \\[1.5ex]
 & + \mbox{5 perm.} 
\end{split}
\end{equation}
We used the fact that after shifting the integration variable we can rewrite the integrand as $b_{321}^I$ itself due to the symmetry of the $F_{2,3}$ kernels. Finally, we symmetrized the integrand in order to explicitly cancel the subleading IR-divergence

For the $B_{222}$ diagram, the situation is a bit more delicate since there are now two IR-divergences that need to be mapped onto a $\qv \rightarrow 0$ divergence. We found it most convenient to undo all $\delta$-functions of the momentum conservation and to start with the full form of the correlator. 

\begin{equation}
(2\pi)^3 \dirac(\kv_1 + \kv_2 + \kv_3) \, B_{222} \eq \langle \, \prod_{i=1}^3 \intqn{i} F_2(\kv_i -\qv_i, \qv_i) \dn{1}(\kv_i -\qv_i) \, \dn{1}(\qv_i) \, \rangle \;.
\end{equation}
The IR-divergences arise when one of the arguments in the $F_2$ kernels goes to zero. Hence, we need to focus on restricting the integration region such that the $\kv_i - \qv_i$ argument in the kernels does not reach zero.
The integrals are symmetric around $|\kv_i - \qv_i| = q_i$, i.e.~we can cut the integration region in half by introducing $\Theta$-functions 

\begin{equation}
\begin{split}
\intqn{i} F_2(\kv_i -\qv_i, \qv_i) \dn{1}(\kv_i -\qv_i) \, \dn{1}(\qv_i) \eq & 2 \intqn{i} F_2(\kv_i -\qv_i, \qv_i) \dn{1}(\kv_i -\qv_i) \, \dn{1}(\qv_i) \\
& \phantom{ \intq} \cdot \Theta(|\kv_i - \qv_i| - q_i) \;. 	
\end{split}
\end{equation}
We could do this for all three $\qv_i$ integrals. However, it would not serve our purpose of getting an explicitly IR-safe integrand which is why we choose to introduce only two $\Theta$-functions, e.g.~for $\qv_1$ and $\qv_2$, rather than for all $\qv_i$. Next we contract all the density fields inside the correlator, integrate out resulting $\delta_D$-functions and symmetrize the integrand. This gives us

\begin{equation}
\begin{split}
\intq \tilde{b}_{222} \defi & \frac{1}{2} \intq \Big\{ \Big[ b_{222}(\qv, \kv_1 ,\kv_2) \, \Theta(|\kv_1+\qv|-q) \Theta(|\kv_2 - \qv |-q) \\[1.5ex]
 & +  b_{222}(-\qv, \kv_1 ,\kv_2) \, \Theta(|\kv_1 - \qv|-q) \Theta(|\kv_2 + \qv|-q) \Big] \\[1.5ex]
 & + \Big[\kv_1 \leftrightarrow \kv_3\Big] + \Big[\kv_2 \leftrightarrow \kv_3\Big] \, \Big\} \;.
\end{split}
\end{equation}
It is now obvious that due to the $\Theta$-functions the integral $\intq \tilde{b}_{222}$ has only IR-divergences in the limit $\qv \rightarrow 0 $. But the shape of this remaining IR-divergence is exactly the sum of the divergences we had before for $\qv \rightarrow \{0,-\kv_1, \kv_2\}$ in Eqs.~\eqref{e:b2220}, \eqref{e:b222k1} and \eqref{e:b222k2}.

Hence, we can write the full one-loop the bispectrum in SPT in an IR-safe form where all IR-divergences cancel exactly at the integrand level

\begin{equation}
\begin{split}
{B}_\textit{SPT} \eq & B_{112} + \intq  \Big\{ \tilde{b}_{222} + \tilde{b}_{321}^I + \big[ b_{321}^{II} + \mbox{5 perm.} \big] + \big[ b_{411} + \mbox{2 cycl. perm.} \big] \Big\} \;. 
\end{split}
\end{equation}


\addcontentsline{toc}{section}{References}

\bibliographystyle{JHEP}

\bibliography{Bispectrum_Bibliography.bib}

\end{document}